\documentclass{aa}
\usepackage{natbib}
\bibpunct{(}{)}{;}{a}{}{,}
\usepackage{epstopdf}
\usepackage{epsfig}
\usepackage{subfig}
\usepackage{url}
\usepackage{rotating}
\usepackage{lscape}
\usepackage{txfonts}
\usepackage{natbib,twoopt}
\usepackage[breaklinks=true]{hyperref} 
\bibpunct{(}{)}{;}{a}{}{,} 
\newcommandtwoopt{\citeads}[3][][]{\href{http://adsabs.harvard.edu/abs/#3}%
{\citealp[#1][#2]{#3}}}
\newcommandtwoopt{\citepads}[3][][]{\href{http://adsabs.harvard.edu/abs/#3}%
{\citep[#1][#2]{#3}}}
\newcommandtwoopt{\citetads}[3][][]{\href{http://adsabs.harvard.edu/abs/#3}%
{\citet[#1][#2]{#3}}}
\newcommandtwoopt{\citeyearads}[3][][]%
{\href{http://adsabs.harvard.edu/abs/#3}{\citeyear[#1][#2]{#3}}}

\begin{document}

\title{Spectral properties of the largest asteroids associated with Taurid Complex}
 
\author
{
 M. Popescu\inst{1,2}
\and
 M. Birlan\inst{1}
\and
 D. A. Nedelcu\inst{2}
\and
 J. Vaubaillon\inst{1}
\and
C. P. Cristescu \inst{3} 
 }
\offprints{M. Popescu, \email{mpopescu@imcce.fr}}
\institute
{
Institut de M\'ecanique C\'eleste et de Calcul des \'Eph\'em\'erides (IMCCE) CNRS-UMR8028, Observatoire de Paris, 77 avenue Denfert-Rochereau, 75014 Paris Cedex, France\\
\and
Astronomical Institute of the Romanian Academy, 5 Cu\c{t}itul de Argint, 040557 Bucharest, Romania\\
\and
Department of Physics, University Politehnica of Bucharest, Romania
}       


\abstract
{
The Taurid Complex is a massive stream of material in the inner part of the Solar System. It contains objects spanning the range of $10^{-6}$--$10^3$m, considered by some authors to have a common cometary origin. The asteroids belonging to Taurid Complex are on  Apollo type orbit, with most of them being flagged as potentially hazardous asteroids. In this context, understanding the nature and the origin of this asteroidal population is not only of scientific interest but also of practical importance.
}
{
We aim to investigate the surface mineralogy of the asteroids associated with Taurid Complex using visible and near-infrared spectral data. Compositional linking between these asteroids and meteorites can be derived based on the obtained spectra.
}
{
We obtained spectra of six of the largest asteroids (2201, 4183, 4486, 5143, 6063, and 269690) associated with Taurid complex. The observations were made with the IRTF telescope equipped with the spectro-imager SpeX. Their taxonomic classification is made using Bus-DeMeo taxonomy. The asteroid spectra are compared with the meteorite spectra from the Relab database. Mineralogical models were applied to determine their surface composition. All the spectral analysis is made in the context of the already published physical data.
}
{
Five of the objects studied in this paper present spectral characteristics similar to the S taxonomic complex. The spectra of ordinary chondrites (spanning H, L, and LL subtypes) are the best matches for these asteroid spectra. The asteroid (269690) 1996 RG3 presents a flat featureless spectrum which could be associated to a primitive C-type object. The increased reflectance above 2.1 microns constrains its geometrical albedo to a value around 0.03.
}
{
While there is an important dynamical grouping among the Taurid Complex asteroids, the spectral data of the largest objects do not support a common cometary origin. Furthermore, there are significant variations between the spectra acquired until now.
}

\keywords{minor planets, asteroids;  techniques: spectroscopic;  methods: observations}

\authorrunning{M. Popescu et al.}
\titlerunning{Spectral properties of six near-Earth asteroids associated with Taurid Complex}
\maketitle
%

\section{Introduction}

The Taurid Complex (hereafter TC) is a massive stream of material in the inner part of the Solar System. It contains objects spanning the range $10^{-6}$--$10^3$m \citepads{1993MNRAS.264...93A}. The incoming dust and particles produce radar and visual meteors, known as the Taurid meteor showers from which the complex derives its name. Referring to large masses, detections, which were associated with Taurids include swarm of meteoroids incident  on the Moon in June 1975 \citepads{1991Icar...91..315O}, many ordinary fireballs, and the Tunguska fireball on June 30, 1908 (e.g. \citeads{1978BAICz..29..129K}). However, \citeads{1998P&SS...46..191S} found important evidence in favor of an asteroidal origin of the Tunguska object. This result was supported by the model of \citeads{1993Natur.361...40C}, which shows that the Tunguska explosion is characteristic for a stony asteroid with a radius of $\approx$30 m entering the Earth's atmosphere at hypersonic velocities. The Farmington  meteorite is also associated with TC. \citeads{1993MNRAS.264...93A} concludes that the objects associated with TC are likely to have a common cometary origin, which can be regarded  as the parent body and also as the source of most of the present day zodiacal complex.

\begin{table*} 
\caption{Log of observations: asteroid designation, date of observation given as Julian Day, the apparent magnitude (V), the phase angle ( $\Phi$), the heliocentric distance (r), the airmass at the mean UT of each observation, the total integration time for each spectrum (ITime), and the corresponding solar analog (S.A.) are presented.}
\label{Circumstances}
\centering
\begin{tabular}{l l c c c c c r}
\hline\hline
Asteroid & Julian Day & V & $\Phi$ ($^\circ$) & r (UA) & Airmass & ITime(s) & S.A. \\ \hline
(2201) Oljato & 2455880.93621 & 16.5 & 4.9 & 1.797 & 1.03 & 1680 & HD 19061  \\
(4183) Cuno & 2455880.90256 & 16.3 & 10.4 & 1.905 & 1.04 & 480 & HD 19061 \\
(4486) Mithra & 2455258.07778 & 14.6 & 55.3 & 1.103 & 1.13 & 1680 & HD 95868 \\ 
(5143) Heracles & 2455880.88450 & 13.4 & 23.5 & 1.326 & 1.15 & 480  & HD 19061 \\
(6063) Jason & 2456556.98773 & 17.7 & 29.0 & 1.632 & 1.15 & 1440 &  HD 224817 \\
(6063) Jason & 2456587.02775 & 14.9 & 25.16 & 1.269 & 1.04 & 1920 & HD 7983 \\
(269690) 1996 RG3 & 2456556.93400 & 17.8 & 3.5 & 1.428 & 1.11 & 1920 & HD 220764 \\ 
(269690) 1996 RG3 & 2456586.86105 & 17.6 & 46 & 1.153 & 1.49 & 480 & HD 215393 \\ 
\hline
\end{tabular}     
\end{table*}

This complex could be constrained by low-inclination (i $<$ 12$^\circ$, semi-major $(a)$ axis having the range 1.8-2.6 a.u., eccentricities $(e)$ in the range of 0.64--0.85, and longitudes of perihelion ($\bar{\omega}$) in the range 100$^\circ$ -- 200$^\circ$ \citepads{1996MNRAS.280..806S}. The Taurid meteor shower has a low inclination (i $<$ 5$^\circ$), and it is seen from the Earth at the epoch of the nodal intersection of the torus of meteoroids and the ecliptic \citepads{2008MNRAS.386.1436B}. Thus, meteoroids from TC interact with the atmosphere near both ascending and descending nodes. These annual periods of interaction revealed by radar and optical phenomena globally between September and December for the pre-perihelion stage and April to July for the post-perihelion stage.

The largest known body of TC is the comet P/Encke which has been regarded as the major source of zodiacal dust \citepads{1952HelOB..41....3W, 2009Icar..201..295W}. For a long time, the association between meteor showers and asteroids was considered with caution. However, the discovery of new asteroid families \citepads{2002Natur.417..720N} in the Main-Belt, new cometary-like asteroids, asteroid outbursts \citepads{2012AJ....143...66J}, or weak cometary activity on NEA (3200) Phaeton \citepads{2013ApJ...771L..36J} changed this opinion. In the case of Taurid meteor shower, the association with asteroids having similar sub-Jovian near-ecliptic orbital elements becomes mandatory mainly because the inclination of comet P/Encke (considered as the parent body) is slightly shifted from the one of Taurid shower.

Several empirical metrics in orbital elements are defined for finding Taurid parent bodies that cluster using asteroids and comets \citepads{1993MNRAS.264...93A, 1999MNRAS.304..743V, 2008EMP..102...73J}. Based on different criteria, like 'D-criterion',  several authors have determined the asteroids belonging to TC \citepads{1993MNRAS.264...93A, 2001A&A...373..329B, 2008MNRAS.386.1436B}.  The size of these asteroids ranges from several kilometers in diameter like (5143) Heracles, (6063) Jason, and (2201) Oljato, down to meter size bodies.

Based on their metrics, \citeads{1993MNRAS.264...93A} clustered 25 asteroids with orbital elements $(a,e,i)$ similar to TC meteoroids. Backward integration based just on gravitational interaction does not allow a convergence of orbits of these bodies toward a unique orbit. The spread in longitude of perihelion for TC is thus interpreted as the chaoticity of orbits due to close encounters with telluric planets  and to the mean-motion resonance 7:2 to Jupiter \citepads{1993MNRAS.264...93A}. 

However, a statistical study of objects based only on orbital elements and backward dynamical integration could not completely solve the TC problem. Indeed, this was expressed by \citeads{1996MNRAS.280..806S}: {\it It is regrettable that colors and hence compositional classification are not yet available for most of the TC asteroids, meaning that there is a dearth of physical evidence to support the dynamical analysis}.

The article presents the results of a near-infrared spectroscopic survey of asteroid members of TC. Physical analysis together with investigation  of their mineralogy bring new insights into the TC cluster. More precisely, spectroscopic observations look into finding similarities in spectral behavior and the derived mineralogical properties which allow quantification of the amount of mineralogical composition of regoliths on the asteroid surface. Globally, this hot topic circumscribes the problem of parent bodies (torus of meteoroids in the inner solar system) associated with the meteor shower. 

The article is organized as follows: Section 2 describes the observing procedure and the data reduction performed to obtain the spectra. Section 3 reviews the methods used to analyze the asteroid spectra. A dedicated discussion is made about merging the visible and the infrared parts of the spectrum. Section 4 presents the spectral results obtained. Each spectrum is analyzed in the framework of the already known physical properties of the asteroid. The discussion of the results, and their implications is made in Section 5. The conclusions end this article. Additional information is given in two annexes in the online material.

\begin{table*}
\caption{Some characteristics of the NEAs studied in this article. The asteroid designations, semi-major axis, aphelion, perihelion, eccentricity, inclination, D parameter, Tisserand Parameter, absolute magnitude, geometric albedo, and taxonomic classification (left: previous classification, right: current work classification) are given. The data was extracted from NEODyS website (\url{http://newton.dm.unipi.it/neodys/}).}
\label{Tab1Prop}
\centering 
\begin{tabular}{l c c c c c c c c c c c}                                                                                                                               
\hline \hline                                                                                                                                                                                                    
Asteroid & a [au] & Q [au] & q [au] & e & i($^\circ$) & D & $T_J$ & H  & pv &\multicolumn{2}{c}{Taxonomic Type}\\ 
 &   &   &   &   &   &   &   &   &   & Previous & This work  \\ \hline
(2201) Oljato & 2.17248 & 3.7201 & 0.6249 & 0.712357 & 2.523 & 0.11 & 3.302 & 16.86 & 0.240 & E, Sq & Q, Sq  \\
(4183) Cuno & 1.98256 & 3.2396 & 0.7255 & 0.634043 & 6.707 & 0.20 & 3.572 & 14.40 & 0.097 & Sq, Q  & Q \\
(4486) Mithra & 2.19961 & 3.6577 & 0.7415 & 0.662906 & 3.04 & 0.17 & 3.337 & 15.60 & 0.297 & S  & Sq \\
(5143) Heracles & 1.83353 & 3.2492 & 0.4179 & 0.772091 & 9.034 & 0.13 & 3.582 & 14.10 & 0.148 & O,Sk & Q  \\
(6063) Jason & 2.21267 & 3.9086 & 0.5167 & 0.766475 & 4.921 & 0.07 & 3.186 & 15.90 & 0.160 & S & Sq  \\
(269690) 1996 RG3 & 2.00001 & 3.2099 & 0.7901 & 0.604961 & 3.571 & 0.20 & 3.587  & 18.50 & - & -  & Cg, Cb \\
\hline                                                                                                                                                                                                               
\end{tabular}                                                                                                   
\end{table*}

\section{The observing procedure and data reduction}

The orbits and the small diameters of the majority of NEAs imply important constraint on the geometries of observations for determining the reflective properties of their surfaces. These conditions are usually satisfied in the case of close approach to Earth, when the apparent magnitude decreases by several units. 

The asteroids studied in this paper were observed with NASA IRTF, a 3m telescope located on the top of Mauna Kea, Hawaii. We used the SpeX instrument in the low resolution Prism mode (R$\approx$100) of the spectrograph, covering the 0.82 - 2.5 $\mu m$ spectral region \citepads{2003PASP..115..362R}.  The observations were performed in remote  mode from Centre d'Observation \`a Distance en Astronomie \`a Meudon (CODAM) in Paris \citepads{2004NewA....9..343B}, and from the Remote Observation Center in Planetary Sciences (ROC) in Bucharest. A 0.8$\times$15 arcsec slit oriented north-south was used. The spectra for the asteroids and the solar analog stars were alternatively obtained on two separate locations on the slit denoted A and B \citep{PopescuUPB2011}. 

Our observations were carried out in four sessions: March 2,2010, November 15,2011, September 21,2013, and October 21,2013. The asteroids (6063) Jason and (269690) 1996 RG3 were observed at different dates with the purpose of monitoring spectral variations due to the heliocentric distance.

We tried to observe all objects as close to the zenith as possible at an airmass smaller than 1.15\footnote{The observation for (269690) 1996 RG3 in October was at an airmass of 1.49 due to weather and schedule constraints.}, as presented in Table~\ref{Circumstances}. Solar analogs in the apparent vicinity of each asteroid were observed for calibration. Thus, HD 95868, HD 19061, HD 224817, HD 220764, and HD 215393 photometric G2V C stars were selected from CDS portal (SIMBAD database)\footnote{\url{http://simbad.u-strasbg.fr/simbad/}} for spectral calibration.

Preprocessing of the CCD images included bias and flat-field  correction. These calibration images were obtained at the beginning or at the end of the observing session. For the wavelength calibration, an Argon lamp spectrum was used.

The data reduction process consists of three steps: 1) obtaining the raw spectra for the object and the solar analog, 2) computation of a normalized reflectance spectrum by dividing the asteroid spectrum by the solar analog spectrum, and 3) performing a correction for telluric  lines. For the first step, the Image Reduction and Analysis Facility - IRAF \citepads{1986SPIE..627..733T} was used, while some IDL routines for the second and third steps were used to diminish the influence of the telluric lines \citepads{2003PASP..115..389V}. For the computation of the normalized reflectance, we took the similar dynamic regimes of the detector \citepads{2004PASP..116..352V, 2003PASP..115..362R}  into account.           
 
The circumstances of observations are presented in Table~\ref{Circumstances}. Generally, the asteroid spectra were obtained by taking images with an integration time of 120s in the nodding procedure for several cycles. The total integration time is given in Table~\ref{Circumstances}. A visual selection of the images was made before introducing them into the data reduction pipeline.

We note that the signal-to-noise ratio (SNR) of spectra is significantly influenced by the precipitable water. For instance, the observations from September 21, 2013 for (269690) 1996 RG3 (at an apparent magnitude 17.8 and in the apparent vicinity of the full Moon) are comparable in terms of the SNR with the data obtained on November 15, 2011 for (4183) Cuno which had the apparent magnitude of 16.3. The zenith opacity at 225 GHz  measured by Caltech Submillimeter Observatory\footnote{\url{http://cso.caltech.edu/tau/}} at the summit of Mauna Kea is related to the precipitable water vapor \citepads{1997Icar..130..387D}. During the observation of 1996 FG3, the value of the zenith opacity was around 0.07 as compared with the value of 0.15 for the time of observations of Cuno.

\section{Methods used to analyze data}

The analysis of spectra was made in the context of previously published physical and dynamical data on these objects. Table~\ref{Tab1Prop} summarizes some physical and dynamical parameters of the objects described in this paper.

We used two online tools to analyze our spectra: 1) M4AST \footnote{\url{http://m4ast.imcce.fr/}} \citepads{2012A&A...544A.130P} for merging visible and near-infrared (NIR) parts of the spectrum, for taxonomic classification, and for comparison with meteorites spectra, and 2) SMASS-MIT website\footnote{{\url{http://smass.mit.edu/}}} (further denoted MIT tool) for Bus-DeMeo taxonomy \citepads{2009Icar..202..160D} and for searching similar spectra obtained by the MIT group.

\subsection{Merging with  near-infrared and visible spectral data}
The mineralogical models derived from meteorite spectra and the new taxonomies (such as Bus-DeMeo) require the visible and near-infrared interval (typical 0.45 - 2.5 $\mu m$, denoted as VNIR). Due to equipment constraints, the visible part and the NIR part of the spectrum are acquired separately and, in general, at different telescopes. Thus, merging the visible part with the NIR part is an important step and was performed, whenever the visible spectrum was available. This was the case for asteroids, (2201) Oljato, (4183) Cuno, and (5143) Heracles, as presented in Fig.~\ref{Spectra}. For each of them, the visible spectrum was merged with our NIR data using a procedure of minimization of data in the common spectral region 0.82-0.9 $\mu m$ (i.e. finding a factor to normalize the visible part to minimize the mean square error between the two spectra in the common interval).

Joining the visible and NIR part leads to a more detailed analysis of the spectra. Using 0.45-2.5 $\mu m$ allows an accurate taxonomic classification and a more accurate mineralogical solution. Moreover, olivine and orthopyroxene, compounds typical on the asteroid surface, have a band minimum around 1 $\mu m$ and a maximum around 0.7 $\mu m$. Thus, a complete analysis of these features could not be made without using the entire VNIR spectrum. Additionally, a VNIR spectrum allows a more consistent curve matching with laboratory spectra.

However, there could be differences in the common interval between the visible part and NIR part that come from different factors, such as the object was observed at different epochs (several years or even more), a different observing geometry (phase angle, airmass), different instruments (the visible detectors have low efficiency around 0.85 $\mu m$, thus, typically, providing a low SNR data), and different solar analogues used. A method to quantify the differences is to measure the slope of the subtraction between the two spectra. This is further referred to as the similarity slope. Ideally this slope should be zero and is independent of the SNR of the spectra. 
A non-zero slope is a consequence either of an inhomogeneous composition of asteroid surface or the spectral changes in the time interval between the moments when the two spectra were obtained. We exclude here the possibility of data acquisition/reduction issues, which could include also artifacts in interpreting spectral data.

When multiple visible spectra were available for the same asteroid, we selected the spectrum for which the absolute value of similarity slope was minimal. The parameters of some of the visible spectra available for our objects analyzed in this paper are given in Table ~\ref{LiteratureSpectra} (online material).

Apart from the advantages, we note that merging the visible and NIR parts could add an additional uncertainty, but one which is taken into account in the evaluation of band center and band area of the 1 $\mu m$ band.

\subsection{Normalization of spectra}
All our spectra were normalized with the reflectance value at 1.25 $\mu m$, which is close to the middle of the interval between the maximum around 0.7 $\mu m$ and the maximum around 1.5 $\mu m$. These features are characteristics of olivine-orthopyroxene mixtures.
 
\subsection{Taxonomy}
Taxonomy is the classification of asteroids into categories. The main goal is to identify groups of asteroids that have similar surface compositions. An accurate taxonomic classification gives important information on the specific mineralogy for each of the defined classes.

To assign a taxonomic type for the spectra presented in this paper we used Bus-DeMeo taxonomy. This taxonomy is based on principal components analysis of VNIR spectra.  We classified the spectra in this taxonomy using two methods: 1) by performing curve matching with the 25 classes defined by the taxonomy (using M4AST website) and 2) by basing on principal components analysis using the MIT tool. Additionally, for obtaining complementary information when ambiguous types were obtained in Bus-DeMeo taxonomy we used the G13 classification \citepads{1996A&A...305..984B}.

Bus-DeMeo taxonomy can also be used with NIR data only, but the solution may not be unique. A particular case is to differentiate between the spectra belonging to S complex. \citeads{2014Icar..227..112D} defined some additional parameters to characterize the band around 1 $\mu m$ to differentiate between S, Sr, Sq and Q types using only the NIR interval. Their accuracy of classification, using these parameters is between 30\% and 75 $\%$. We consider that the curve matching methods can provide a better approach in this case, because it takes the aspect of the whole spectrum and not just a particular feature into account. Furthermore, if we consider the uncertainty bars for S, Sr, Sq, and Q taxonomic types, they span almost the same range of relative reflectance  in the NIR (when normalizing to 1.25 $\mu m$).

\subsection{Curve fitting with meteorites spectra}

Confronting the spectral data derived from telescopic observations with laboratory measurements is an important step for studying the asteroid physical properties \citepads{1992Icar...99..153B, 2011A&A...535A..15P}. Among the laboratory samples, meteorites can provide the most fruitful results for understanding asteroid composition. This is due to the reason that meteorites, prior to their arrival,  are themselves small bodies of the solar system. Thus, spectra comparison represents a direct link for our understanding of asteroid-meteorite relationships \citepads{2012A&A...544A.130P}. 

We compared our asteroid spectra with the laboratory spectra from the Relab\footnote{\url{http://www.planetary.brown.edu/relabdocs/relab.htm}} database using all the curve matching methods available in M4AST (mean square error, chi-square, correlation coefficient, and standard deviation of the error).  We selected the first three matches from the solutions found by all curve matching methods. The results are plotted in the online material, while Table~\ref{CHIT} summarizes the results.

\subsection{Mineralogical models}
\citeads{1986JGR....91..641C} introduced an analytical approach that permits the interpretation of visible and near-infrared spectral reflectance to determine the olivine-pyroxene composition. These parameters are the wavelength position of the reflectance minima around 1 $\mu m$ and 2 $\mu m$ ($BI_{min}$ and $BII_{min}$), the band centers (around 1 $\mu m$ - denoted as BIC, and 2 $\mu m$ - denoted BIIC), and the band area ratio ($BAR = \frac{BII}{BI}$), which is the ratio of the areas of the second absorption band relative to the first absorption band. These parameters were calculated as suggested by \citeads{1986JGR....91..641C}.

The band center is the wavelength of maximum absorption. If there is no overall continuous slope in the spectrum, the band center and the band minimum are coincident. \citeads{1986JGR....91..641C} outlined that "if there is a continuum slope in the spectral region of the absorption feature, the band center will be displaced in the downslope direction by an amount related to the slope of the continuum and the shape of the absorption feature".

\citeads{2010Icar..208..789D} reviewed the formulas for determining the mineral compositions and abundances based on VNIR spectra of S- and Q- type asteroids. Using 48 ordinary chondrites spanning the subtypes H, L, and LL and the petrologic types 4-6, they derived new calibrations for determining mineral abundances and mafic silicate composite. The accuracy of their determination was that they correctly classified H, L and LL chondrites based on spectrally-derived mineralogical parameters in $\approx 80\%$ cases.

The equation which describes the olivine (ol) to pyroxene ratio (px) is given as a function of BAR:
\begin{equation}
\frac{ol}{ol+px} = -0.242 \times \frac{BII}{BI} + 0.728; (R^{2} = 0.73).
\label{OLPX}
\end{equation}

The correlation between fayalite (Fa) in olivine and Band I center is described by a second order polynomial:
\begin{equation}
\begin{split}
mol\% Fa[ol] = -1284.9 \times BIC^{2} + 2656.5 \times BIC - 1342.3;\\
		      (R^{2} = 0.92).
\end{split}
\label{Fa}
\end{equation}

They draw a similar conclusion for the ferrosilite:
\begin{equation}
\begin{split}
mol\% Fs[px] = -879.1 \times BIC^{2} + 1824.9 \times BIC - 921.7;\\
		      (R^{2} = 0.91),
\end{split}
\label{Fs}
\end{equation}
where $R^{2}$ is the coefficient of determination \citepads{2010Icar..208..789D}.

The application of mineralogical models requires the characterization of the band around 0.9 $\mu m$, which is not completely characterized with only the NIR part of the spectrum available. Thus, for the objects studied in this paper, only the composite VNIR spectra of (2201) Oljato, (4183) Cuno, and (5143) Heracles allow the use of these mineralogical approaches of which the results are summarized by Fig.~\ref{BIIBI}. The computation of BIC, BIIC and BAR is made using M4AST, which implements the standard procedures as described by \citeads{1986JGR....91..641C}. The values resulting from Eq.~\ref{OLPX}, ~\ref{Fa} and \ref{Fs} are provided within the accuracy given by $R^{2}$.

\section{Results}

\begin{figure*}
\begin{center}
\includegraphics[width=6cm]{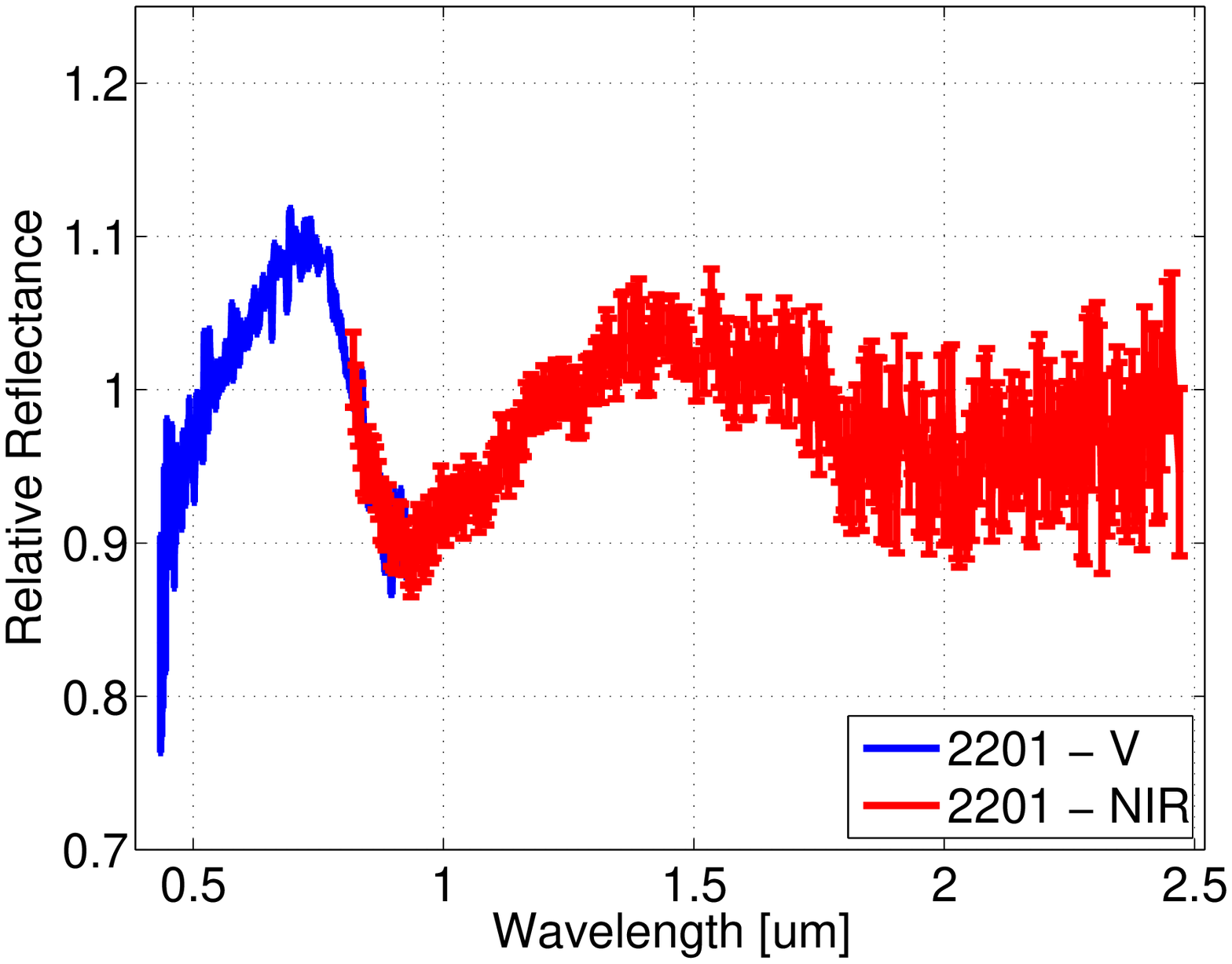}
\includegraphics[width=6cm]{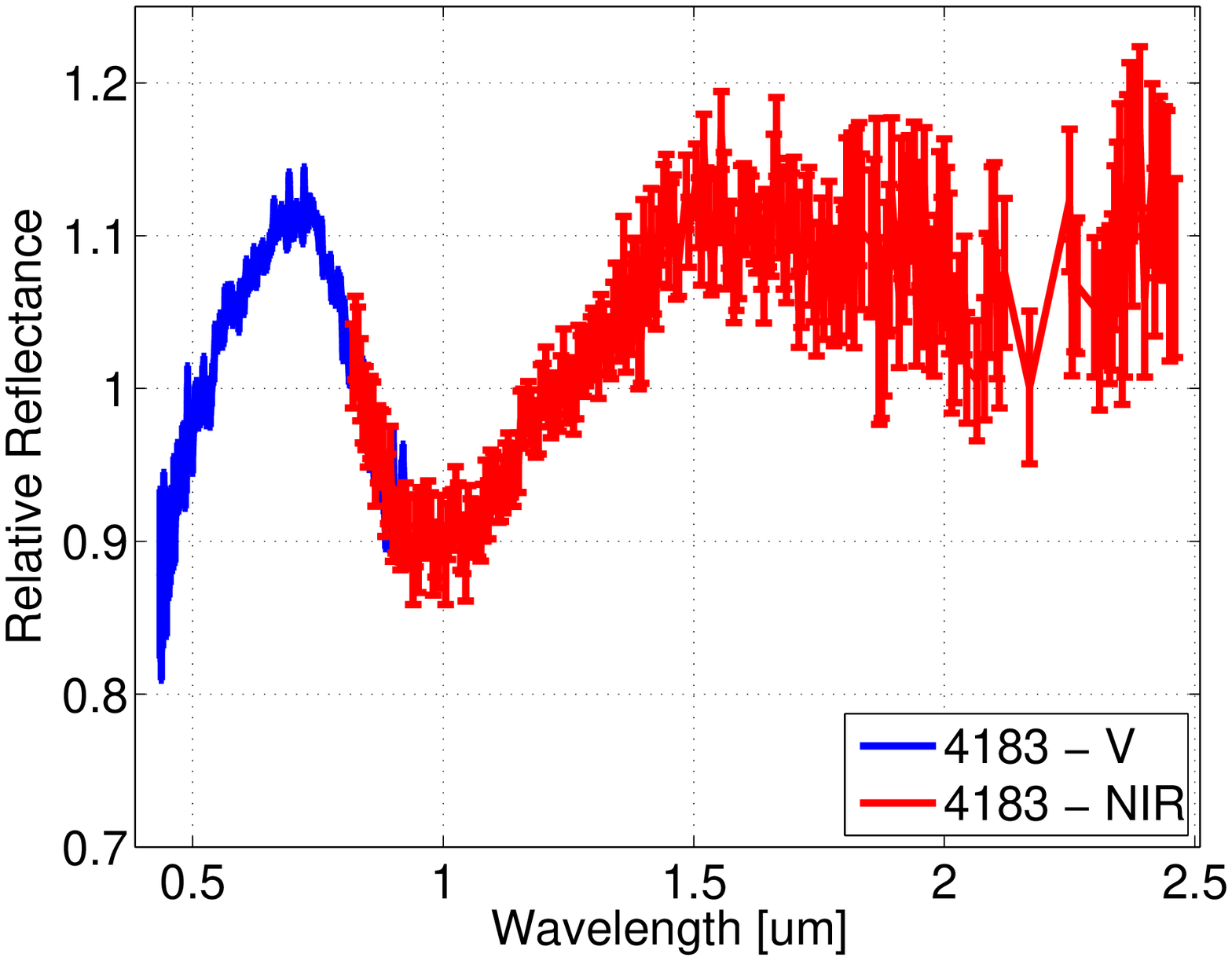}
\includegraphics[width=6cm]{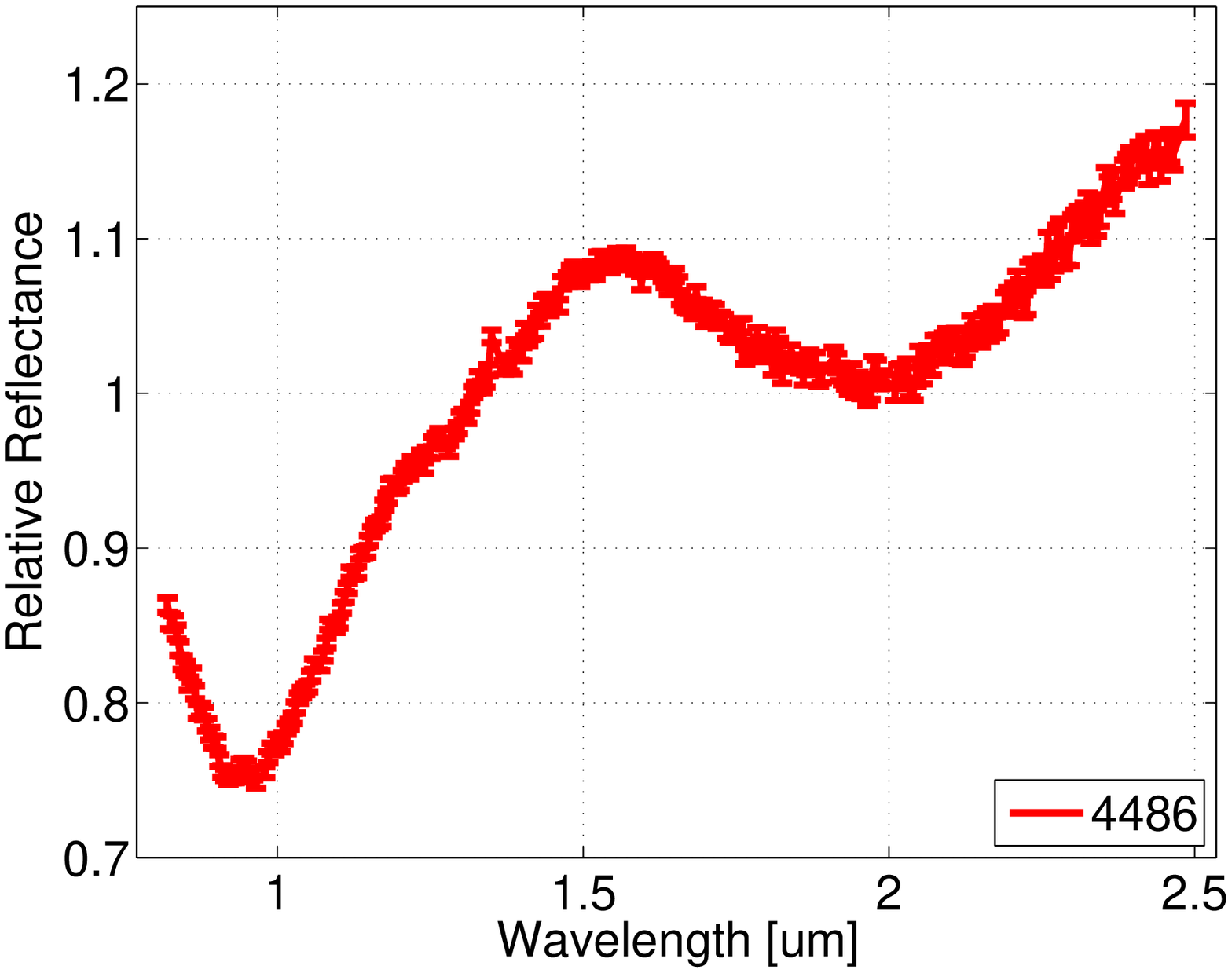}
\includegraphics[width=6cm]{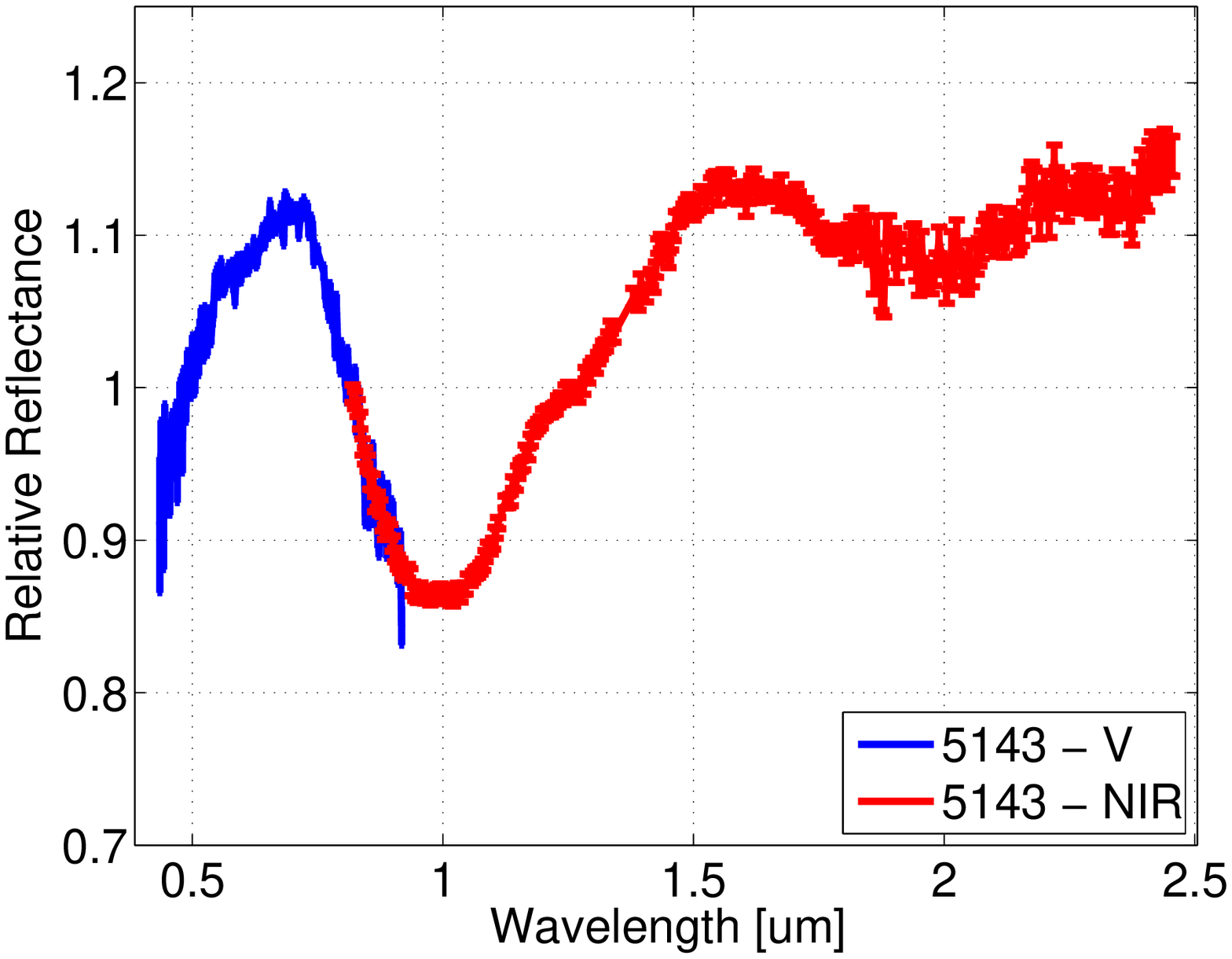}
\includegraphics[width=6cm]{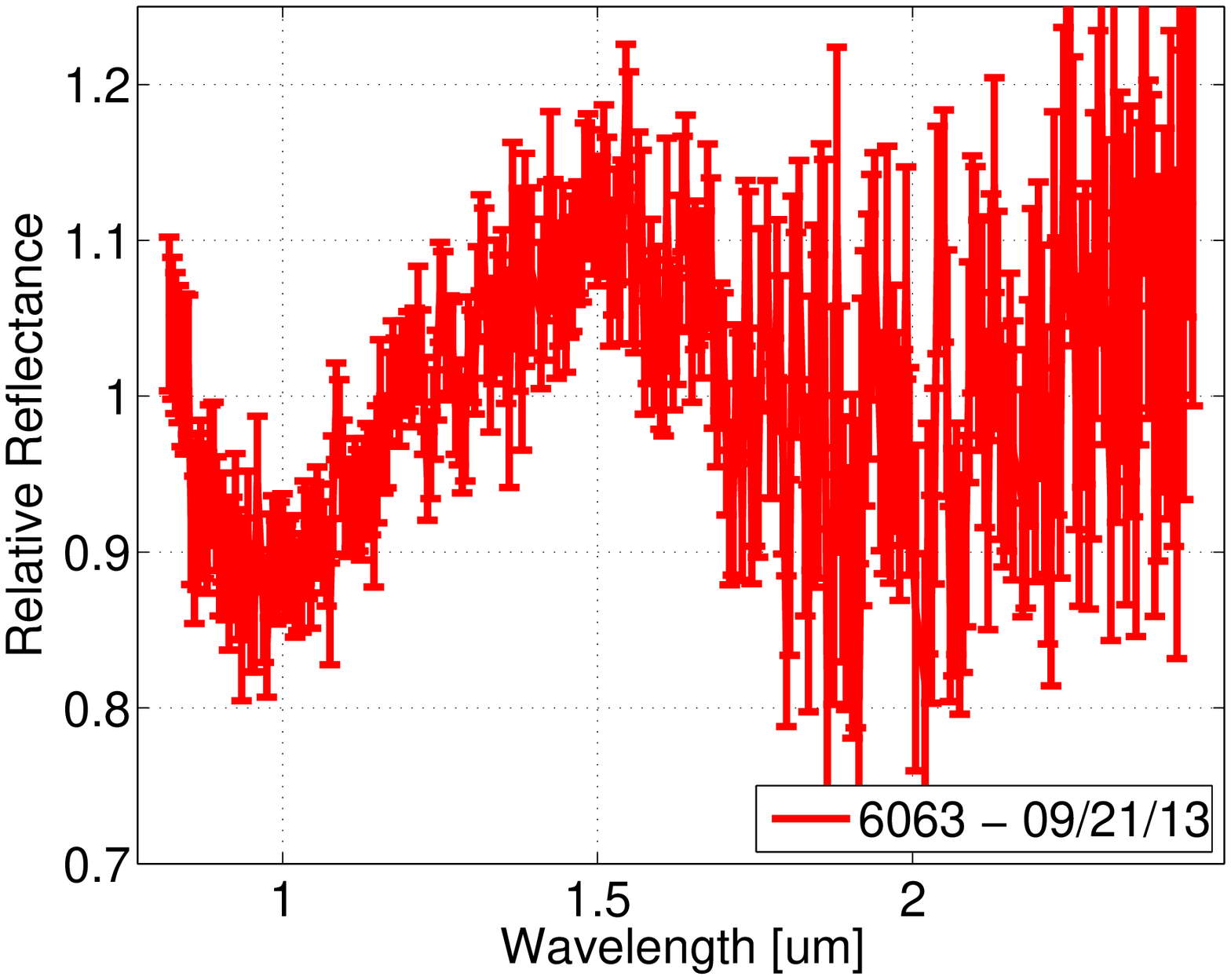}
\includegraphics[width=6cm]{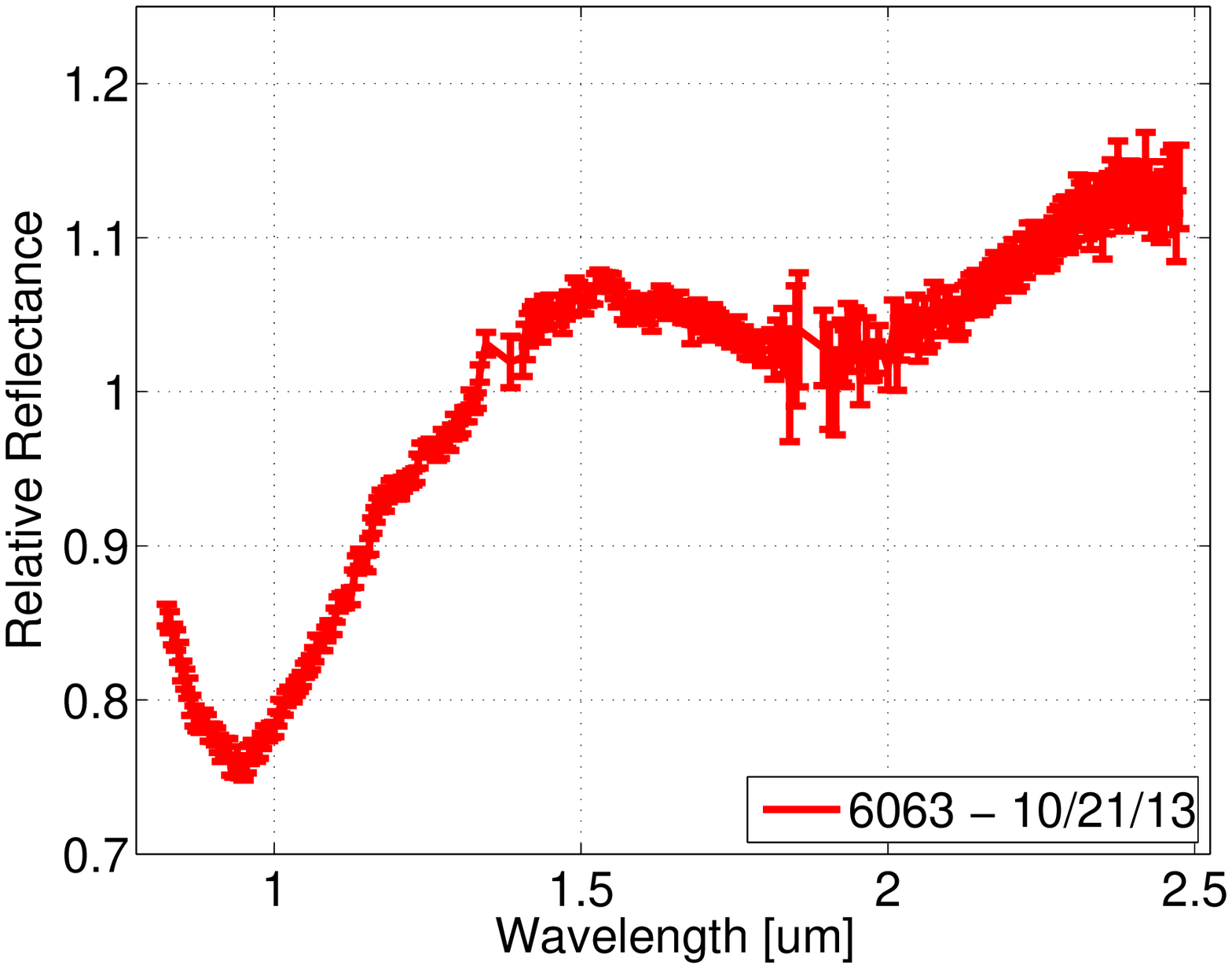}
\includegraphics[width=6cm]{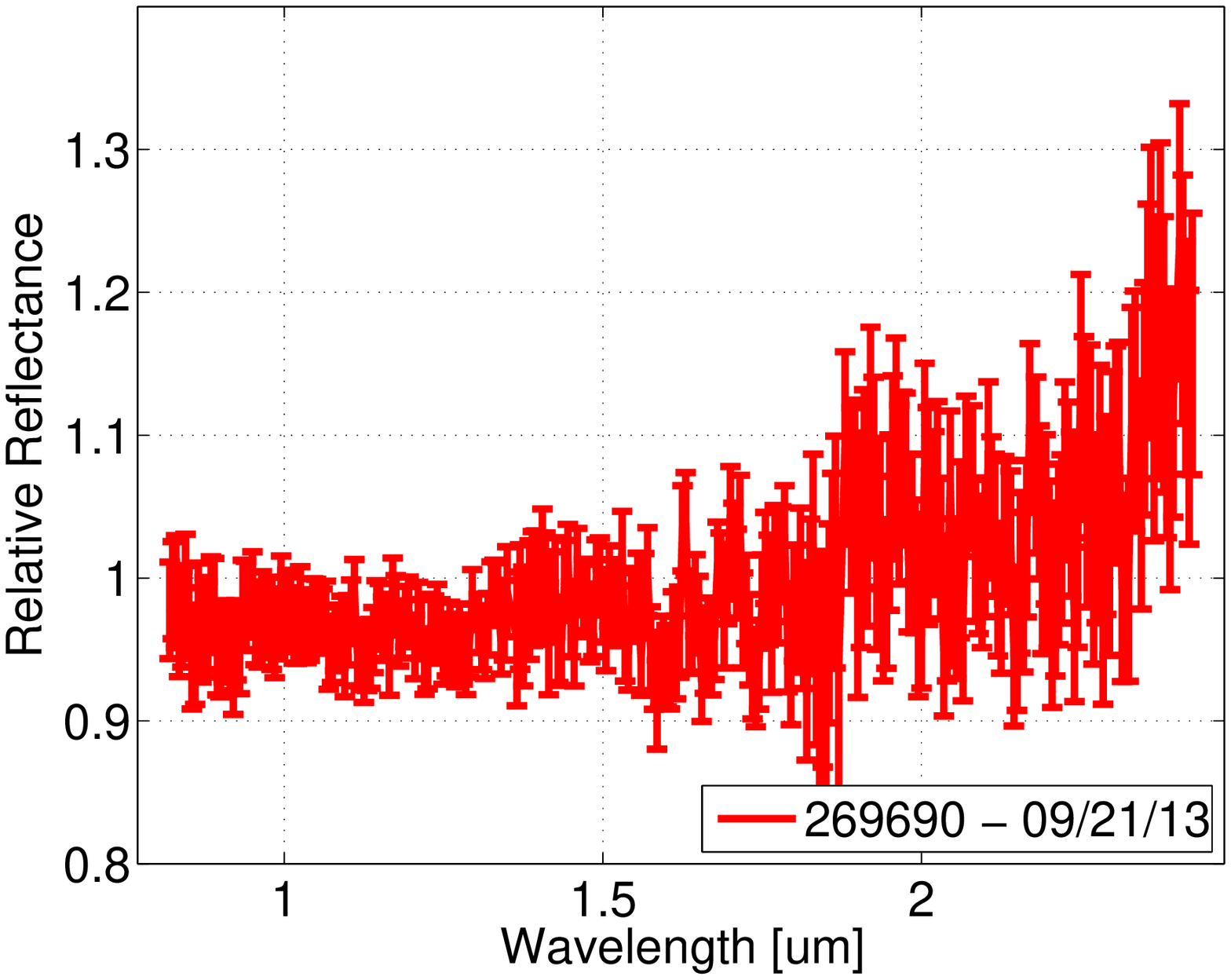}
\includegraphics[width=6cm]{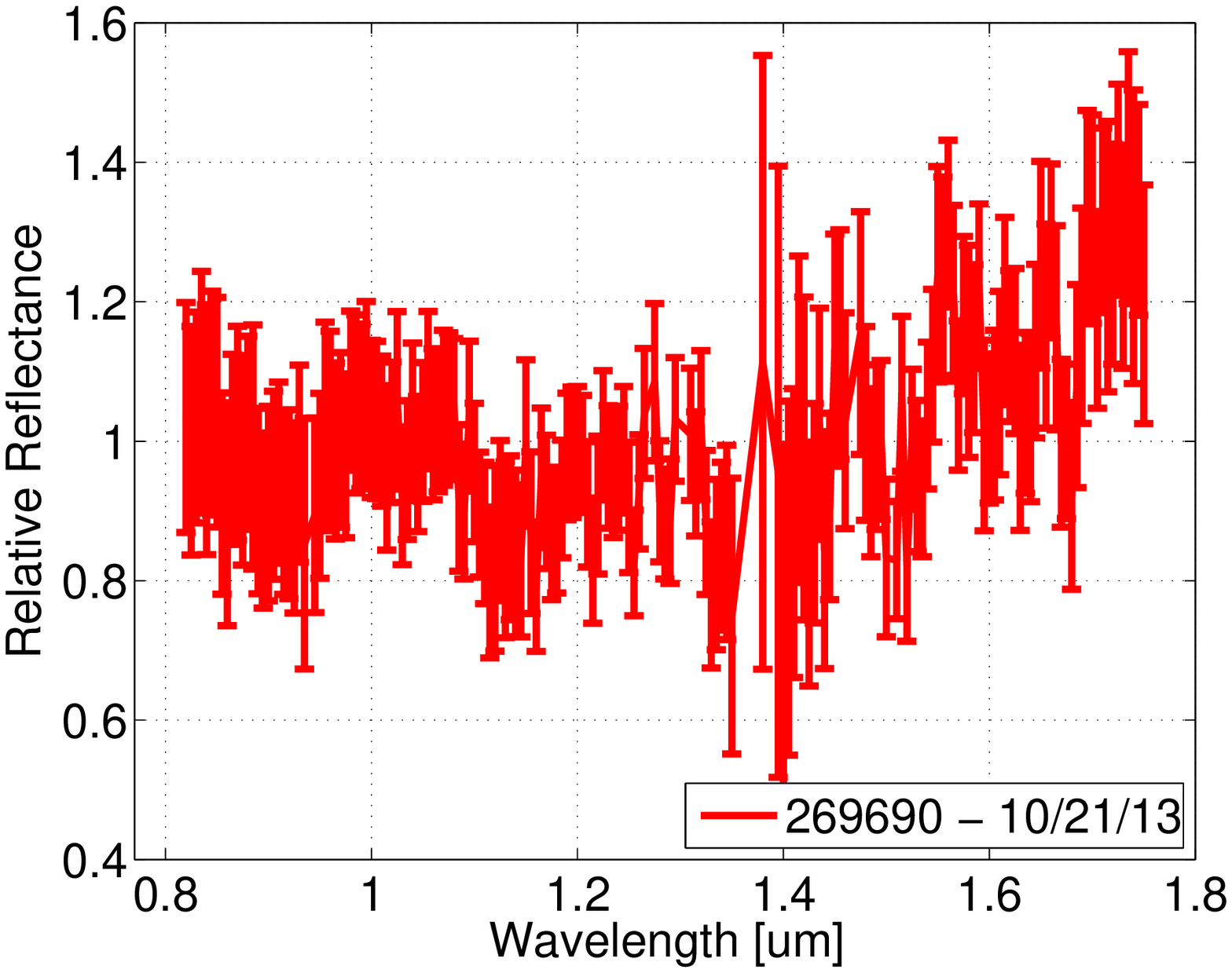}
\end{center}
\caption{Spectra of (2201) Oljato, (4183) Cuno, (4486) Mithra, (5143) Heracles, (6063) Jason, and (269690) 1996 RG3  with error-bars. All spectra are normalized to 1.25 $\mu m$. The NIR spectra (red) were merged with the visible counterpart (blue) when this was available in the literature. The visible spectra shown in this figure were obtained by \citeads{ 2002Icar..158..146B} for Oljato and by \citeads{2004Icar..170..259B} for Cuno and Heracles.}
\label{Spectra}
\end{figure*}

This section describes the results obtained for the observed asteroids: \object{(2201) Oljato}, \object{(4183) Cuno}, \object{(4486) Mithra}, \object{(5143) Heracles}, \object{(6063) Jason}, and \object{(269690) 1996 RG3}. The spectra are plotted in Fig.~\ref{Spectra} with error bars and merged with the visible range when available.

\begin{figure*}
\begin{center}
\includegraphics[width=6cm]{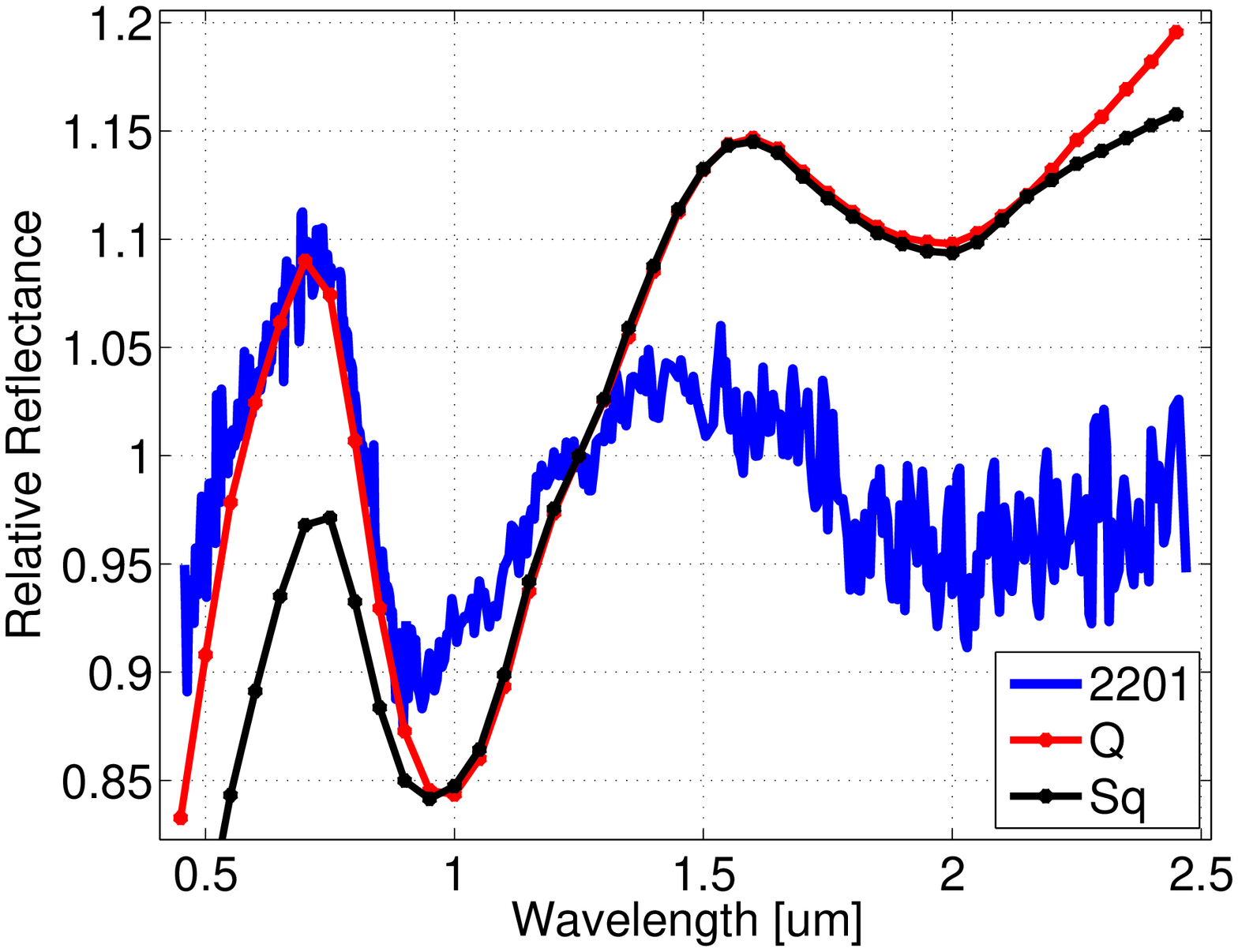}
\includegraphics[width=6cm]{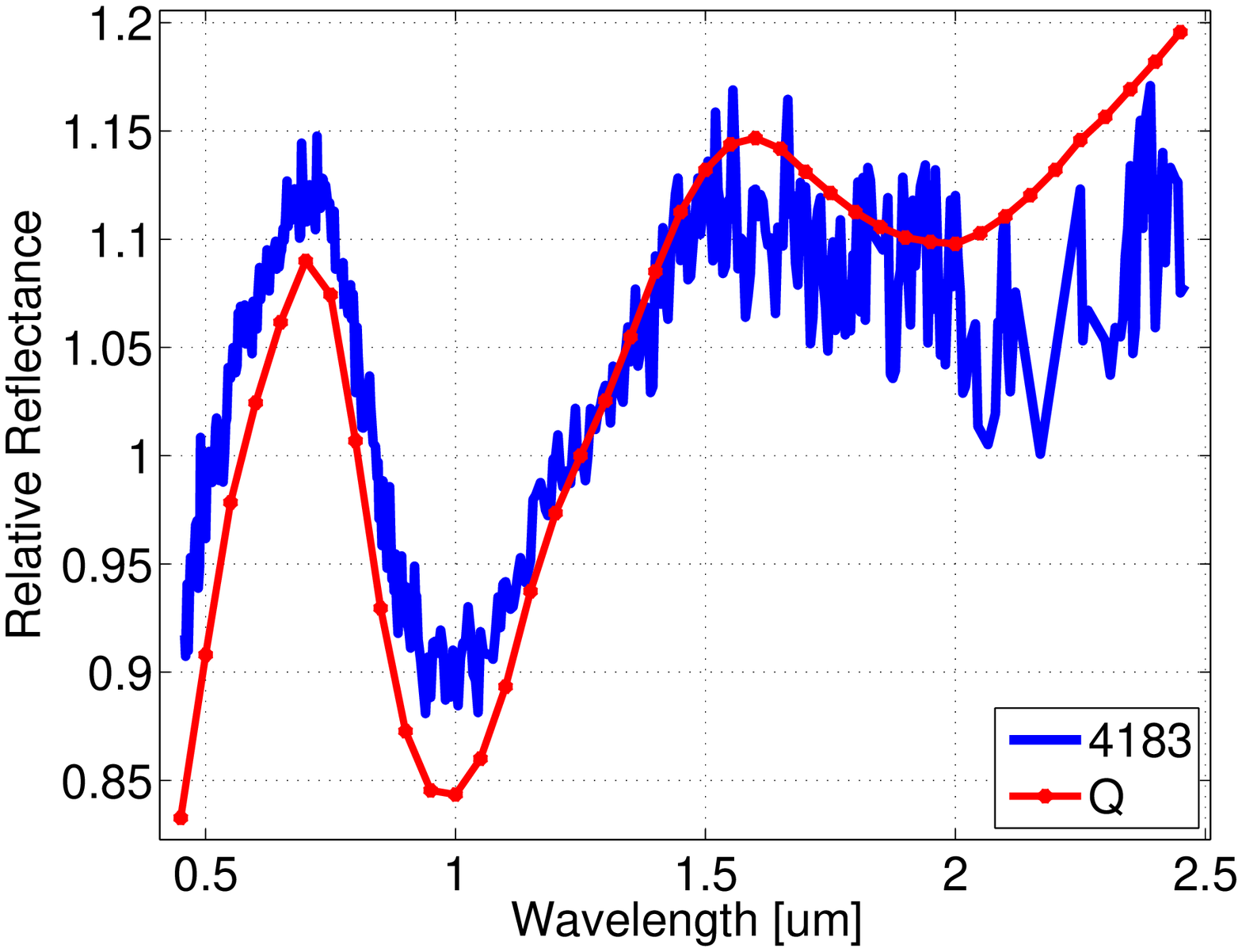}
\includegraphics[width=6cm]{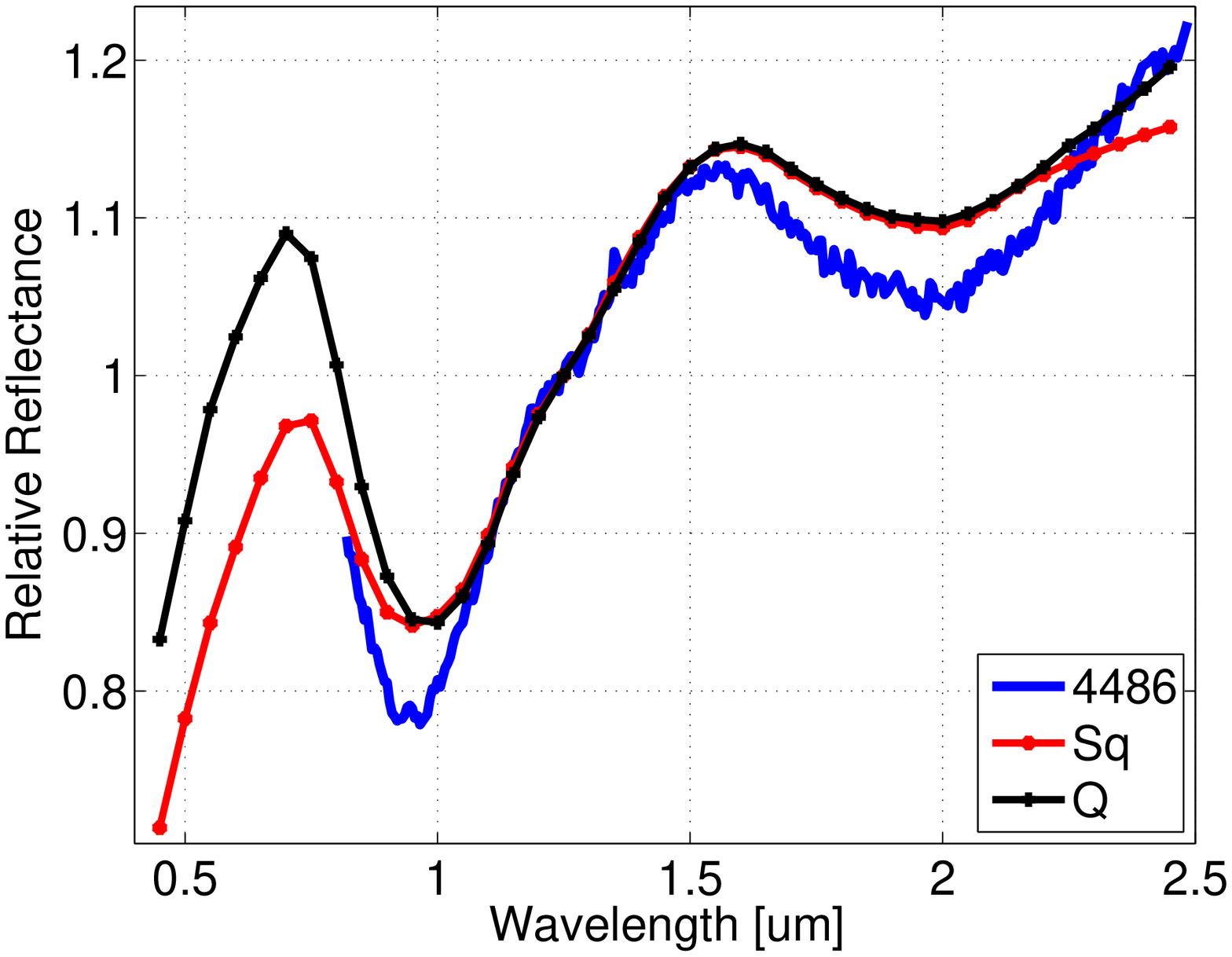}
\includegraphics[width=6cm]{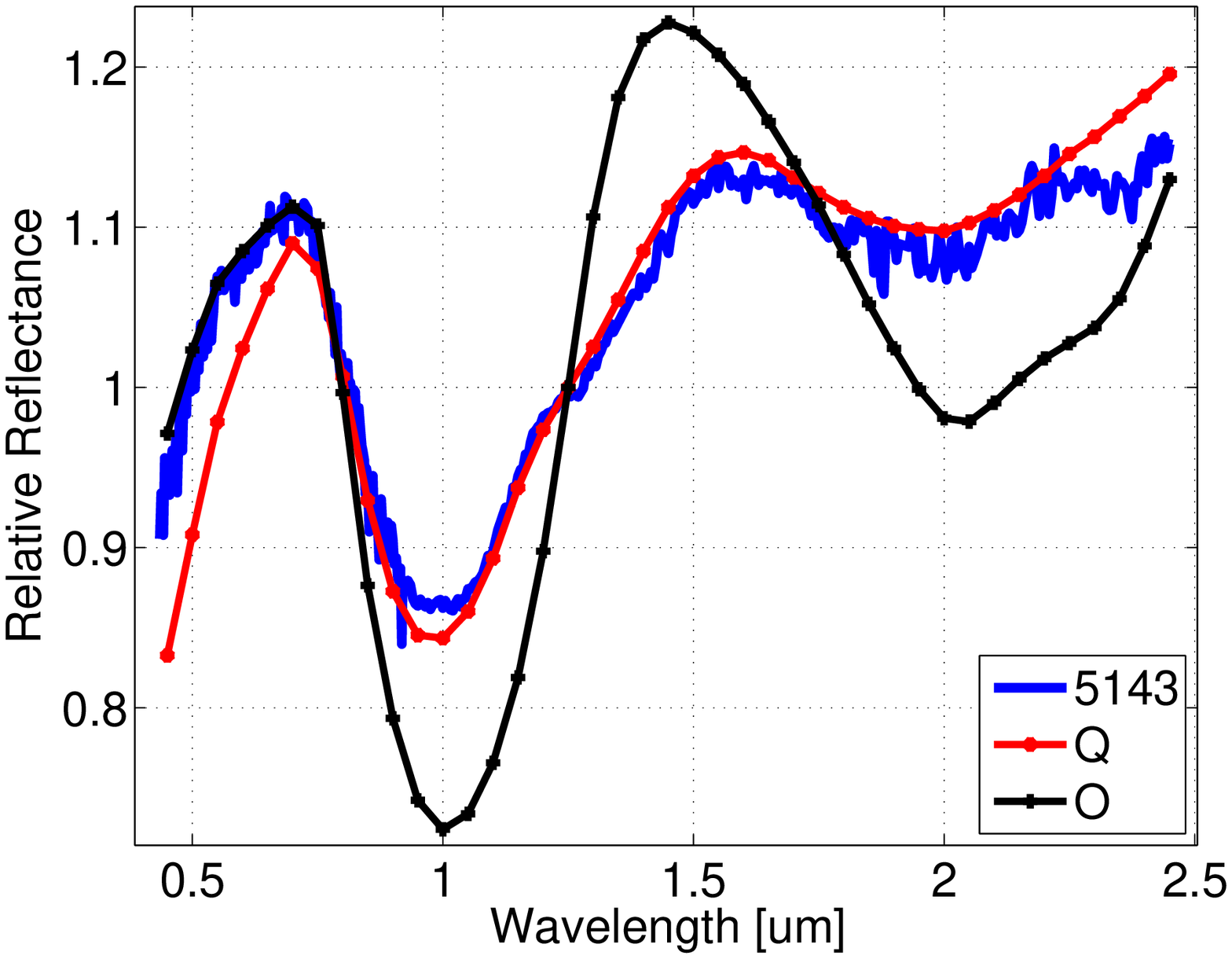}
\includegraphics[width=6cm]{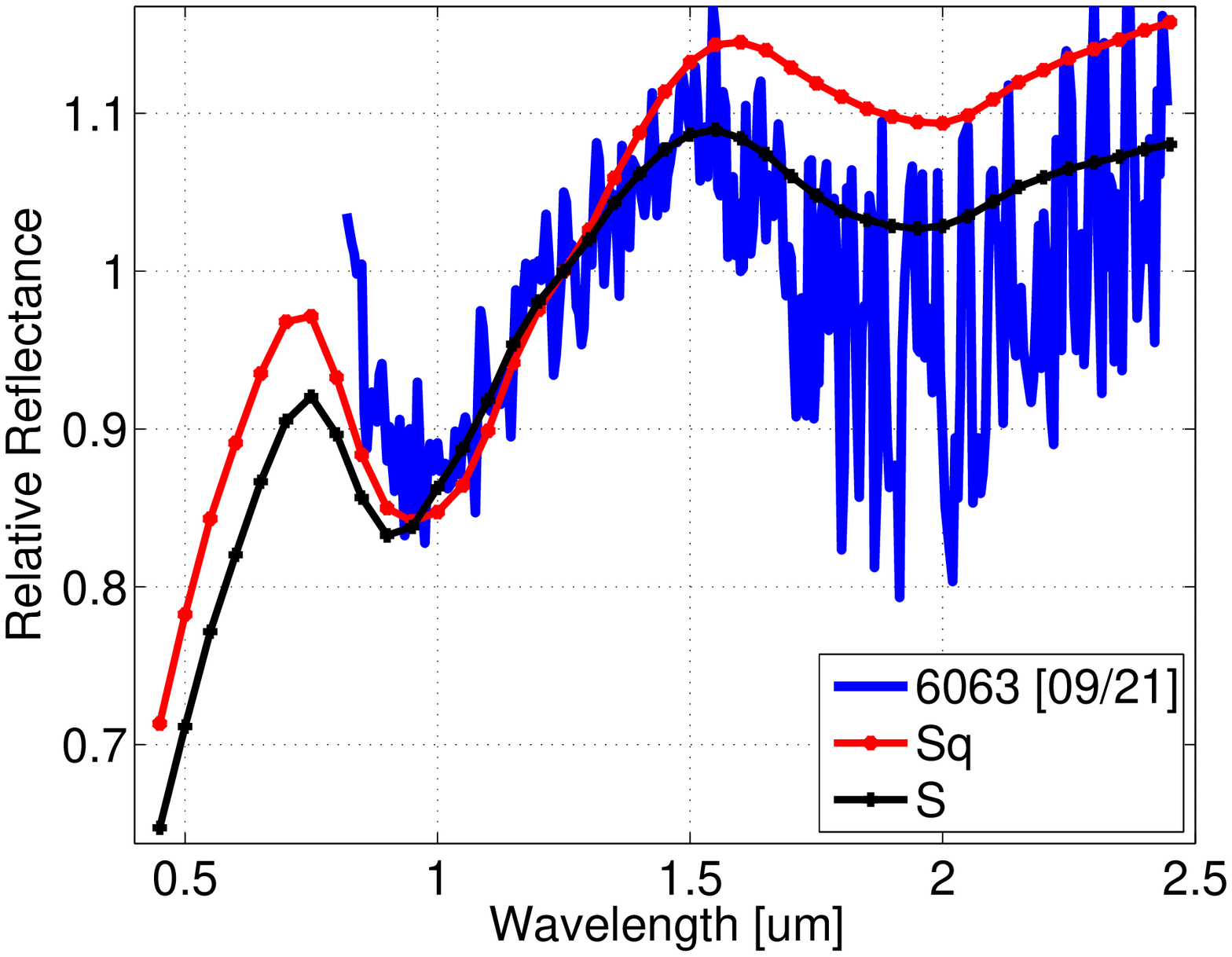}
\includegraphics[width=6cm]{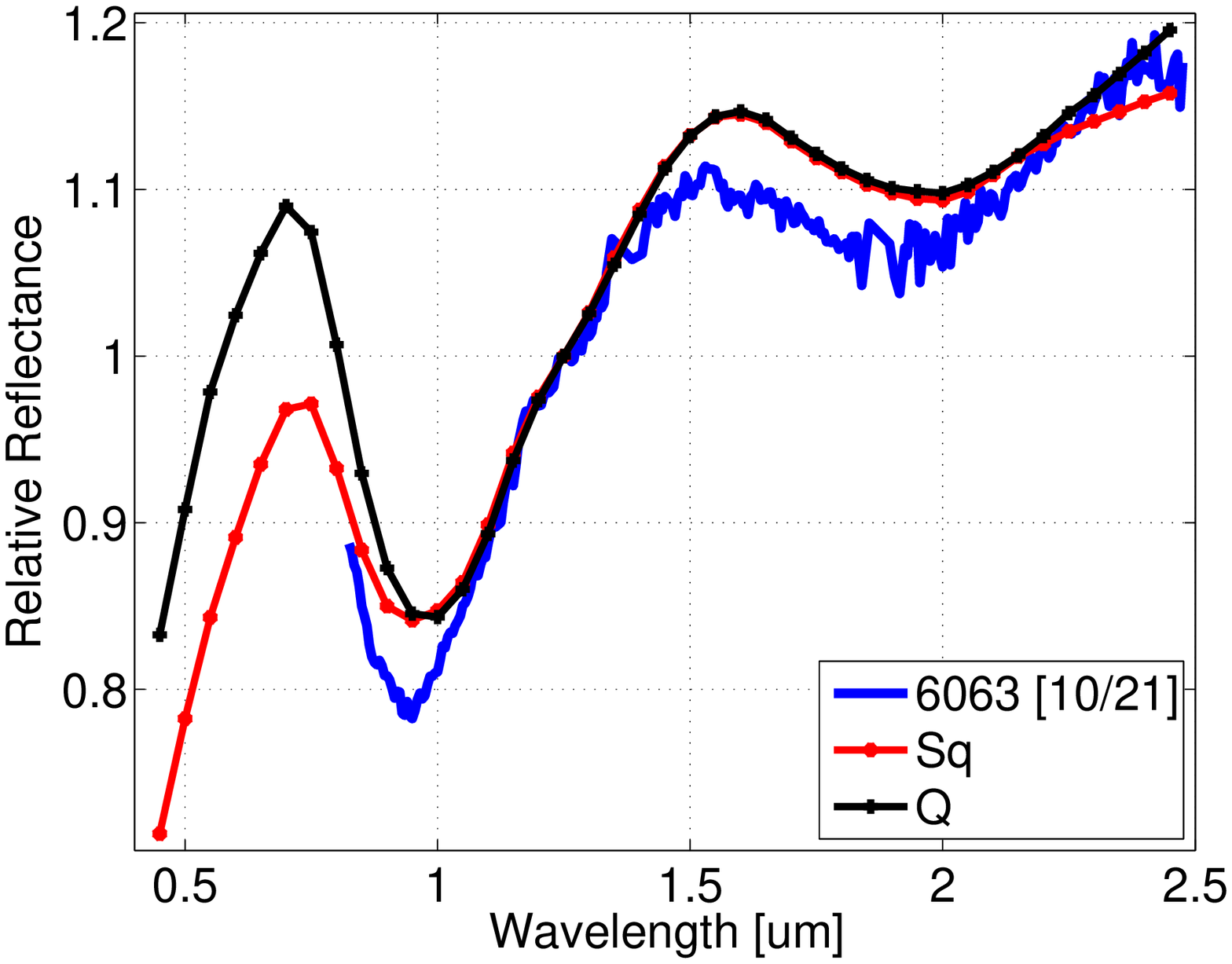}
\includegraphics[width=6cm]{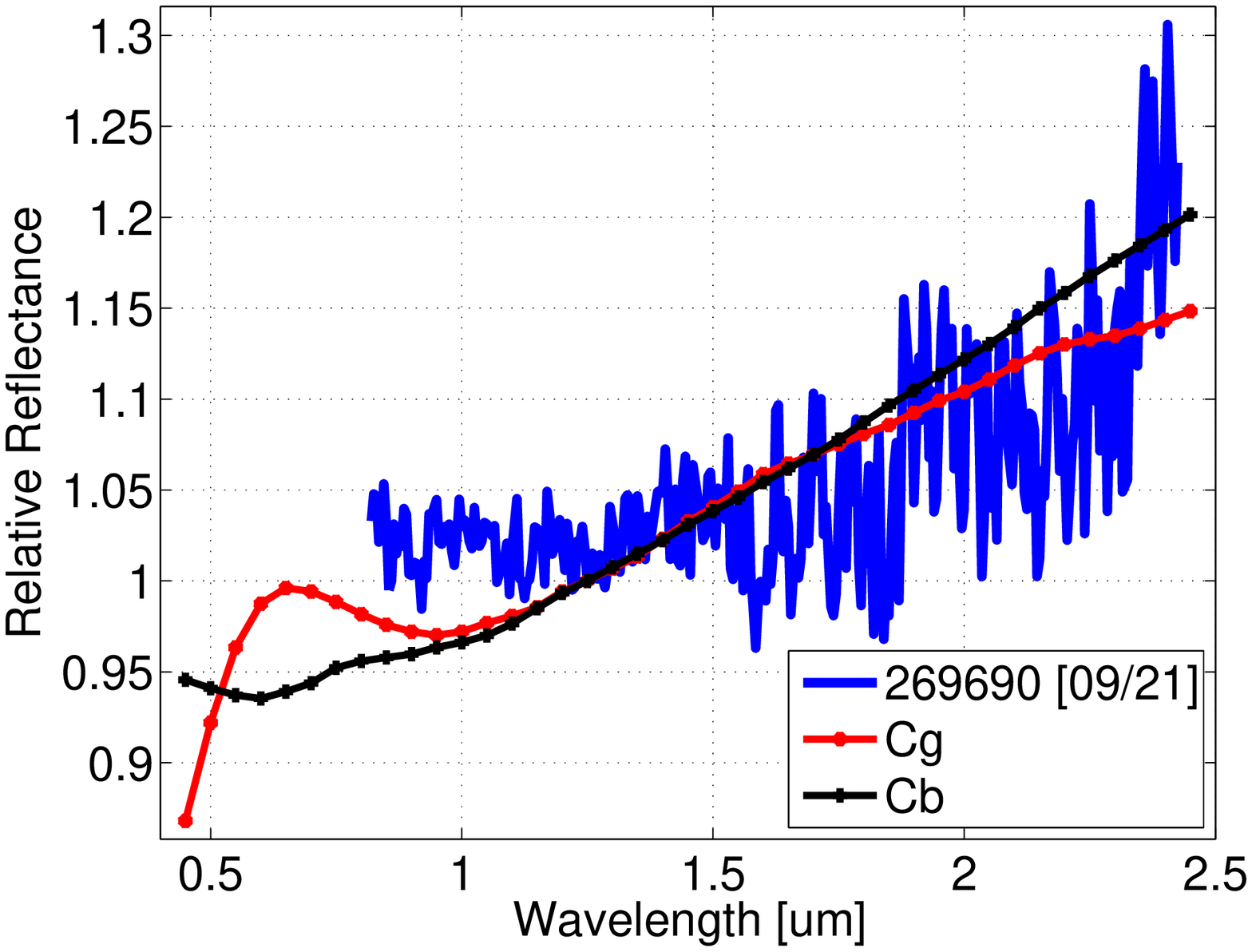}
\end{center}
\caption{Classification in Bus-DeMeo taxonomic system for (2201) Oljato, (4183) Cuno, and (5143) Heracles. The data are normalized at 1.25 $\mu m$. The asteroid spectra are plotted against the curves for the resulting classes obtained by \citeads{2009Icar..202..160D}.}
\label{Taxonomy}
\end{figure*}

\begin{figure*}[!ht]
\begin{center} 
\includegraphics[width=8cm]{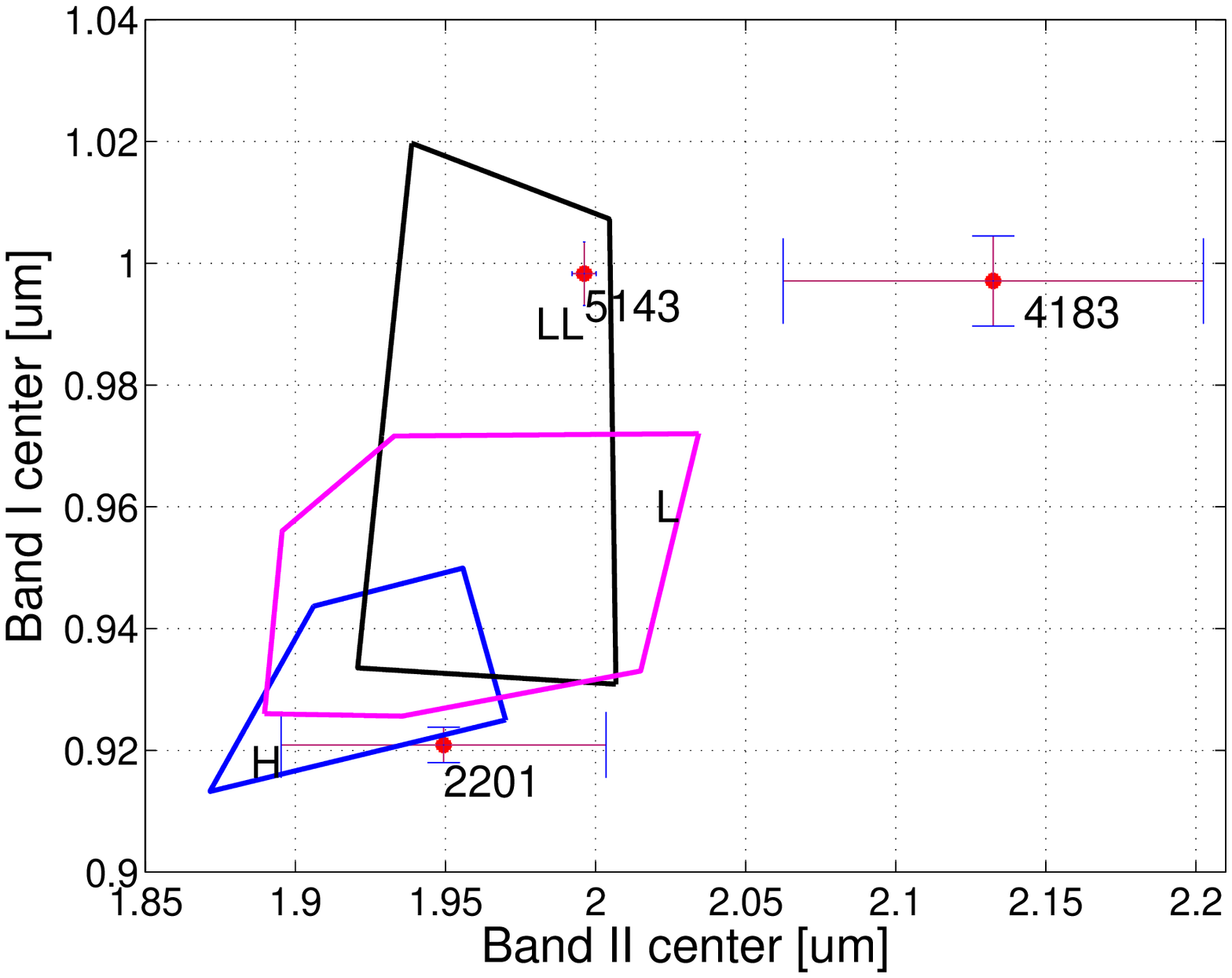}
\includegraphics[width=8cm]{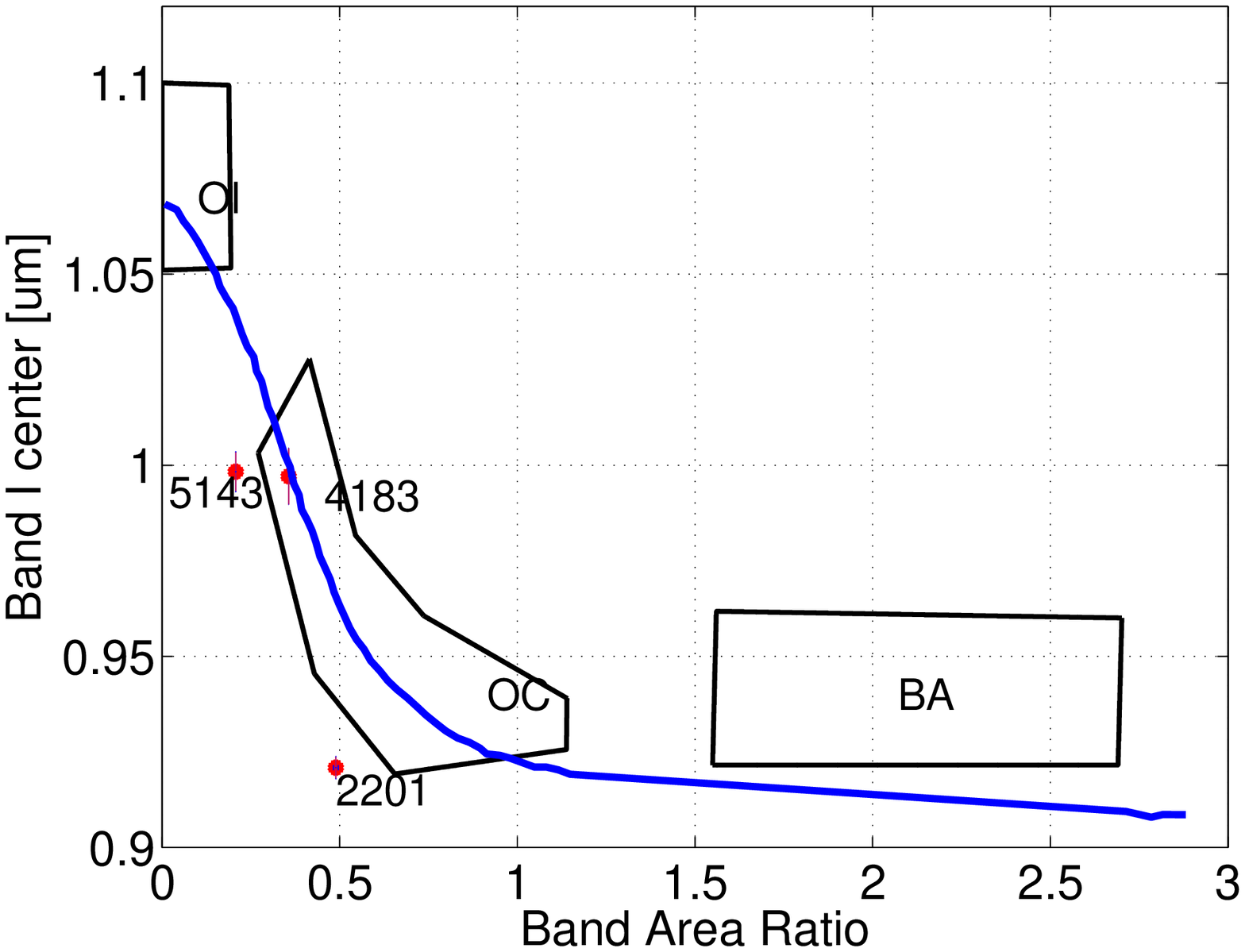}
\end{center}
\caption{(Left) Wavelength position of the centers of the two absorption bands computed using \citeads{1986JGR....91..641C}. The enclosed regions correspond to the band centers computed for the H, L, and LL chondrites, respectively \citepads{2010AA...517A..23D}. (Right) Band area ratio (BAR) versus band I centers. The regions enclosed by continuous lines correspond to the values computed for basaltic achondrites, ordinary chondrites (OC), and olivine-rich meteorites (Ol) \citepads{1993Icar..106..573G}.}
\label{BIIBI}
\end{figure*}

The discussion about taxonomic type of each object is made with reference to Fig.~\ref{Taxonomy}. The results for taxonomic classification of spectra are compared with the physical properties and previously taxonomic classifications (Table~\ref{Tab1Prop}). The results of comparison with meteorite spectra are shown in Tabel~\ref{CHIT} and in Appendix~\ref{Anexa1}.

\begin{table*}
\caption{Summary of the results obtained by matching the asteroids spectra with spectra from Relab database. The comparison  was made using M4AST. The first three matches are presented by taking the results of all curve matching methods into account.}
\label{CHIT}
\centering
\begin{tabular}{l l l c c c c c}
\hline\hline
Spectrum & Sample ID  & Meteorite & Fell & Type & SubType & Texture & Size [ $\mu m$]\\
\hline
(2201) Oljato & TB-TJM-073 & Dwaleni & 12-10-1970 & OC & H6  & Particulate & 0-150 \\
 & TB-TJM-141 & L'Aigle & 26-04-1803 & OC & L6 & Particulate & 0-150 \\
 & MB-CMP-003-D & Dwaleni & 12-10-1970 & OC & H6 & Particulate & 25-250 \\
\hline
(4183) Cuno & TB-TJM-067 & Bandong & 10-12-1871 & OC & LL6 & Particulate & 0-150 \\
 & TB-TJM-077 & Karatu & 11-09-1963  & OC & LL6 &  Particulate & 0-75\\
 & TB-TJM-144 & Wethersfield (1971) & 08-04-1971 & OC & L6 &  Particulate & 0-150 \\
\hline
(4486) Mithra &  TB-TJM-139  & Kunashak  & 11-06-1949 & OC & L6 & Particulate & 0-150 \\
 & MR-MJG-057 & Colby (Wisconsin) &  04-07-1917 & OC & L6 & - & -  \\
 & MH-FPF-053-D & Nuevo Mercurio & 15-12-1978 & OC & H5 & Particulate & 0 - 350\\
\hline 
(5143) Heracles & TB-TJM-067 & Bandong & 10-12-1871 & OC & LL6 &  Particulate & 0-150 \\
 & MR-MJG-057 & Colby (Wisconsin)  & 04-07-1917 & OC & L6 & - & -  \\
 & MR-MJG-072 & Jelica & 01-12-1889 & OC & LL6 & - & - \\
\hline
(6063) Jason & MB-CMP-002-L & Paragould  & 17-02-1930 & OC & LL5 &  Particulate & 25-250 \\
(09/21/13) & TB-TJM-107 & Mabwe-Khoywa  &  17-09-1937  & OC & L5 & Particulate & 0-150  \\
 & MH-JFB-022 & Gifu & 24-07-1909 & OC & L6 & Slab & - \\
\hline
(6063) Jason & MR-MJG-057 & Colby (Wisconsin) & 04-07-1917 & OC & L6 &  - & - \\
(10/21/13) & MB-TXH-086-A & Y74442 &  -  & OC & LL4 & Particulate & 0-25  \\
 & MH-CMP-003 & Farmington & 25-06-1980 & OC & L5 & Particulate & 20-250 \\
\hline
(269690) 1996 RG3 & MB-CMP-019-1 & Y-82162,79 &  -  & CC & CI Unusual &  Particulate & 0-125 \\
 & MB-TXH-064-F & Murchison heated& 28-09-1969 & CC & CM2 & Particulate & 0-63  \\
 & MC-RPB-002 & LEW90500,45 & - & CC & CM2 & Particulate & 0-500 \\
\hline
\end{tabular}
\end{table*}

\newpage
\subsection{(2201) Oljato}

Since its discovery, (2201) Oljato has been considered a good candidate to be an extinct cometary nucleus \citepads{1989aste.conf..880W}. Over time, it was analyzed using photometric and spectroscopic observations in visible, near-infrared, and radio regions (e.g.\citeads{1983PhDT.........2M}, \citeads{1996Icar..122..122L}, \citeads{2004Icar..170..259B}). The first observations show a puzzling object, in the sense that the high radiometric ratio \footnote{Infrared flux, given as ratio to a stellar standard, with the purpose of estimating the albedo and the temperature of asteroids based on the equilibrium between absorbed Sun light and emitted energy, as in \citeads{1977Icar...31..456H}} does not fit with some spectral features \citepads{1983PhDT.........2M} that are characteristic for dark primitive objects.

\citeads{2002aste.conf..205H} determined the geometrical albedo $p_{V} = 0.24$, using the NEA thermal model (NEATM). Other models, like STM (standard thermal model) and FRM (fast rotating model) give a higher albedo (0.63 and 0.49, respectively). Having the absolute magnitude $H =  16.86 $ and considering $p_{V} = 0.24$  the effective diameter of this object is 2.1 km. The high value of the albedo, typical for S-type asteroids, makes it unlikely that (2201) Oljato is an extinct comet nucleus, as suggested by its highly eccentric orbit and the first spectral observations.

According to Ondrejov NEO Photometric Program webpage\footnote{\url{http://www.asu.cas.cz/~ppravec/newres.txt}}, the rotation period of (2201) Oljato is $\approx$ 26 h, while the lightcurve amplitude is around 0.1 magnitudes .

Pioneer Venus Orbiter observed many interplanetary field enhancements (rare, but very distinct interplanetary magnetic field structures) in the longitude sector where the orbit of Oljato lies inside the Venus orbit \citepads{1987GeoRL..14..491R,2013MPS..tmp..327L}. \citeads{2013MPS..tmp..327L} attributed these observed phenomena to interactions between the material co-orbiting Oljato and the solar wind. They concluded that this co-orbiting material was most probably produced by an earlier collision with Oljato or due to its internal activity.  The last data from Venus Express (between 2006 - 2012) show that the rate of interplanetary field enhancements was reduced substantially during the recent Oljato return, suggesting that the co-orbiting material has drifted out, or was destroyed by collisions \citepads{2013MPS..tmp..327L}.

Data in the 0.3 - 0.9 $\mu m$  region were first obtained by \citeads{1984Icar...59...25M} in December 1979, using intermediate band spectroscopy. The spectrum shows a high reflectance in the spectral region between 0.33-0.4 $\mu m$ and a feature (a local maxima) around 0.6 $\mu m$.  The spectrum obtained by \citeads{1993JGR....98.3031M} on July 6, 1983 did not confirm the ultraviolet dominant feature found in 1979.

The visible spectrum of Oljato obtained by \citeads{1996Icar..122..122L} in October 11, 1994 shows an absorption feature centered around 0.9 $\mu m$, which is attributable to the presence of aqueous alteration products. This spectrum also shows an absorption band for $\lambda < 0.5 \mu m$, and any evidence of emission around 0.388 $\mu m$ due to the radical CN, which is an important feature on cometary spectra \citepads{1996Icar..122..122L}.

Different authors have classified (2201) Oljato as an S, C, or E taxonomic type depending on the spectral interval they considered \citepads{1993JGR....98.3031M, 1998Icar..133...69H, 2002Icar..158..146B}. Appendix~\ref{Anexa2} (online material) shows some of these visible spectra as found in the literature. There is an important difference, which can be illustrated by the position of maximum around 0.7 $\mu m$ between the spectrum of \citeads{1996Icar..122..122L} (max. position at 0.659 $\mu m$) obtained on October 1994 and the one obtained one year later (December 1995) by \citeads{2004Icar..170..259B} (max. position at 0.724 $\mu m$).

We merged our NIR part of the spectrum with the one obtained by \citeads{ 2002Icar..158..146B}. This choice was made by considering the similarity in the common spectral interval (Fig.~\ref{Spectra}) with the similarity factor being 0.013. The obtained VNIR is characterized by a high reflectance around 0.7 $\mu m$ as compared with the rest of the spectrum, which has a large and deep band around 0.9 $\mu m$ and a wide absorption band around 2 $\mu m$. The first band minimum is $BI_{min} = 0.926 \pm 0.003 \mu m$ while the second band minimum is $BII_{min} = 2.015 \pm 0.041 \mu m$.  The slope in the spectral interval 0.82 - 2.45 $\mu m$ is 0.01 $\mu m^{-1}$, while the composite VNIR spectrum has a slope of -0.018 (when normalizing to 1.25 $\mu m$). These slopes are atypical for S-complex. By inspecting the NIR spectra obtained by MIT-UH-IRTF Joint Campaign\footnote{$http://smass.mit.edu/minus.html$}, we found less than 2\% of the total spectra belonging to S-complex with a slope lower than 0.01 in the NIR spectral region.

The spectrum is classified in Bus-DeMeo taxonomy as Sq-type (Fig.~\ref{Taxonomy}) by the SMASS-MIT tool. The first two principal components PC1 = -0.3994 and PC2 = 0.1326 place the spectrum at the border between Sq and Q type. According to \citeads{2009Icar..202..160D} the Sq-type is characterized by a wide 1-micron absorption band with evidence of a feature near 1.3 $\mu m$ like the Q-type, except the 1-micron feature is more shallow for the Sq. The $BI_{min}$ is in the range of S-Sr types\citepads{2014Icar..227..112D}. However, the spectrum is fitted by the Q type in the visible region. The G13 taxonomy place the object in group 1 with a spectrum similar with (5) Astraea and (6) Hebe, thus belonging to S-complex. However the matching with this type is only in the NIR part. Classifying this spectrum  using curve matching methods gives ambiguous results because there is a large difference in the NIR part compared to S-complex typical spectrum. Thus, we cannot assert a certain type between Q and Sq to this spectrum.

The meteorite spectra similar with this spectrum are those of ordinary chondrites (OC) with petrologic classes 5 and 6. Most of the matches are with OC-H meteorite spectra. The spectra that fit most are those of two samples from the Dwaleni meteorite, an OC H6 which fell on October 12, 1970, and a spectrum of a sample from L'Aigle meteorite, an OC L6 (Fig.~\ref{OljatoMet}).  The spectrum of an olivine basalt ilmenite mixture also shows similarities with this asteroid spectrum.

On the composite spectrum we can apply the mineralogical model, thus obtaining the $\frac{ol}{ol+px} = 0.520$ ratio. The position of the first band center corresponds to a molar percentage of Fa of 14.4$\%$ and to a molar percentage of Fs of 13.3$\%$. Considering the plots BIC versus BAR and BIIC versus BIC (Fig.~\ref{BIIBI}), we found this spectrum in the OC H region (relatively close to the limit of OC L region), which agrees with spectral matching results.

The slope of the VNIR spectrum corresponds to a very fresh surface. This can be correlated with a mass loss from Oljato, as suggested by \citeads{2012AJ....143...66J} as a possible cause for the observed repetitive magnetic disturbances. The spectrum similar with OC-H meteorites also supports this hypothesis. \citeads{2012AJ....143...66J} considered that this object is not inert but did not find sufficient evidence to identify a mechanism for explaining the mass loss.

The differences in the visible spectra obtained by several authors support the hypothesis of material loss or inhomogeneous surface composition. A good opportunity to re-observe this object will arise in July 2015, when it will reach an apparent magnitude of 16.5.

\subsection{(4183) Cuno}

With an estimated diameter D = 5.38 km, (4183) Cuno is one of the largest potentially hazardous asteroids (PHA). According to the Warm Spitzer data \citepads{2011AJ....141...75H}  it has a relatively low albedo, $p_{V} = 0.10$. Radar images reveal an elongated object with a prominent feature that may indicate the presence of one or more concavities \citepads{2001AAS...198.8907B}.

According to the Ondrejov webpage, this object rotates with a period  of 3.55 h, which is the shortest period among the objects studied in this paper.

Based on orbital similarity, this asteroid was associated with the December Aurigids meteoroid stream \citepads{2004EM&P...95..697P}. \citeads{2004JIMO...32...60T} associated this object with the Tagish Lake meteorite and with the $\mu$ Orionid fireball stream. This dynamical association  may suggest a cometary origin, but the  spectra obtained in the visible by \citeads{2004Icar..170..259B} and \citeads{2007Icar..188..175F} show a spectrum similar with ordinary chondrites. Based on the NIR part of the spectrum obtained with SpeX instrument on the IRTF on October 24, 2011, \citeads{2014Icar..227..112D} classified this object as a Q type with a fresh surface.

We obtained NIR data on November 15, 2011. To perform the analysis on a larger spectral interval, we merged our NIR with the visible part obtained by \citeads{2004Icar..170..259B}. The similarity slope on the common interval (defined in Section 3) is $0.085 \mu m^{-1}$, as compared with $0.918 \mu m^{-1}$ for the spectral data obtained by \citeads{2007Icar..188..175F}.

The VNIR spectrum has the $BI_{min}$ located at 0.998$\pm 0.007 \mu m$, while the $BII_{min}$ is located at 2.124$\pm 0.061 \mu m$. The amplitude of the first maximum (at $\sim$0.7 $\mu m$) is equal (within the error bars) with the amplitude of the second maximum (at $\sim$ 1.5 $\mu m$). The slope of the VNIR spectrum is 0.05  $\mu m^{-1}$, which corresponds to a fresh surface. Both MIT and M4AST tools classify this object as Q type (Fig.~\ref{Taxonomy}), a result that agrees with the one found by \citeads{2014Icar..227..112D}.

The comparison with meteorite spectra gives a good fit with the Bandong meteorite (Table~\ref{CHIT}). This fit is obtained by using all curve matching methods available in M4AST. Overall, the fitting shows good agreement with OC LL6 meteorites.

Applying the mineralogical model on the composite VNIR spectrum, the $\frac{ol}{ol+px} = 0.64 $ is obtained. The position of the BIC indicates a molar percentage of Fa of 29.0$\%$ and a molar percentage of Fs of 23.9$\%$. The plot (Fig.~\ref{BIIBI} - Right) of BAR versus BIC reveals an object in the OC region with high olivine content. On the plot of BIC versus BIIC (Fig.~\ref{BIIBI} - Left), this object is placed in the LL region.

Similar results are obtained using the spectrum from MIT-UH-IRTF spectrum. The main difference between the two spectra is the position of the second band center: 1.98 $\pm$ 0.02 $\mu m$ for MIT spectrum compared with 2.12$\pm 0.061$ $\mu m$ for our data.

\subsection{(4486) Mithra}

(4486) Mithra satisfies the D criterion for the similarity with the TC (D = 0.17), but its orbital longitude of perihelia ( $\bar{\omega} = 250^o$) places this object in a separate group, for which the orbital longitude of perihelia lie in the range $220^o<\bar{\omega}<260^o$. Another five objects were identified with similar orbital parameters. The largest asteroid of this group is (2212) Hephaistos \citepads{1993MNRAS.264...93A, 2001A&A...373..329B}. A possible association between (4486) Mithra and $\beta$ Taurids daytime fireballs is suggested by \citeads{1996EM&P...72..311H}.

Using radar images, \citeads{2010Icar..208..207B} found that (4486) Mithra has one of the most bifurcated and irregular shapes seen in the NEA population. They reveal a double-lobed object with a valley between the lobes $\approx$380 m with the maximum dimensions being X = 2.35$\pm$0.15 km, Y=1.65$\pm$0.10 km, Z = 1.44$\pm$0.10 km. Thus, for an absolute magnitude H = 15.6 and the equivalent diameter D = 1.69 km, the geometrical albedo is $p_{v} = 0.36$ \citepads{2010Icar..208..207B}. With an estimation of 67.5 h rotation period, (4486) Mithra is a slow rotator \citepads{2010Icar..208..207B}.

The broadband colors (B-R=1.242$\pm$0.010 mag; V-R=0.428$\pm$0.010 mag; R-I=0.271$\pm$0.013 mag), published by \citeads{2010ATel.2488....1H}, are consistent with  Sk and Sq taxonomic types.

We obtained the NIR spectrum of (4486) Mithra on March 2, 2010 (Table~\ref{Tab1Prop}). This spectrum can be classified between Q and Sq type. The difference between the two classes is mostly given by the visible part (Fig.~\ref{Taxonomy}). If we consider the algorithm suggested by \citeads{2014Icar..227..112D}, the $BI_{min}$ = 0.946$\pm 0.002 \mu m$ may suggest an S-type. However, the curve matching methods show a spectrum similar with the Q and Sq type (e.g. mean square error less than half as compared with S type). Thus, by considering also the result of \citeads{2010ATel.2488....1H}, we can conclude to an Sq type for this object.

The comparison with the spectra from the Relab database shows a match with OC LL6 meteorites spectra. Among the best fits, there are those with the spectra of a particulate samples from Kunashak and Colby (Wisconsin) meteorites. An intruder among the matching solutions is a spectrum of a particulate sample of the OC H5 Nuevo Mercurio meteorite, which can be explained by the degeneracy of the mineralogical solutions.

\subsection{(5143) Heracles}

In late 2011, (5143) Heracles made its closest approach to Earth, since its discovery. According to the radar observations performed from Arecibo Observatory on December 10-13, 2011 \citepads{2012CBET.3176....1T} this asteroid is a binary system. The primary component has a diameter of $3.6 \pm 1.2$ km, while the secondary component has a diameter of $0.6 \pm 0.3$ km. The orbital period of the satellite is between 14 and 17 hours, which implies a separation of at least 4 km.

A total of 41 lightcurves \citepads{2012MPBu...39..148P} were obtained by different observers in the interval October 21 - December 11, 2011. Based on these data a synodic period of $2.706 \pm 0.001$ hrs with three maxima and minima per cycle was found. The amplitude of the lightcurves varied between 0.08 and 0.18 magnitudes, depending on phase angle.

The absolute magnitude reported by \citeads{2012Icar..221..365P} is H = 14.270, which is close to the value derived from the lightcurve observations, H =  14.10.  Very different values are reported for the albedo: based on the  WISE thermal observations, \citeads{2012Icar..221..365P} revised the geometric albedo and found $p_{V} = 0.1481$, while based on the data from Warm Spitzer  \citepads{2011AJ....142...85T} found a value of $p_{V} \approx 0.40$. The effective diameters computed based on these albedo values are $D_{eff} \approx 4.83$ for $p_{V} = 0.1481$ and $D_{eff} \approx 2.94$ for $p_{V} = 0.40$. Both values are in the limit of error bars obtained via radar observations.

There are several spectra available for this object. Table ~\ref{LiteratureSpectra} summarizes the parameters of some of the available spectra in the visible region. The main difference between these spectra is the slope.

Based on these spectral data, the taxonomic classifications V, Sk, O, and Q were proposed. Considering a visible spectrum, \citeads{2002Icar..158..146B} noted that it is similar with spectra of two other NEAs, (4341) Poseidon and 1997 RT, concluding an O taxonomic type, though the 1 $\mu m$ band is not as deep as it is for  Bo\v{z}n\v{e}mcov\'{a}. However, considering the  composite VNIR spectrum, \citeads{2009Icar..202..160D} reclassified  this asteroid as a Q type, because its spectrum does not show the distinct "bowl" shape of the 1 $\mu m$ absorption band.

We merged the visible spectrum from \citeads{2004Icar..170..259B} with our NIR data (Fig.~\ref{Spectra}). Our choice of the visible spectrum was based on the similarity slope in the common spectral interval - 0.135 $\mu m^{-1}$ as compared with $ 0.301 \mu m^{-1} $ for the other two spectra obtained by \citeads{1995Icar..115....1X} and \citeads{2004Icar..169..373L}. Furthermore, the visible spectrum was observed at similar phase angle ($\Phi = 13^{\circ}$ compared to 23.5$^{\circ}$) and has the best SNR.

The obtained VNIR spectrum is classified as a Q-type using both M4AST and  MIT tools, which agree with  \citeads{2009Icar..202..160D}. The spectrum is characterized by a large and deep 1 $\mu m$ absorption band and has also a small feature around 1.3 $\mu m$ (Fig.~\ref{Taxonomy}). While the match in the NIR part is almost perfect with the Q-type, there are several differences in the visible part where it is closer to the O-type.

The comparison with the spectra from the Relab database shows that the closest spectral fit is obtained with LL6 and L6 OC meteorites. All the curve matching methods give the first ten matchings as OC LL6 and L6. Most of the spectra that fit (5143) Heracles are those of the meteorites Bandong, Colby (Wisconsin) and Jelica (Table~\ref{CHIT}). The best-fit solution was obtained with a spectrum of a particulate sample (0-150 $\mu m$) from the Bandong meteorite (Sample ID: MH-FPF-050-B). Bandong\footnote{\url{http://www.nhm.ac.uk/}} is an LL6 ordinary chondrite fallen on December 10, 1871 in Java, Indonesia. All three of these meteorites with spectra similar to Heracles spectrum fell at the end of the Taurid meteor shower period.

Applying the mineralogical analysis on the composite VNIR spectrum, a $\frac{ol}{ol+px} = 0.68$ is obtained. The position of the first band center, which is located at BIC = 0.998 $\pm 0.005 \mu m$, allows the computation of a molar Fa of 29.1$\%$ and a molar Fs of 24$\%$. These data fully agree with the results found by \citeads{2013Icar..222..273D}. The plot of BIC versus BIIC place the object in the OC LL region, while it is placed in the OC region with high olivine content in the plot BAR versus BIC (Fig.~\ref{BIIBI}). 

The VNIR spectrum has a small slope (0.07$/\mu$m), which suggests an unweathered surface. \citeads{2010Natur.463..331B} explained the unweathered spectra by close encounter with major planets (closer than the Earth - Moon distance) within the past 0.5 Myears. 

\subsection{(6063) Jason}

(6063) Jason is a large Apollo type asteroid that most fit the D-criterion with the TC, D = 0.07. This value is comparable to D = 0.04 for the comet P/Encke, which is considered as the primary source of the Taurid meteor shower.

Observations in the NIR spectral region for this asteroid were carried out by \citeads{1988Icar...73..482B} on June 2, 1984 using the IRTF Telescope. They performed photometry in 52 passbands from 0.8 to 2.5 $\mu m$. Their spectrophotometric results correspond to an S type object.

Applying the standard  thermal model, \citeads{1988Icar...73..482B} suggested an albedo around 0.16, as derived from observation using broad band N (10.1 $\mu m$). This value is close to the average value for S-type asteroids.

According to the NEODyS website, the absolute magnitude H derived from astrometric observation via orbit determination is 15.9 (comparable with H = 16.7 $\pm$ 0.2 found by \citeads{1988Icar...73..482B}). Thus, the effective diameter can be estimated to $D_{eff} \approx 2.2$ km.

We observed this object at two different heliocentric distances:
\begin{itemize}
\item {on September 21, 2013 when the object was at heliocentric distance 1.632 AU and $\Phi = 29.0^{\circ}$ (the obtained spectrum will be later denoted as SpecJ1). 
Because it was observed at an apparent magnitude of 17.7, this spectrum has a low  SNR;}
\item{ on October 21, 2013 when the object was at heliocentric distance of 1.269 AU and $\Phi = 25.2^{\circ}$) 
(hereafter denoted as SpecJ2).}
\end{itemize}

Both spectra exhibit features characteristic of the S-type complex with two bands around 
1 $\mu m$ and 2 $\mu m$ and a small feature around 1.3 $\mu m$. However, there are some distinguishable differences between the two spectra:
\begin{itemize}
\item {the slope (computed in the 0.82-2.5 $\mu m$) of the SpecJ1 is 0.038 $\mu m^{-1}$ as compared with 0.205 $\mu m^{-1}$ for SpecJ2, which is redder than SpecJ1;}
\item {the first band minimum for SpecJ1 is at 0.968 $\pm$ 0.009 $\mu m$, while it is at 0.939 $\pm$ 0.003 $\mu m$ for SpecJ2;}
\item {different band depths.}
\end{itemize}

Classifying these two spectra in Bus-DeMeo taxonomy, several choices can be assigned within the S complex. Thus, SpecJ1 can be classified between the S and Sq subtype (Fig.~\ref{Taxonomy}). The S-type is the best ranked by the MIT tool and is also among the first solutions obtained using curve matching methods. However, the position of the first $BI_{min}$, which is larger than 0.96 $\mu m$ suggests the Sq type (which has the second rank after using MIT tool but is not found by curve matching). The spectrum SpecJ2 is classified in Bus-DeMeo taxonomy between Sq and Q types using both M4AST and MIT tools. While according to MIT tool, the Q type has average residuals less than the Sq type (0.027 compared with 0.031), the Sq type better fits this spectrum. Also, most of the parameters (except $BI_{min}$ which is typical for S type), which characterize first band \citepads{2014Icar..227..112D}, are in the range of Sq type \citepads{2014Icar..227..112D}.

The comparison with spectra from the Relab database with SpecJ1 shows very good similarities with some experimental samples  such as  481/652-C3 Ilm, which are particulate shocked ilmenite with a size less than 250 $\mu m$. Another spectrum that also fits this asteroid spectrum is of a particulate lunar sample "2014 Luna 20 soil 250-1000 $\mu m$". Among the meteorite spectra, the majority of matchings are with OC L and LL, petrologic types 4-6. The meteorite spectra with the best matching coefficients is of a particulate sample from the OC LL5 Paragould (sizes between 25-250 $\mu m$).

The spectra from Relab that fit SpecJ2 are those of OCs with low iron content (most of them L5 or L6 subtypes). The best fit of SpecJ2 is the spectrum of a sample from the OC L6  Colby (Wisconsin) meteorite. The spectrum SpecJ2 is also matched by the Farmington meteorite, which is already associated to the Taurid shower \citepads{1993MNRAS.264...93A}.

The taxonomy and comparison with meteorite spectra shows a surface composition similar with the general type of OC meteorites. However, the small differences seen in spectra could not be explained just by the observing geometry, since the phase angles are almost identical, and the asteroid temperature has varied between 214 K to 242 K \citepads{2013Icar..222..273D}. Thus, we speculate on an inhomogeneous composition of the surface. The different positions of the $BI_{min}$ imply different Fa and Fs ratios (which cannot be accurately computed without the visible part).

The spectrum of (6063) Jason obtained in November 12, 2013 when the asteroid was at 0.98 AU is available in the MIT-UH-IRTF database. While the positions of the two band minima ($BI_{min}$ and $BII_{min}$) are the same (within the error bars) as for SpecJ2, the band around 2 $\mu m$ ($BII$) of SpecJ2 is more shallow than that of the spectrum from the MIT-UH-IRTF. This could be an argument to speculate on the  heterogeneity of the asteroid's surface.

\subsection{(269690) 1996 RG3}

Few data are known about (269690) 1996 RG3, a PHA discovered in September 1996 by the Spacewatch survey at Kitt Peak. It has an absolute magnitude H = 18.5, which suggests an object with the diameter around 1 km. The delta-V required for a spacecraft rendezvous with this object is 6.859 km/s, making it accessible for a space mission.

To investigate the influence of the object temperature and phase angle ($\Phi$) on spectral data, we observed this object twice(Table~\ref{Circumstances}). The object was observed when it was at the heliocentric distances r = 1.428 AU ($\Phi = 3.5^{\circ}$) on September 21 and r = 1.153 AU ($\Phi = 46^{\circ}$) on October 21. Both spectra are plotted in Fig.~\ref{Spectra}.

The spectrum from September 21 was obtained in better atmospheric conditions than the one from October 21. The zenith opacity measured by the Caltech Submillimeter Observatory  in September was three times smaller than the one in October. Both nights were bright (Full Moon), and the target was in the apparent vicinity  of the Moon. The spectrum taken in October was cut at 1.75 $\mu m$ because no reliable data were obtained in the 1.75 - 2.5 $\mu m$  spectral region due to precipitable water from the atmosphere.

\begin{figure}[!ht]
\begin{center}
\includegraphics[width=9cm]{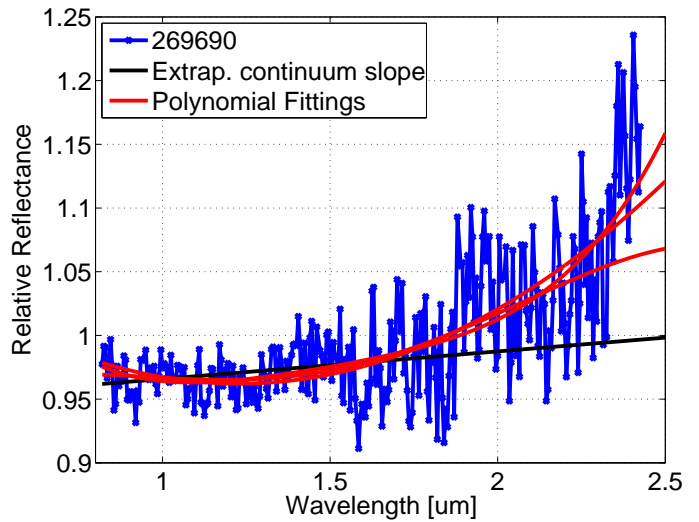}
\end{center}
\caption{Estimation of thermal flux in the spectrum (SpecJ1) of (269690) 1996 RG3. The black line indicates where a linearly extrapolated continuum would fall, and the red lines are different polynomial fittings (illustrating the interval for the extrapolated reflectance at 2.5 $\mu m$) showing the presence of thermal flux.}
\label{TE269690}
\end{figure}

Both spectra are featureless with increasing reflectance at the end of spectrum, which can be interpreted as thermal flux and is typical for a primitive object. The taxonomic classification also supports the hypothesis of a primitive object. According to the MIT online tool, the spectrum obtained in September can be classified as Cb with an average residual 0.033, or C with an average residual 0.042 (other possible solutions could be B, L, and X). Using M4AST curve matching methods to determine the taxonomic type, Cg, Cb and C classes are suggested. Since no feature can be observed in the 1 - 1.3 $\mu m$ interval (which is characteristic to C type) and a small positive slope begins around 1.3 $\mu m$, Cg is the most plausible taxonomy for this object (Fig.~\ref{Taxonomy}). The class Cg contains spectra similar with (175) Andromache. We applied G13 classification on the spectrum obtained on September 21 and found that it belongs to the group 3, which contains objects, such as (1) Ceres, (10) Hygiea, and (106) Dione.

We used M4AST to assign a taxonomic type for the spectrum obtained in October, since we have only the 0.82 - 1.75 $\mu m$ spectral interval. Within a reliability factor of 46.3\% (given that the computation is performed only on half of the NIR spectral interval), the types that match this spectrum are C, Cg, and Cb. This confirms the classification already presented.

Even if the level of noise is significant, it can be assumed that the tail, which starts at 2.1 $\mu m$ is caused by asteroid thermal emission, which characterizes primitive NEAs when near perihelion \citepads{2005Icar..175..175R}.  Thus, to estimate the albedo of this object, we computed the "thermal excess" ($\gamma$) as defined by  \citeads{2005Icar..175..175R} (see Fig.~\ref{TE269690}). The $\gamma$-value (eq.~\ref{Texces}) is computed by fitting polynomial curves (degree between 2 and 4) on different spectral intervals.

\begin{equation}
\gamma = \frac{R_{2.5} + T_{2.5}}{R_{2.5}}-1 = 0.123 \pm 0.053,
\label{Texces}
\end{equation}

where $R_{2.5}$ is the reflected flux at 2.5 $\mu m$ and $T_{2.5}$ correspond to thermal flux at 2.5 $\mu m$ (eq.~\ref{Texces}).

Considering the Figs. 2 and 3 of \citeads{2005Icar..175..175R}, a very low value of the albedo less than 0.04 results, which suggests a diameter around 1.5 km (using $p_V = 0.03$ and $H = 18.5$).

A NIR spectrum of this object was obtained on November 12, 2013 in the MIT-UH-IRTF survey, when the object was at $\approx$ 0.84 AU and $\Phi = 109.7^{\circ}$. It also shows a thermal excess tail but is not as high as expected based on our estimation. This can be explained by the high phase angle that is outside the range described in Fig. 4 of \citeads{2005Icar..175..175R}, which reduces significantly the value of thermal excess.

A comparison with the meteorite spectra from the Relab database (even though it is limited to a noisy and featureless spectrum) gives spectra similar to those of carbonaceous chondrites CM2 and CI. In Fig. ~\ref{FG3Met}, we plotted three of the meteorite spectra, which best fit the spectrum of (269690) 1996 RG3.

\begin{table*}
\caption{The intervals of the orbital intersection distances computed for each close approach between Earth and asteroid orbit for the period  1850-2010. The table shows two intervals of dates for the year when the two orbital close approaches occur. These intervals are computed for the same period (1850-2010).}
\label{t1}
\centering
\begin{tabular}{l c l c l}
\hline \hline
\noalign{\smallskip}
Asteroid & 1st min. (A.U.) & Date & 2nd min. (A.U.) & Date \\
\hline
\noalign{\smallskip}
(2201) Oljato & 0.000250 - 0.006931 &  Jun 5 - 10 & 0.000296 - 0.011437 &   Dec 17 - 22 \\
(4183) Cuno & 0.028469 - 0.045370 &  Dec 29 - Jan 1 &  0.089301 - 0.103171 &  May 25 - 29 \\
(4486) Mithra & 0.043581 - 0.046179 &  Aug 9 - 14  & 0.051784 - 0.052446 &   Mar 20 - 24 \\
(5143) Heracles & 0.058617 - 0.082383 &  Jul 7 - 11 &  0.134340 - 0.155174 &   Nov 30 - Dec 2 \\
(6063) Jason & 0.055263 - 0.074456 &  Nov 12 - 14 & 0.063650 - 0.083464 &   May 20 - 25 \\
(16960) 1998QS52 & 0.000191 - 0.062847 &  Jun 12 - 15 &  0.242405 - 0.305853 &   Oct 21 - 23 \\
(269690) 1996RG3 & 0.000033 - 0.004146 &  Feb 25 - Mar 2 &  0.047266 - 0.052012 &   Oct 27 - 29 \\
\noalign{\smallskip}
\hline
\end{tabular}
\label{pa8t}
\end{table*}

\section{Discussions}

Recently, the topic of meteors and superbolids and their relation to parent bodies was reviewed after the explosion of a massive body over Chelyabinsk, Russia in February 15, 2013. This air-burst estimated at  500$\pm$100 kilotons of TNT \citepads{2013Natur.503..238B} could be associated with (86039) 1999 NC43\footnote{(86039) 1999 NC43 is a 2 km-size NEA} in terms of dynamical parameters \citepads{2013Natur.503..235B}. The most probable origin of the bolide is from the inner belt $\nu_{6}$ region. The ejection velocity of the Chelyabinsk bolide from its parent body is estimated between 0.7 and 2 km/s and is consistent with a collision with another asteroid.

\begin{figure}[!ht]
\begin{center}
\includegraphics[width=6cm,angle=270]{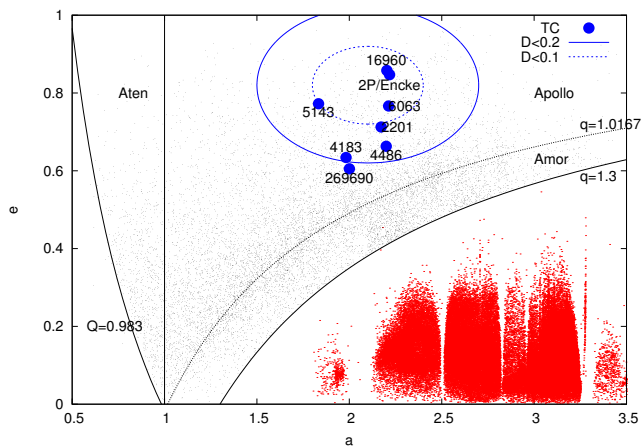}
\end{center}
\caption{The grouping of TC asteroids in a representation ($a,e$) of all asteroids, where $a$ is the semi-major axis and $e$ the eccentricity.}
\label{aeplot}
\end{figure}

All the asteroids belonging to TC have Apollo type orbits (Fig.~\ref{aeplot}).  Moreover, (2201) Oljato, (4183) Cuno, (4486) Mithra, (16960) 1998 QS52, and (269690) 1996 RG3 are some of the largest PHAs. Thus, the probability that some meteoroids and meteorites originate in these objects is not negligible. This hypothesis is supported by the low values of minimum orbital intersection (MOID), see Table~\ref{pa8t}. Furthermore, based on the data shown by \citeads{2000came.book.....G}, we found that there is a certain peak in the number of falls of OC meteorites  corresponding to May, June, and July, which can be correlated with the dates of closest Earth, these include asteroid orbits (2201, 4183, 5143, 6063, 16960), as shown in Table~\ref{pa8t}.

For our discussion, we also consider the spectral properties of (16960) 1998 QS52, which also belongs to TC, having a D factor of  0.24 \citepads{2008MNRAS.386.1436B}. Its spectrum was already published by \citeads{2011A&A...535A..15P}. Based on the VNIR it was classified as a Sr type asteroid. This asteroid has spectral properties similar to L4, LL4, and L5 subtypes of OC meteorites. This spectrum is unreddened, corresponding to a fresh surface. The values of  BIC = 0.97, BIIC = 2.03, and BAR = 0.232 correspond to $\frac{ol}{ol+px} = 0.63$, to molar percentage of Fa of 25.5$\%$ and to molar percentage of Fs of 21.3$\%$, respectively recomputed using the \citeads{2010Icar..208..789D} model.

A unique generator at the origin of TC was sustained and argued by introducing non-gravitational forces into the backward 
integration approach \citepads{1996MNRAS.280..806S}. Thus, if the objects of TC cluster are indeed of common origin (same progenitor), this could be explained only if non-gravitational forces are used for the efficient dispersion of fragments over the longitude of perihelion corresponding to observational evidences of Taurid showers\footnote{D-criterion used by \citeads{1996MNRAS.280..806S} and \citeads{1993MNRAS.264...93A}, as defined in the space (a,e,i).}. Besides the enlargement of the number of objects inside TC, justifying the evolution of objects using gravitational and observed non-gravitational forces could correlate the common origin of both TC and Hephaistos group of asteroids \citepads{1996MNRAS.280..806S}. Thus, \citeads{1996MNRAS.280..806S} conclude that  TC are debris produced by the disintegration of a large comet, which occurred 20,000-30,000 years ago.

In the hypothesis of the cometary progenitor, P/Encke and TC could have similarities in terms of mineralogy and chemical structure. Thus, a chondritic porous matrix impregnated with carbonaceous volatiles is the most probable structure of TC objects. This implies that the original source was large enough to sustain an internal/local differentiation \citepads{1993QJRAS..34..481A}. The development of an insulating crust of TC will not be enough for maintaining a fragment nucleus of a comet for a long time \citepads{1992CeMDA..54....1W}. Sooner or later, these {\it dormant comets} are reactivated either by internal factors, radionuclides heating, cracking, and dehydration at extreme perihelion temperatures, or by impacts with other debris/meteoroids.

Among this sample of seven objects, representing the largest asteroids from the TC, only (269690) 1996 RG3 has a flat featureless spectrum, which can be associated with a primitive C-type object. The other six asteroids present spectra similar to OC meteorites with high petrologic classes (typically 6), which suggest an evolved surface. By looking to the comparison with meteorite spectra, we can observe that there is a certain spread among the solutions found: (2201) Oljato has a "blue" spectrum, which is similar with OC H; (4183) Cuno and (5143) Heracles are similar with OC LL, while (4486) Mithra, (6063) Jason, and (16960) 1998 QS52 have spectra similar with OC L type. Even if spectral comparison has its limitations \citepads{2010Icar..209..564G}, it provides clues about the surface compositional variation among these objects.

The asteroids belonging to TC are on very eccentric orbit, with Tisserand parameter close to the limit of three, which marginally indicates an association with Jupiter family comets.  From our sample, (2201) Oljato, (5143) Heracles, (6063) Jason  and (16960) 1998 QS52 cross the orbits of Venus, Earth, and Mars, while (4183) Cuno, (4486) Mithra, and (269690) 1996 RG3 cross the orbit of Earth and Mars and approach the orbit of Venus. The close encounters to telluric planets in the recent past can explain their fresh surface \citepads{2010Natur.463..331B, 2014Icar..227..112D}, as described by the small spectral slope values of (2201) Oljato, (4183) Cuno, (5143) Heracles and (16960) 1998 QS52. These slopes are not frequent among NEAs belonging to the S-complex, being typical for OC meteorites (e.g. Fig.~7 from \citepads{2010AA...517A..23D}). They noted that the small spectral slope is frequent among small NEAs, explaining that they may have lost their regoliths during the collision that most likely created them and were unable to develop and retain new regolith, preserving preferentially larger grains on their surface. Experiments in the laboratory with meteorite samples found that reflectance spectra are darker and bluer for coarser grain sizes, supporting their hypothesis. 

\citeads{2008Natur.454..858V} made a histogram of the $\frac{ol}{ol+px}$ ratios of 57 ordinary chondrites. This histogram has a peak around 0.65 which corresponds to OC-L meteorites that are in contrast with a similar histogram of 38 NEAs belonging to the S-complex, which has a peak around 0.8. Compared with our results, we observe that the spectra of Cuno, Heracles, and 1998 QS52 have $\frac{ol}{ol+px}$ ratios of 0.64, 0.68, and 0.63, respectively, which are close to the peak computed by \citeads{2008Natur.454..858V}.

Radar observations of Cuno, Mithra, and Heracles were made. While Heracles is a binary object, Cuno and Mithra have a very elongated shape with concavities, which may suggest a violent history and are most likely a rubble pile structure.

The asteroids studied in this paper did not show the spectral characteristics similar with other asteroids associated with meteor showers like (3200) Phaeton. \citeads{2007A&A...461..751L} noted that the asteroids associated with meteor showers are B-type asteroids with a featureless spectrum in the NIR, a slightly blue gradient, and are curved over the whole 0.43 - 2.5 $\mu m$ spectral region.

While the cluster of dynamical parameters could be an indicator of a common origin, the spectral data of the largest asteroids from the TC do not support the hypothesis of a common cometary origin. Furthermore, there are significant variations between the spectra acquired until now. With the exception of (269690) 1996 RG3, which has the characteristics of a primitive object, the other six objects have spectra similar with OC. These asteroid spectra are also different, spanning all subtypes of ordinary chondrites.

\section{Conclusions}

Six asteroids of TC were spectroscopically investigated using NIR low resolution spectra, as obtained using SpeX/IRTF. Our observed  asteroids span diameters larger than 1 km. Our spectra were extended with visible spectra from the literature, when available.

Several conclusion are derived from our observational data and analysis:
\begin{itemize}
\item {Five of our objects, (2201) Oljato,(4183) Cuno, (4486) Mithra, (5143) Heracles, and (6063) Jason, present spectral characteristics similar to the  S taxonomic complex with spectral bands around 1 and 2 $\mu m$. Their spectra are similar with OC meteorites. For each of these five asteroids, we found good fits with some meteoritic samples associated to Taurid shower falls;}
\item {(269690) 1996 RG3 presents a flat featureless spectrum with an increasing reflectance after 2.1 $\mu m$; this thermal excess constrains the geometrical albedo to values around 3$\%$;}
\item {The asteroid (269690) 1996 RG3 is the only target of our sample which could be associated to a primitive C-type object. 
Its spectral characteristics could be associated to one of cometary material. Thus, we can speculate on the 
common origin of this asteroid and the comet P/Encke, also a member of TC; }
\item {The spectral observations are obtained only for objects larger than 1 km. However, the number of primitive asteroids belonging to TC could be size dependent. Thus, mineralogical resemblance between comet P/Encke and TC asteroids should continue to be investigated.}
\end{itemize}


\begin{acknowledgements} 
This work was supported by a grant of the Romanian National Authority for Scientific Research, Program for research - Space Technology and Advanced Research - STAR, project number 67. This research utilizes spectra acquired with the NASA RELAB facility at Brown University. The asteroid spectra where acquired using NASA Infrared Telescope Facility as well as the CODAM and AIRA ROC remote facilities. We thank all the telescope operators for their contribution. AIRA ROC was funded by a grant of the Romanian National Authority for Scientific Research, CNCS-UEFISCDI project number PN-II-RU-TE-2011-3-0163. We thank Nick Hammar and Andreea Popescu for spelling corrections.
\end{acknowledgements}

\bibliographystyle{aa}
\bibliography{AST3TC}

\Online
\newpage
\clearpage
\onecolumn

\begin{appendix}
\section{Asteroid spectra vs meteorites spectra}
\label{Anexa1}

The closest spectral matches between the reflectance spectra of the asteroids analyzed in our article and the laboratory spectra of different meteorites. Additional details related to meteorite samples are given in Table ~\ref{CHIT}. 

\begin{figure*}[!ht] 
\begin{center}
\centering
\subfloat[ ]{\label{fig:2201a2met}}\includegraphics[width=6cm]{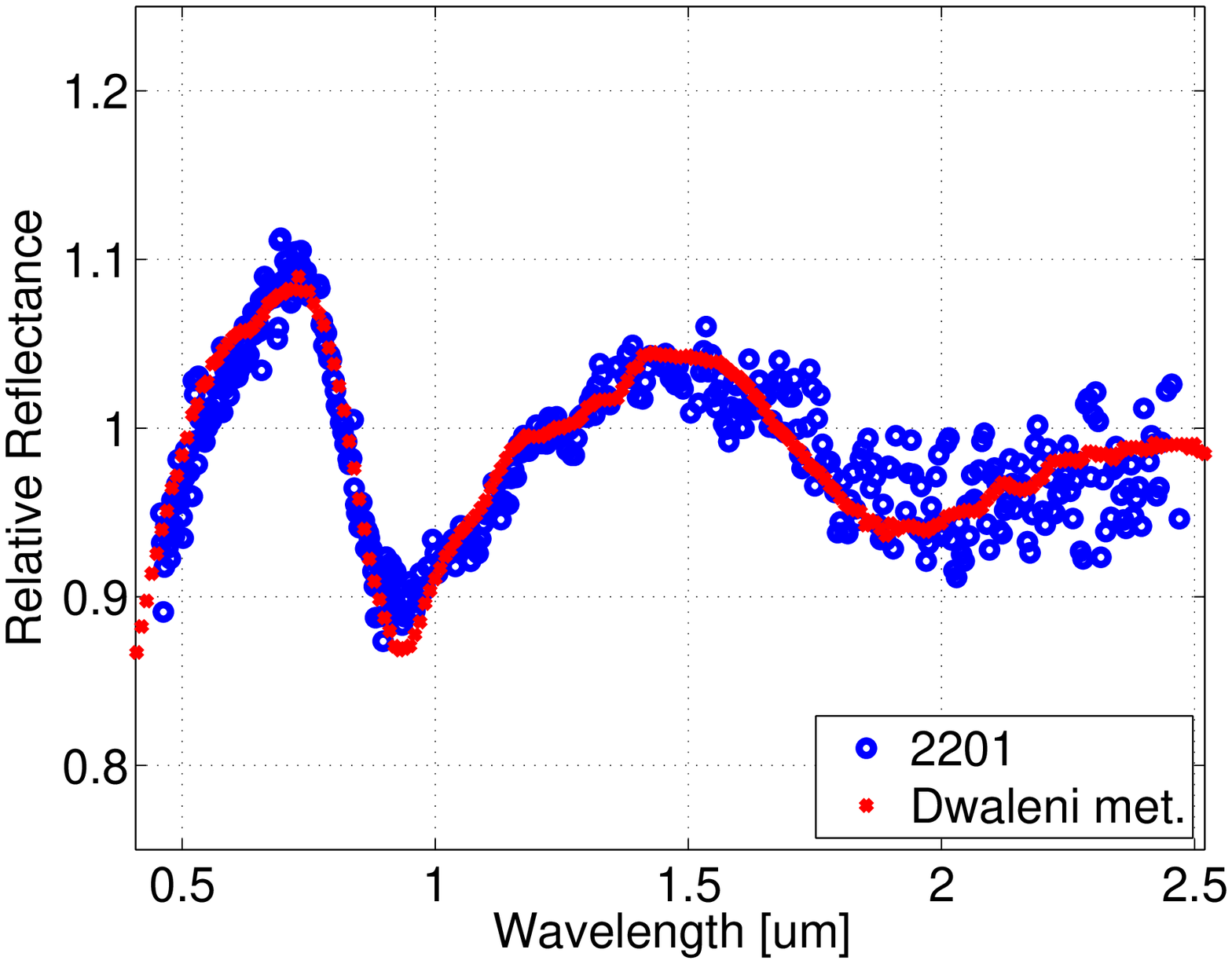}
\subfloat[ ]{\label{fig:2201b2met}}\includegraphics[width=6cm]{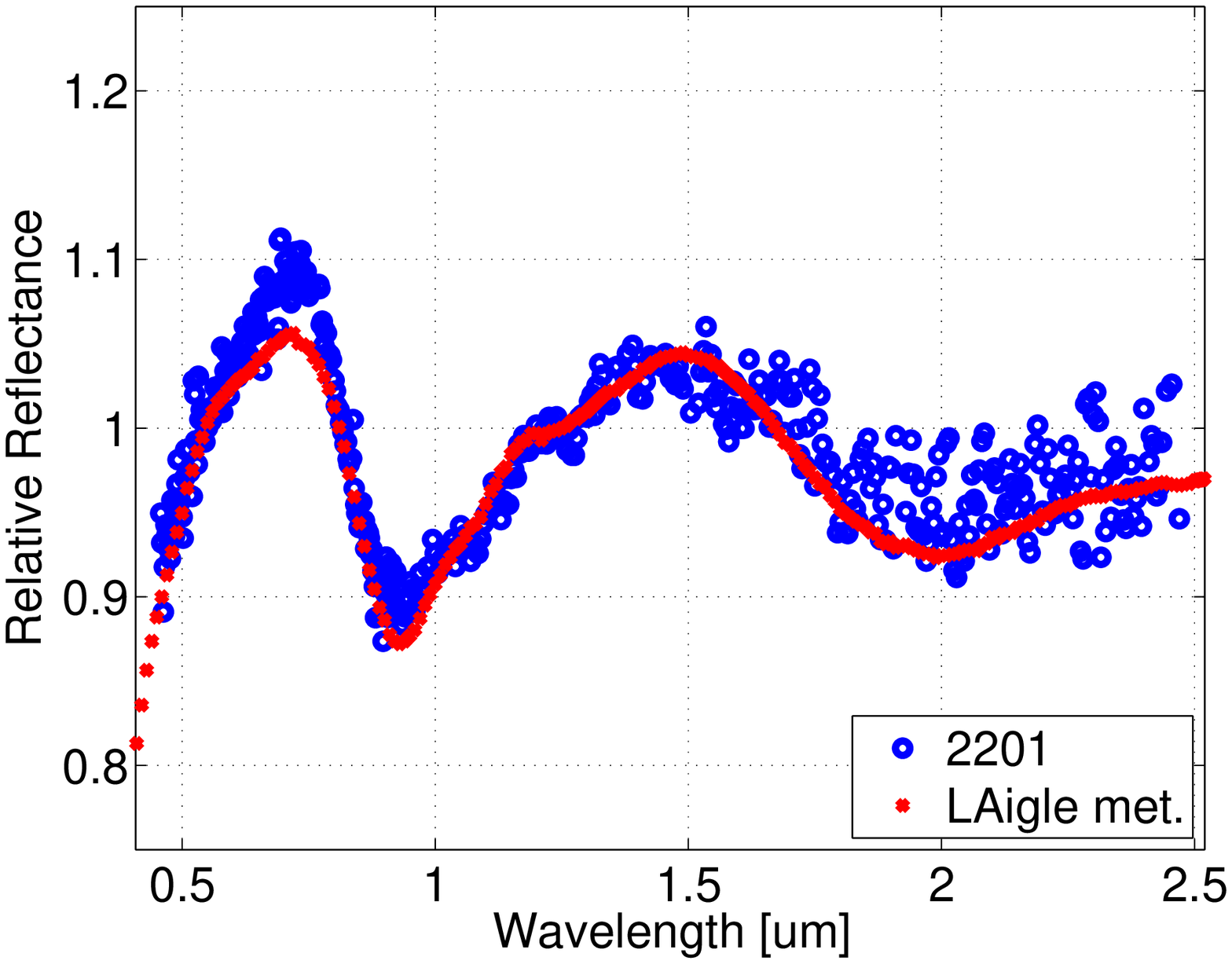}
\subfloat[ ]{\label{fig:2201c2met}}\includegraphics[width=6cm]{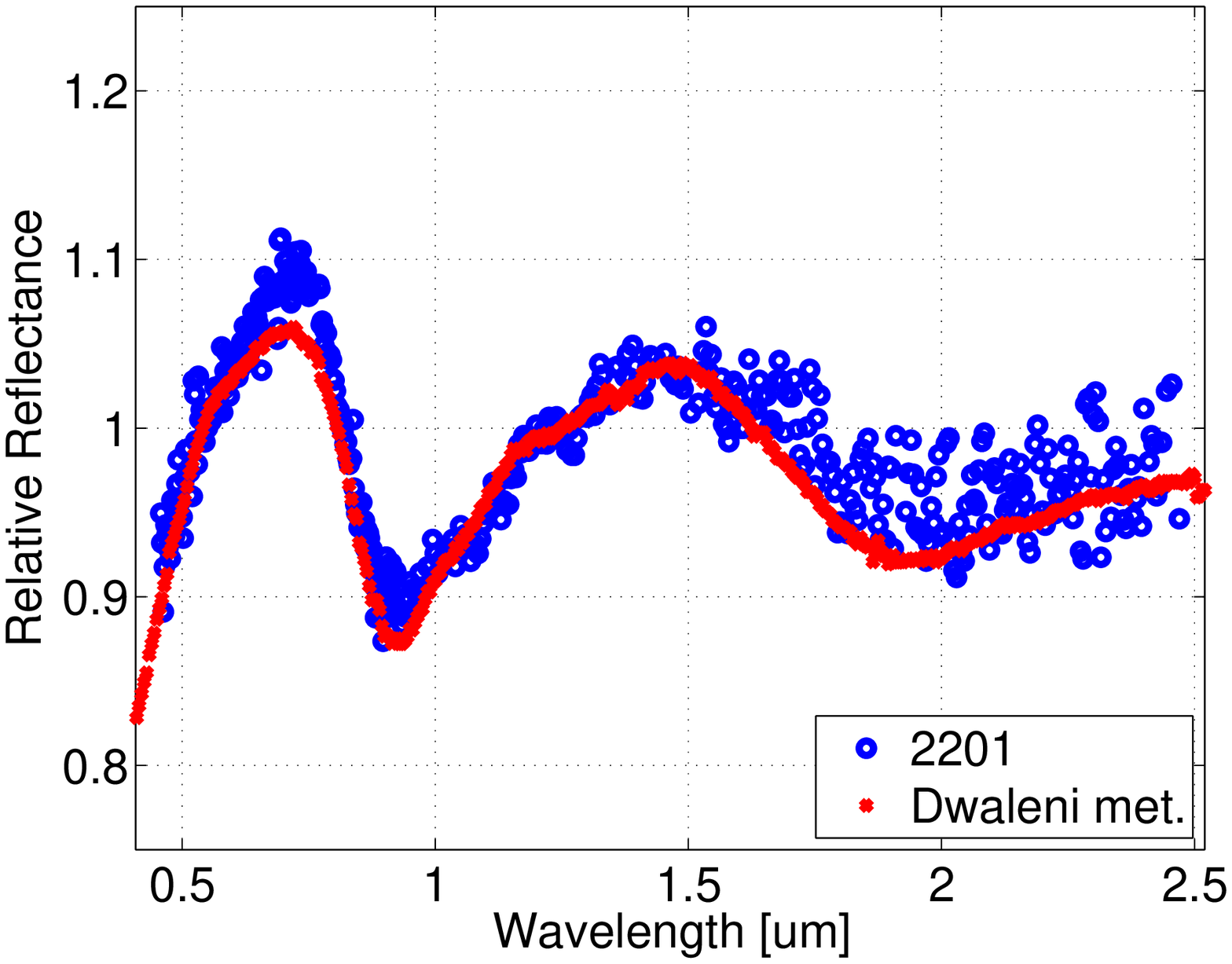}
\end{center}
\caption{Reflectance spectrum of (2201) Oljato and the closest three matches resulting from meteorite comparison: (a) H6 ordinary chondrite Dwaleni (Sample ID: TB-TJM-073); (b)  L6 ordinary chondrite L'Aigle (Sample ID: TB-TJM-141); (c) H6 ordinary chondrite Dwaleni (Sample ID: MB-CMP-003-D).}
\label{OljatoMet}
\end{figure*}

\begin{figure*}[!ht] 
\begin{center}
\centering
\subfloat[ ]{\label{fig:4183a2met}}\includegraphics[width=6cm]{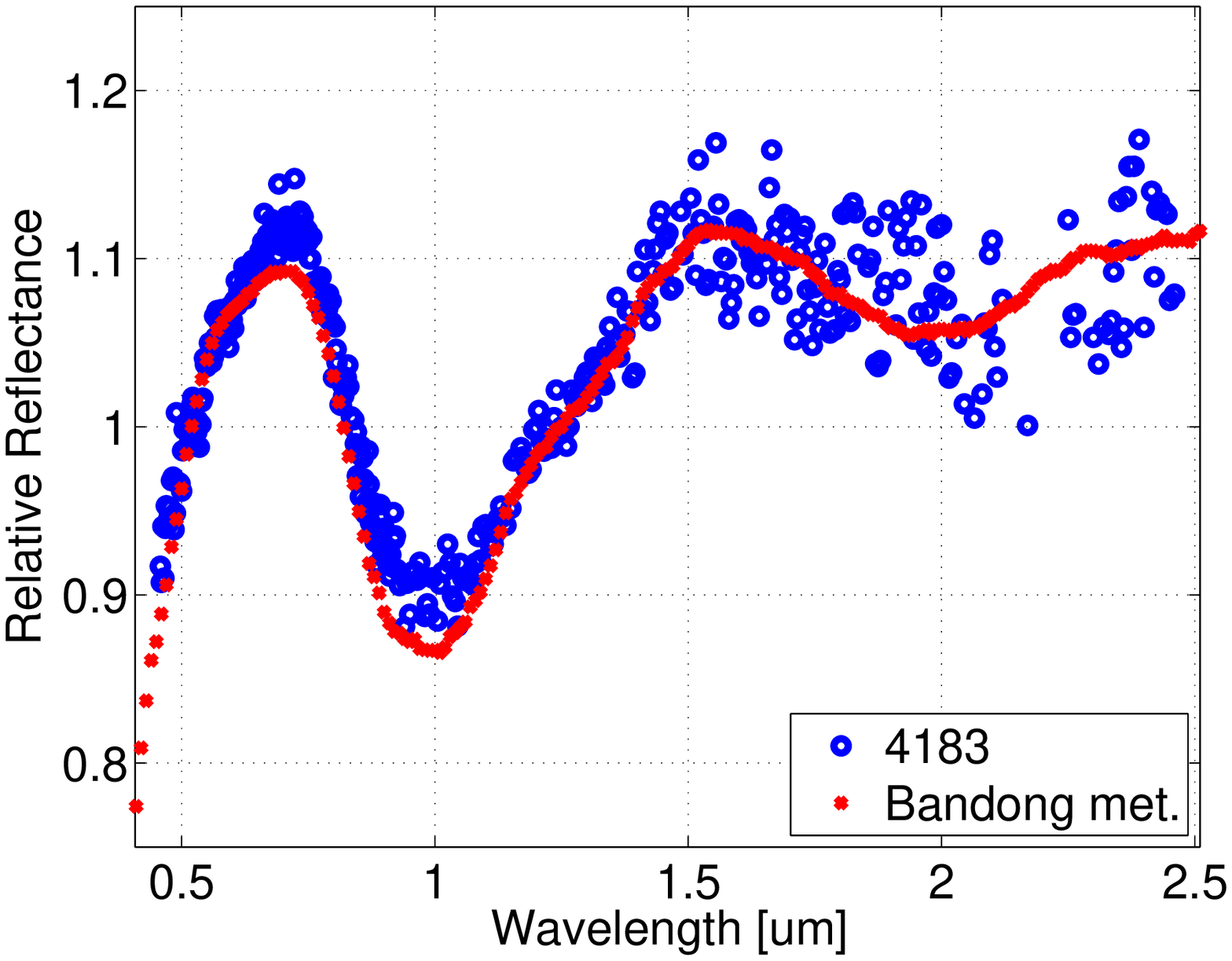}
\subfloat[ ]{\label{fig:4183b2met}}\includegraphics[width=6cm]{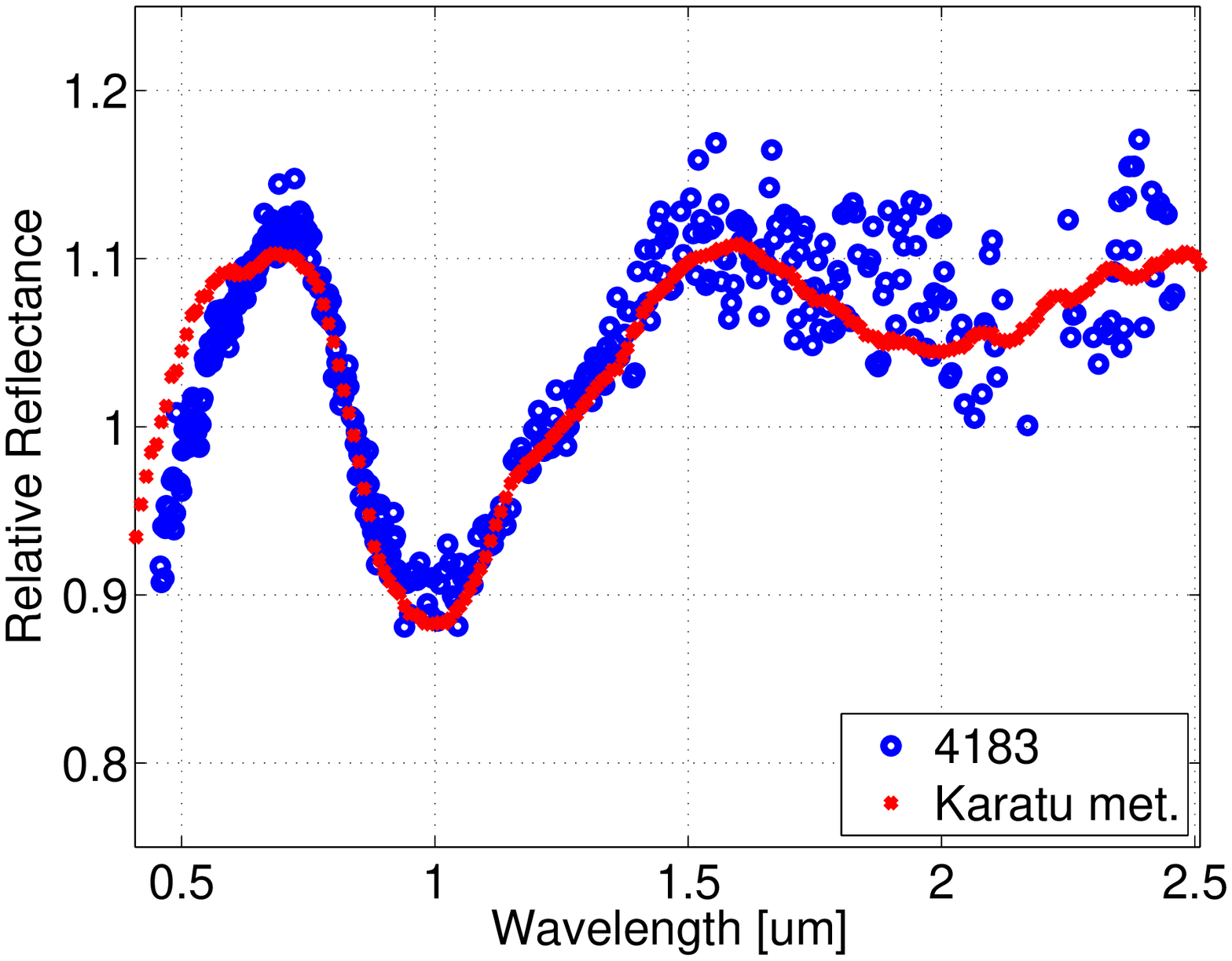}
\subfloat[ ]{\label{fig:4183c2met}}\includegraphics[width=6cm]{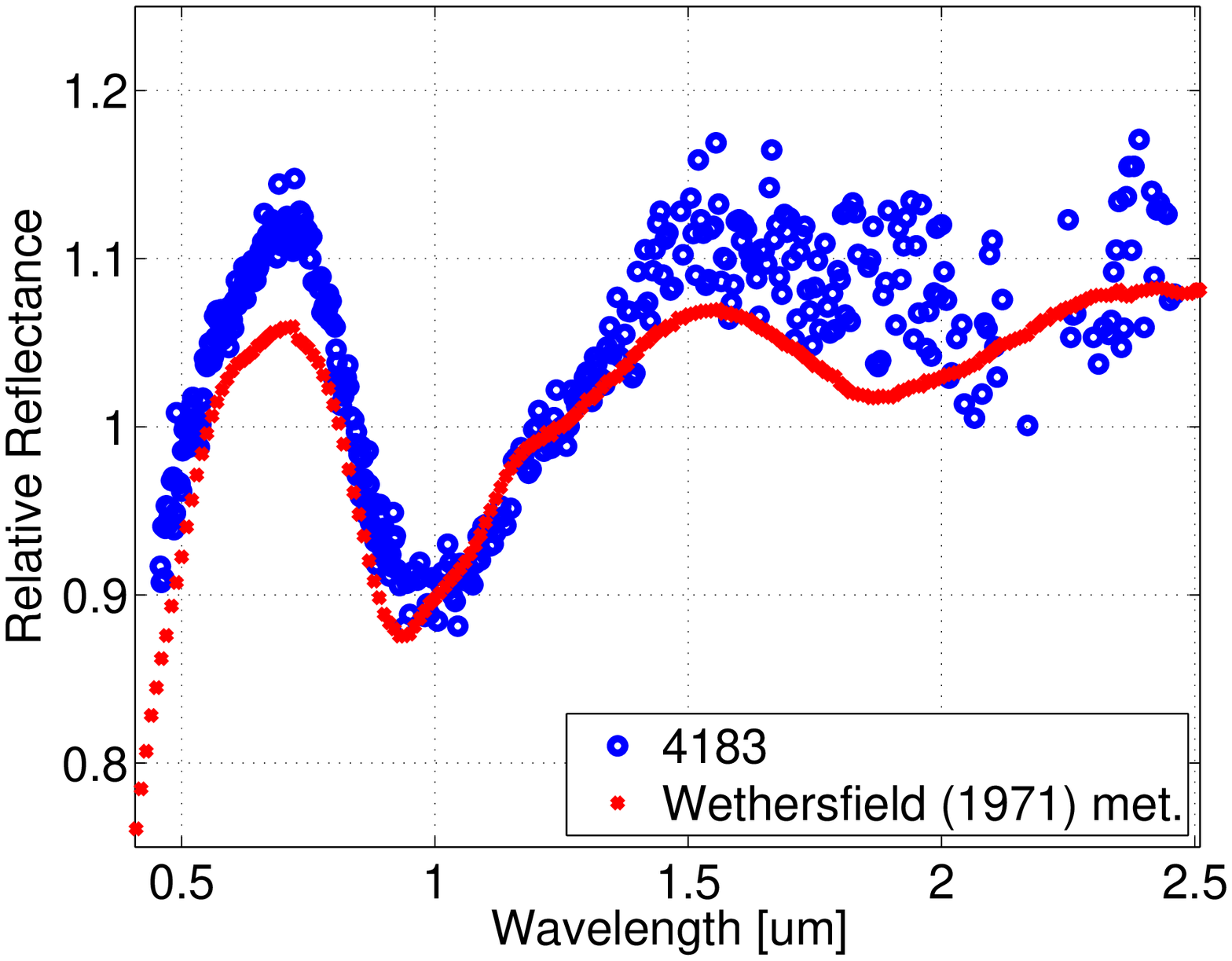}
\end{center}
\caption{Reflectance spectrum of (4183) Cuno and the closest three matches resulting from meteorite comparison: (a) LL6 ordinary chondrite Bandong (Sample ID: TB-TJM-067); (b) LL6 ordinary chondrite Karatu (Sample ID: TB-TJM-077); (c) L6 ordinary chondrite Wethersfield/1971 (Sample ID: TB-TJM-144).}
\label{CunoMet}
\end{figure*}

\begin{figure*}[!ht] 
\begin{center}
\centering
\subfloat[ ]{\label{fig:4486a2met}}\includegraphics[width=6cm]{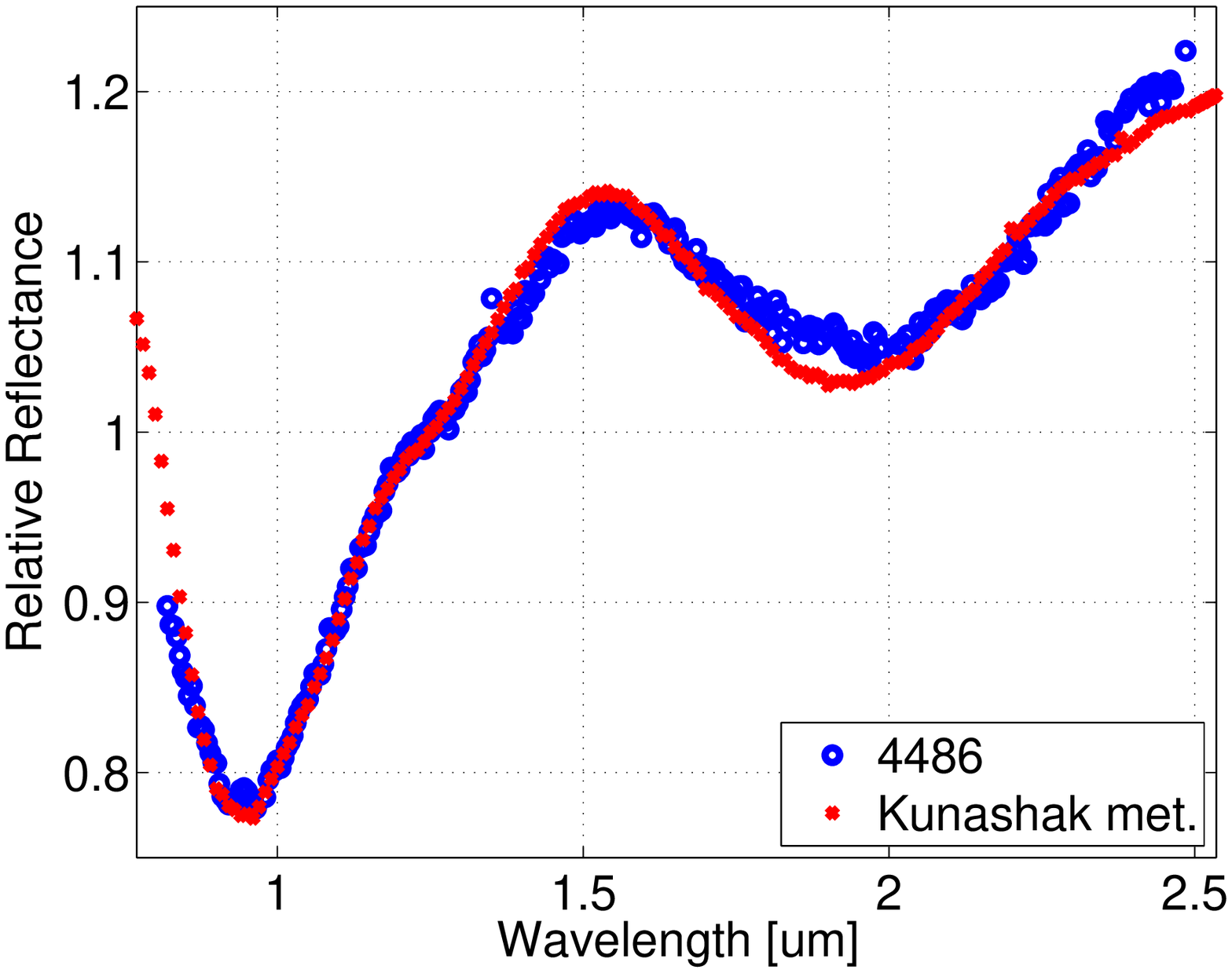}
\subfloat[ ]{\label{fig:4486b2met}}\includegraphics[width=6cm]{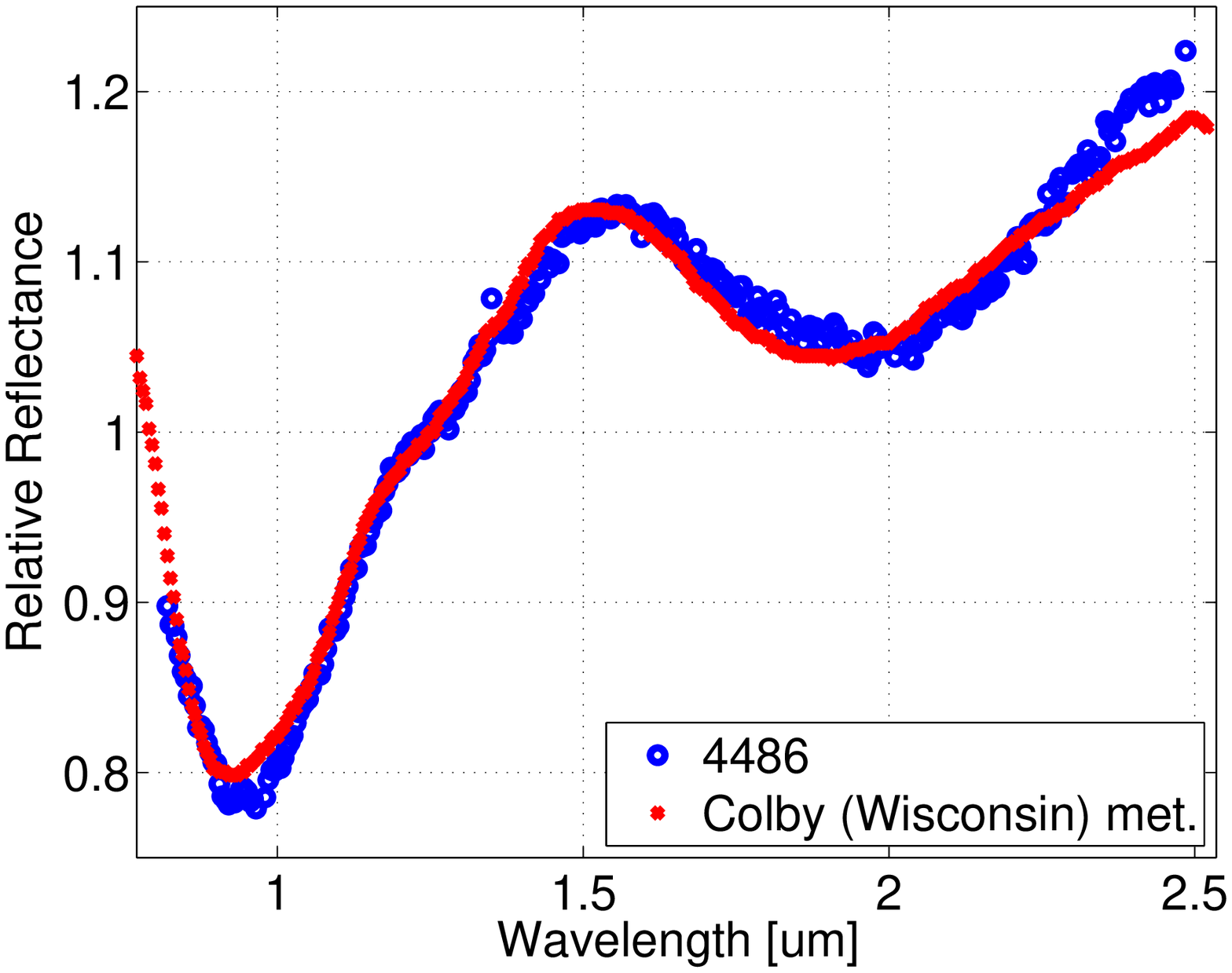}
\subfloat[ ]{\label{fig:4486c2met}}\includegraphics[width=6cm]{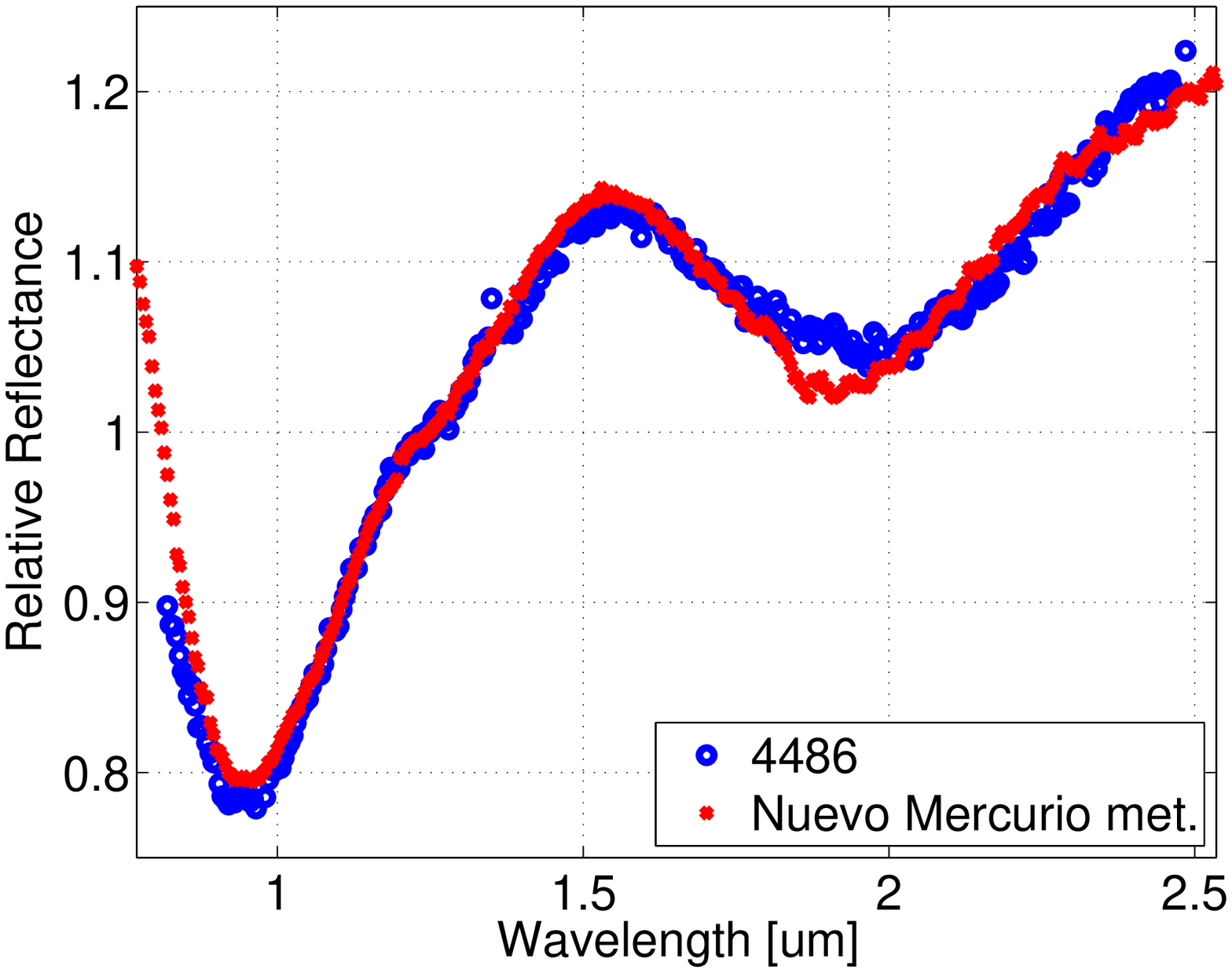}
\end{center}
\caption{Reflectance spectrum of (4486) Mithra and the closest three matches resulting from meteorite comparison: (a) L6 ordinary chondrite Kunashak (Sample ID: TB-TJM-139); (b) L6 ordinary chondrite Colby (Wisconsin) (Sample ID: MR-MJG-057); (c) H5 ordinary chondrite Nuevo Mercurio (Sample ID: MH-FPF-053-D).}
\label{MithraMet}
\end{figure*}

\begin{figure*}[!ht] 
\begin{center}
\centering
\subfloat[ ]{\label{fig:5143a2met}}\includegraphics[width=6cm]{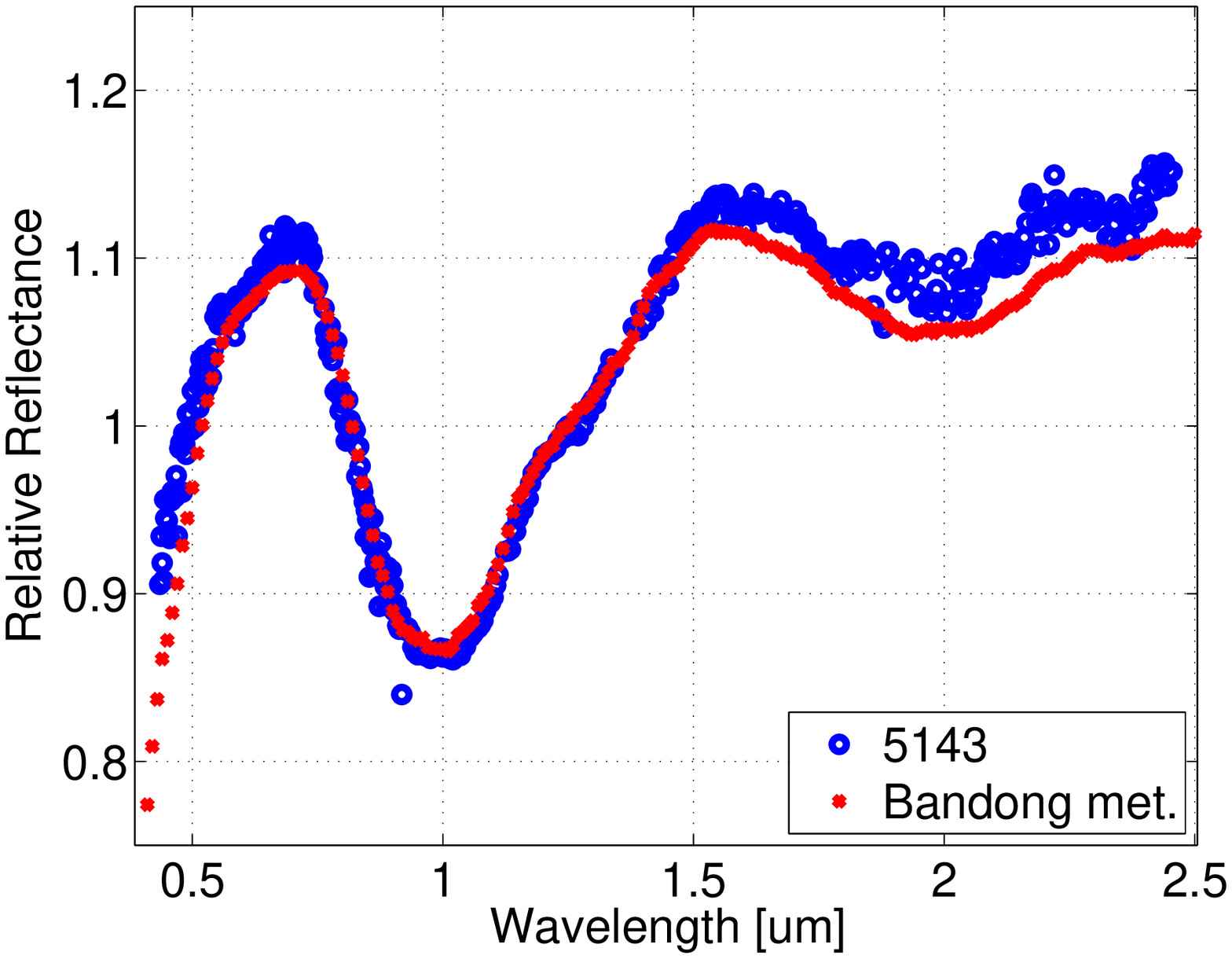}
\subfloat[ ]{\label{fig:5143b2met}}\includegraphics[width=6cm]{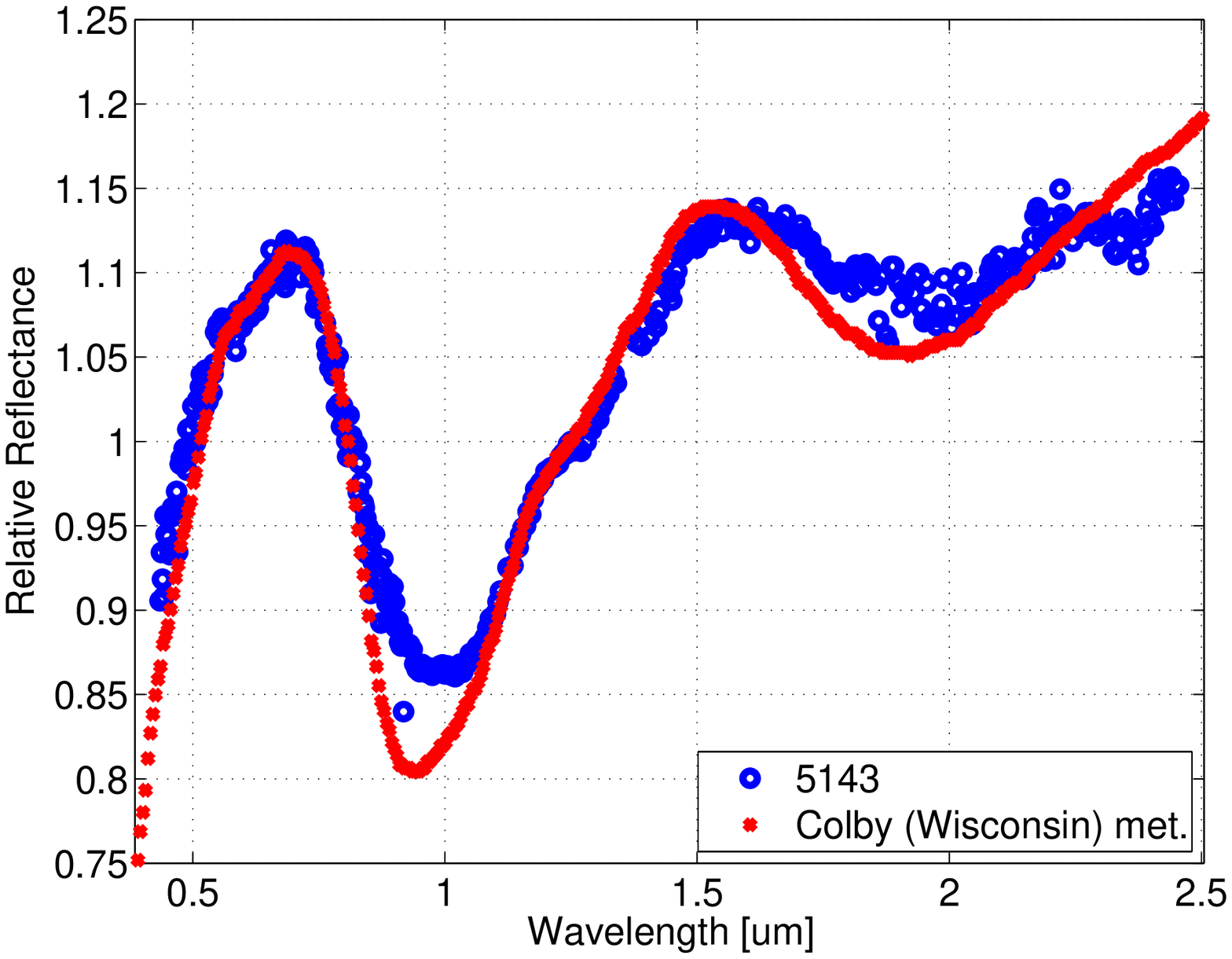}
\subfloat[ ]{\label{fig:5143c2met}}\includegraphics[width=6cm]{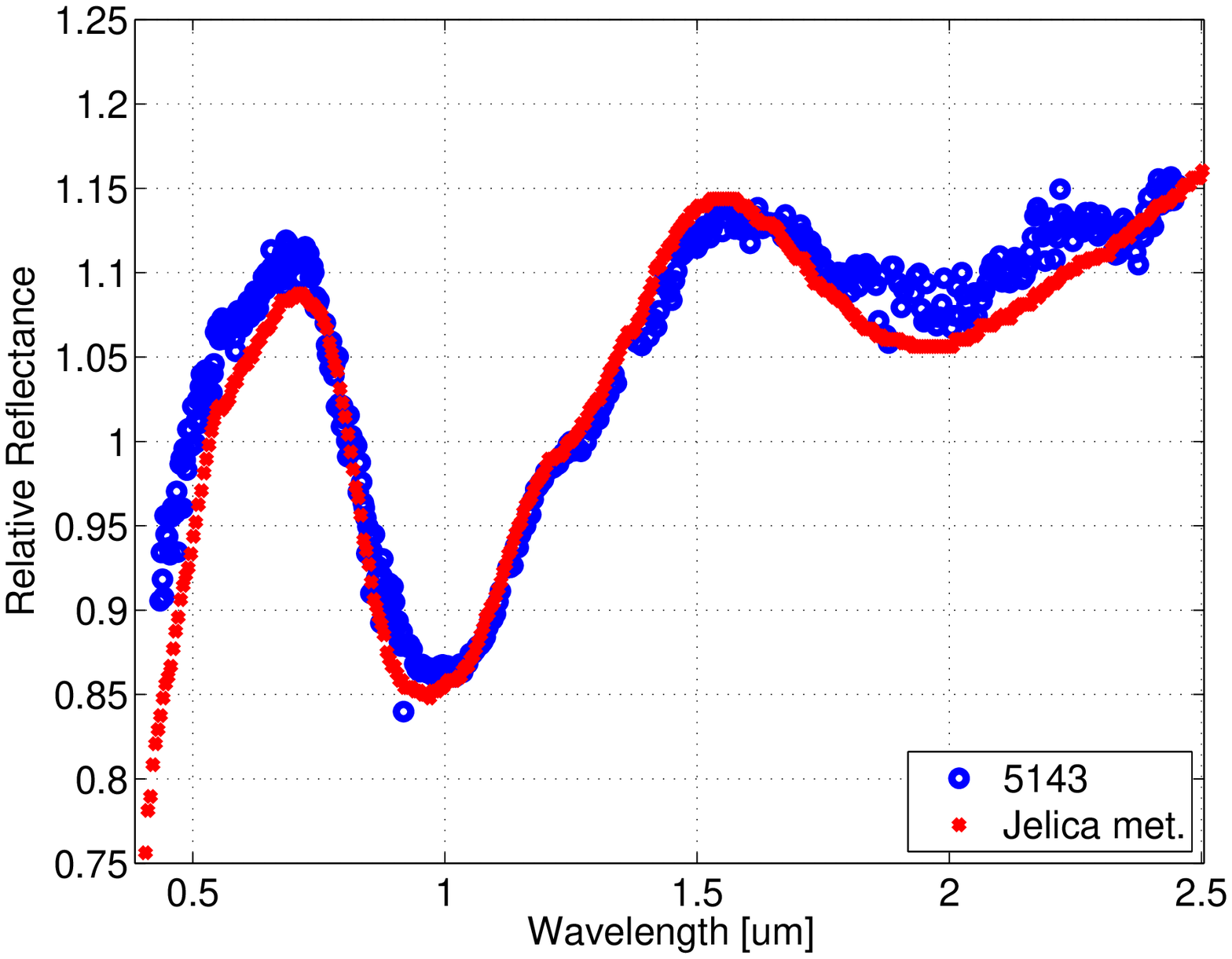}
\end{center}
\caption{Reflectance spectrum of (5143) Heracles and the closest three matches resulting from meteorite comparison: (a) LL6 ordinary chondrite Bandong (Sample ID: TB-TJM-067); (b)  L6 ordinary chondrite Colby (Wisconsin) (Sample ID: MR-MJG-057); (c) LL6 ordinary chondrite Jelica(Sample ID: MR-MJG-072).}
\label{HeraclesMet}
\end{figure*}

\begin{figure*}[!ht] 
\begin{center}
\centering
\subfloat[ ]{\label{fig:6063sepa2met}}\includegraphics[width=6cm]{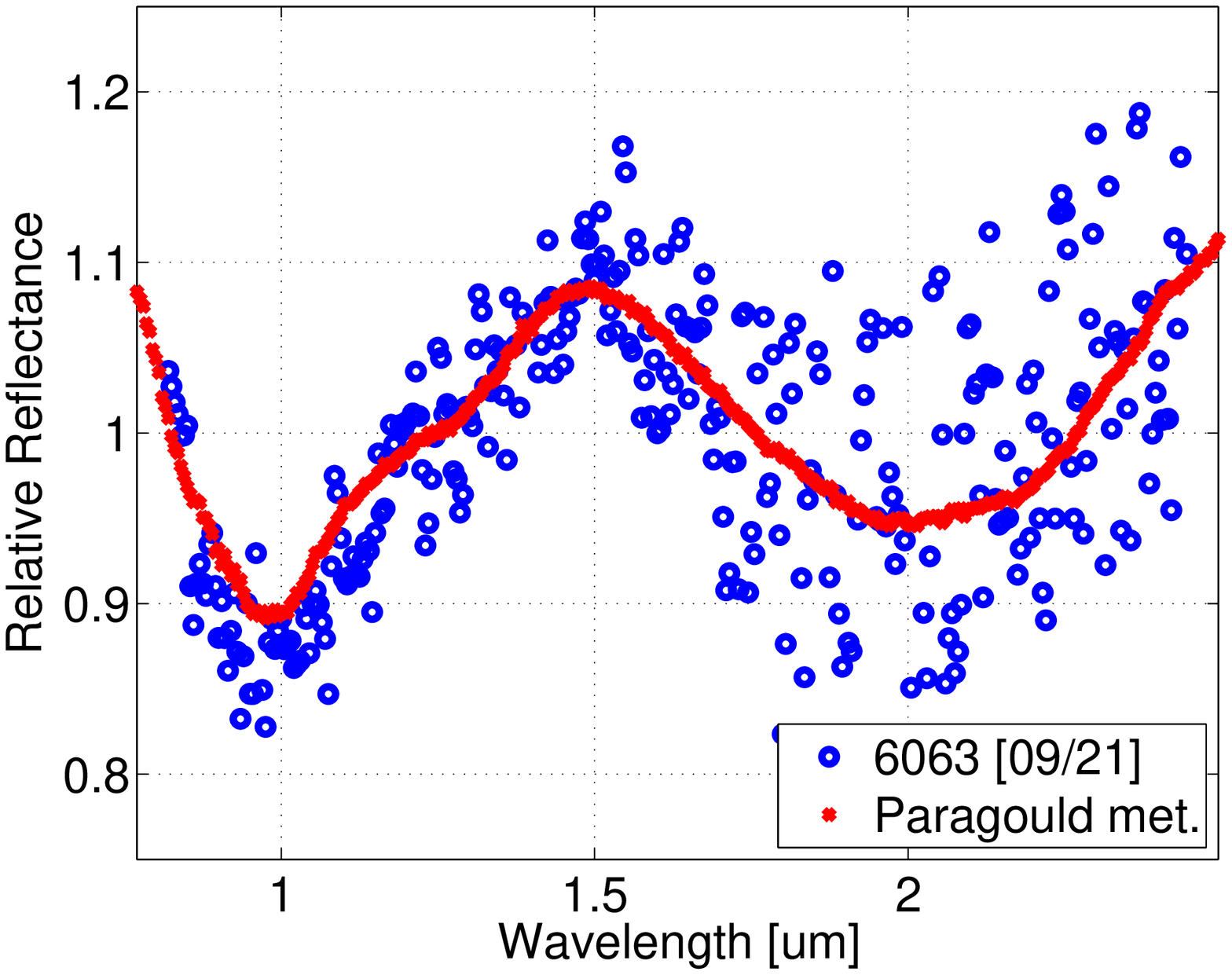}
\subfloat[ ]{\label{fig:6063sepb2met}}\includegraphics[width=6cm]{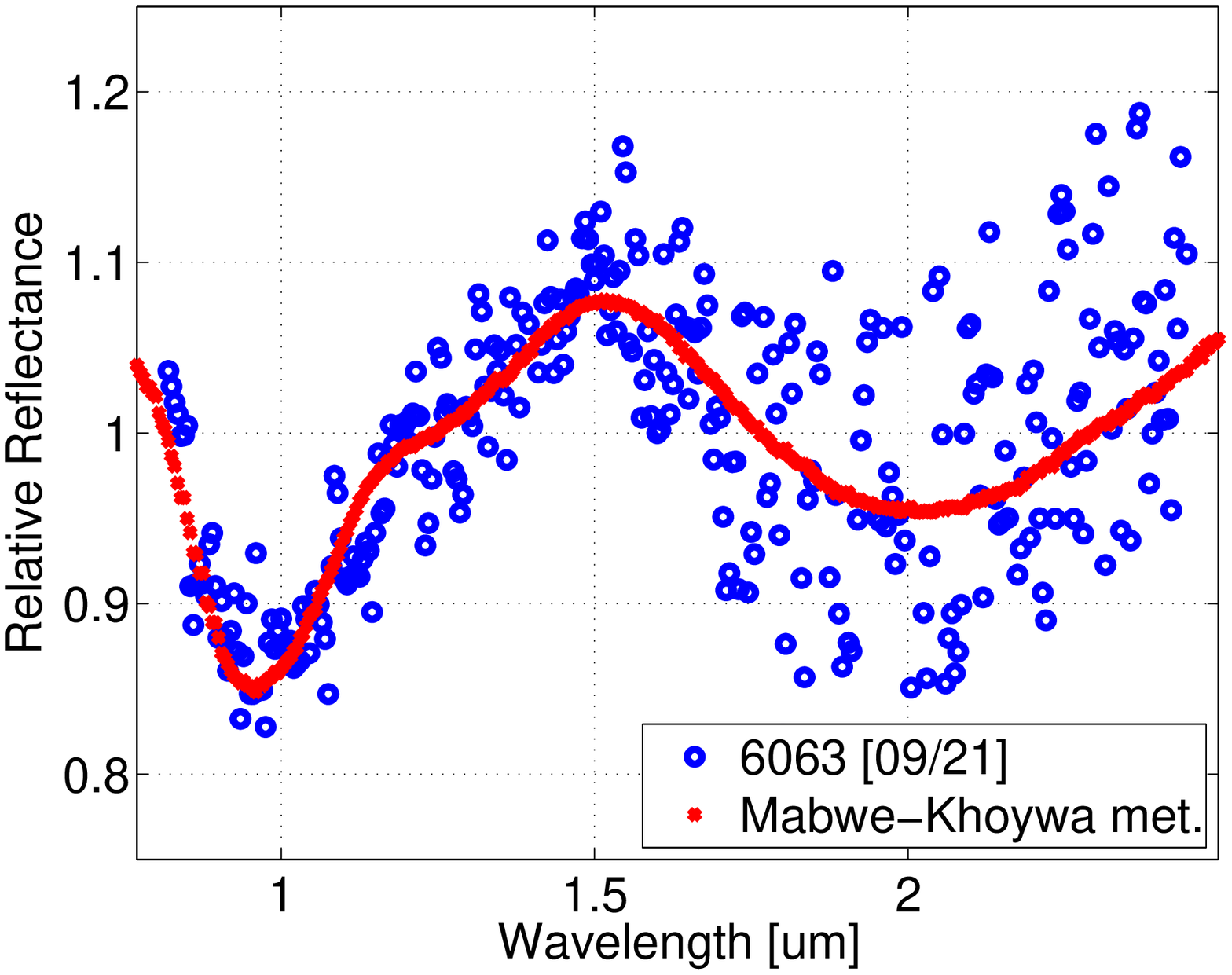}
\subfloat[ ]{\label{fig:6063sepc2met}}\includegraphics[width=6cm]{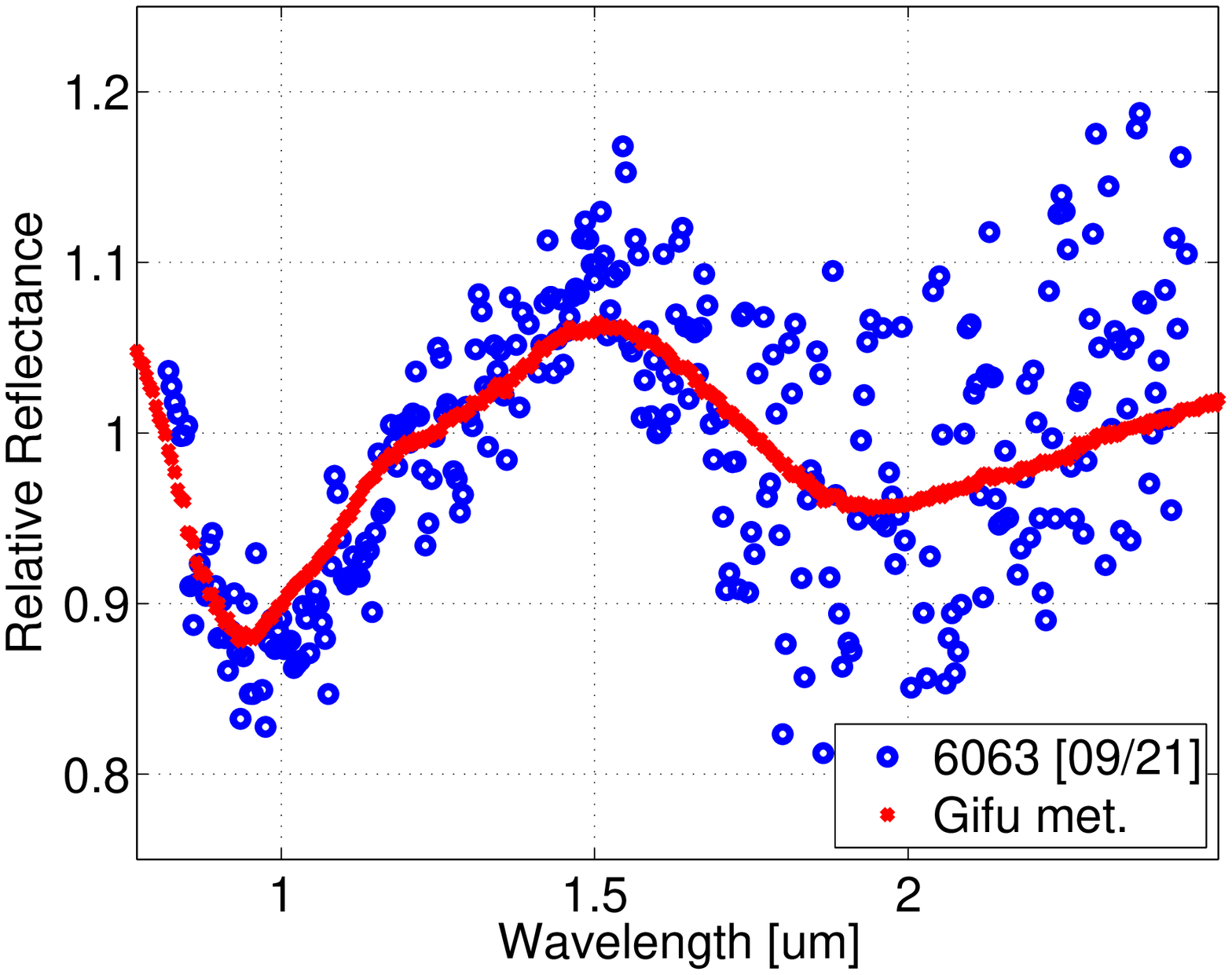}
\end{center}
\caption{Reflectance spectrum of (6063) Jason (spectrum obtained on September 21, 2013)  and the closest three matches resulting from meteorite comparison: (a) LL5 ordinary chondrite Paragould(Sample ID: MB-CMP-002-L); (b)  L5 ordinary chondrite Mabwe-Khoywa (Sample ID: TB-TJM-107); (c) L6 ordinary chondrite Gifu(Sample ID: MH-JFB-022).}
\label{Jason1Met}
\end{figure*}

\begin{figure*}[!ht] 
\begin{center}
\centering
\subfloat[ ]{\label{fig:6063octa2met}}\includegraphics[width=6cm]{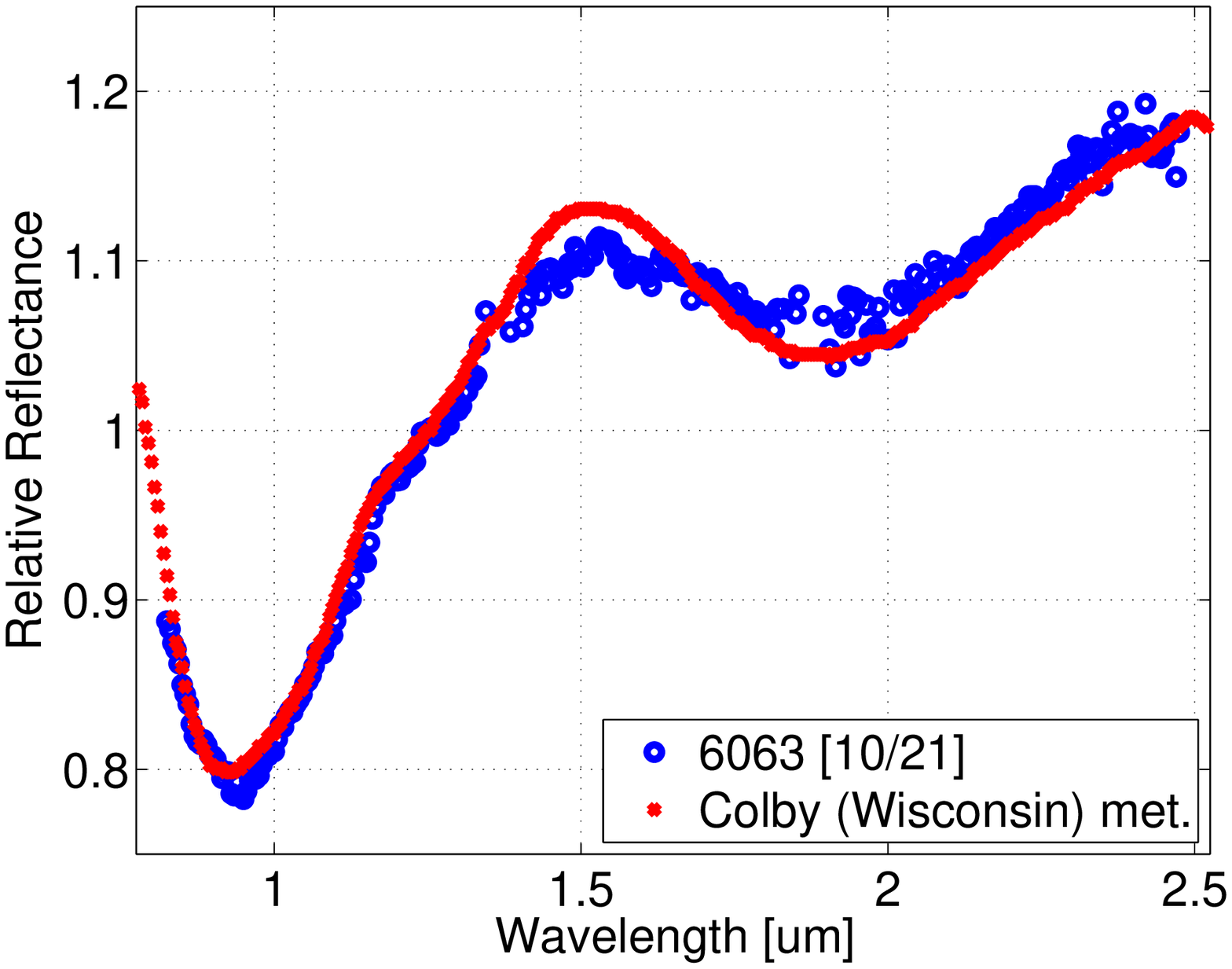}
\subfloat[ ]{\label{fig:6063octb2met}}\includegraphics[width=6cm]{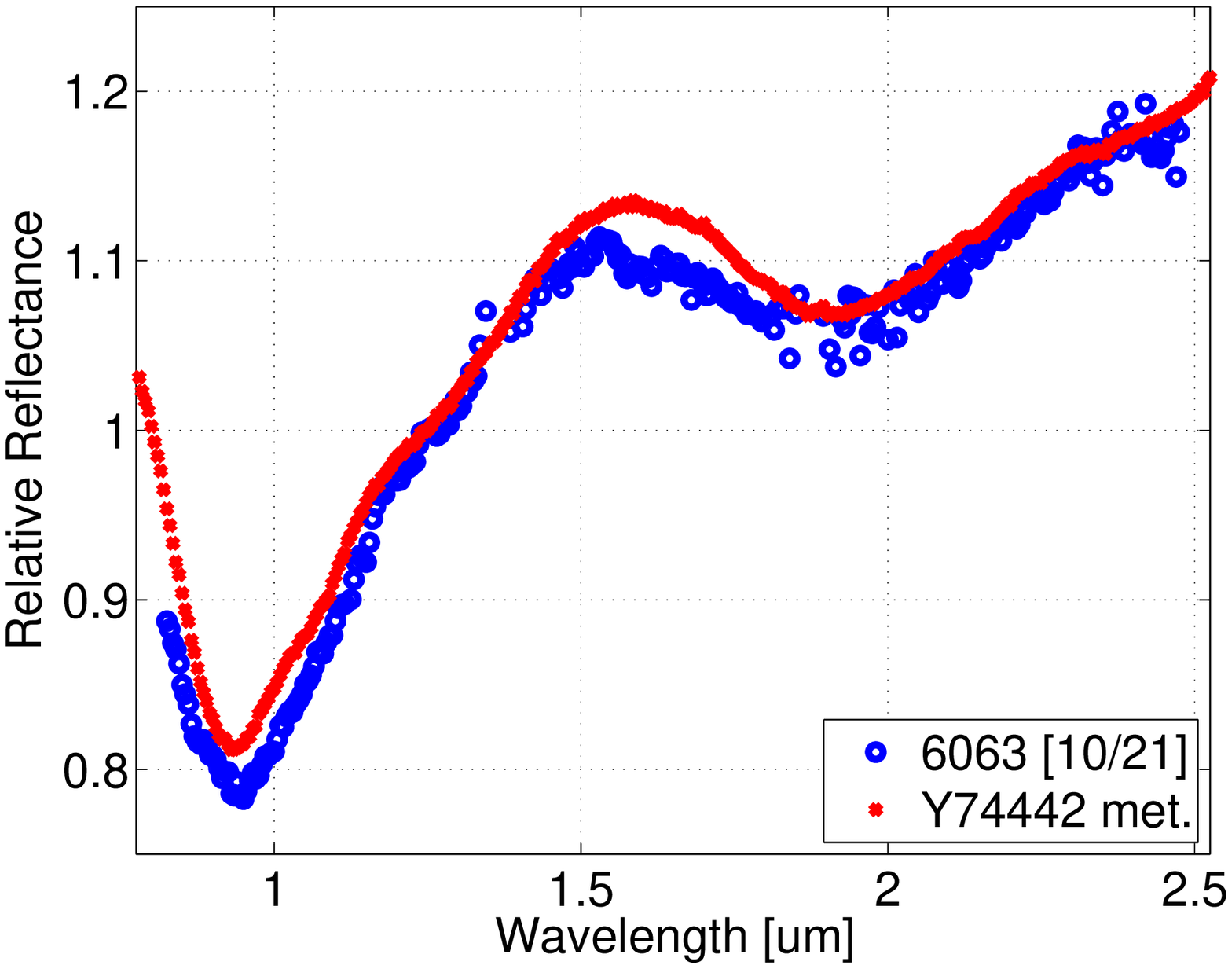}
\subfloat[ ]{\label{fig:6063octc2met}}\includegraphics[width=6cm]{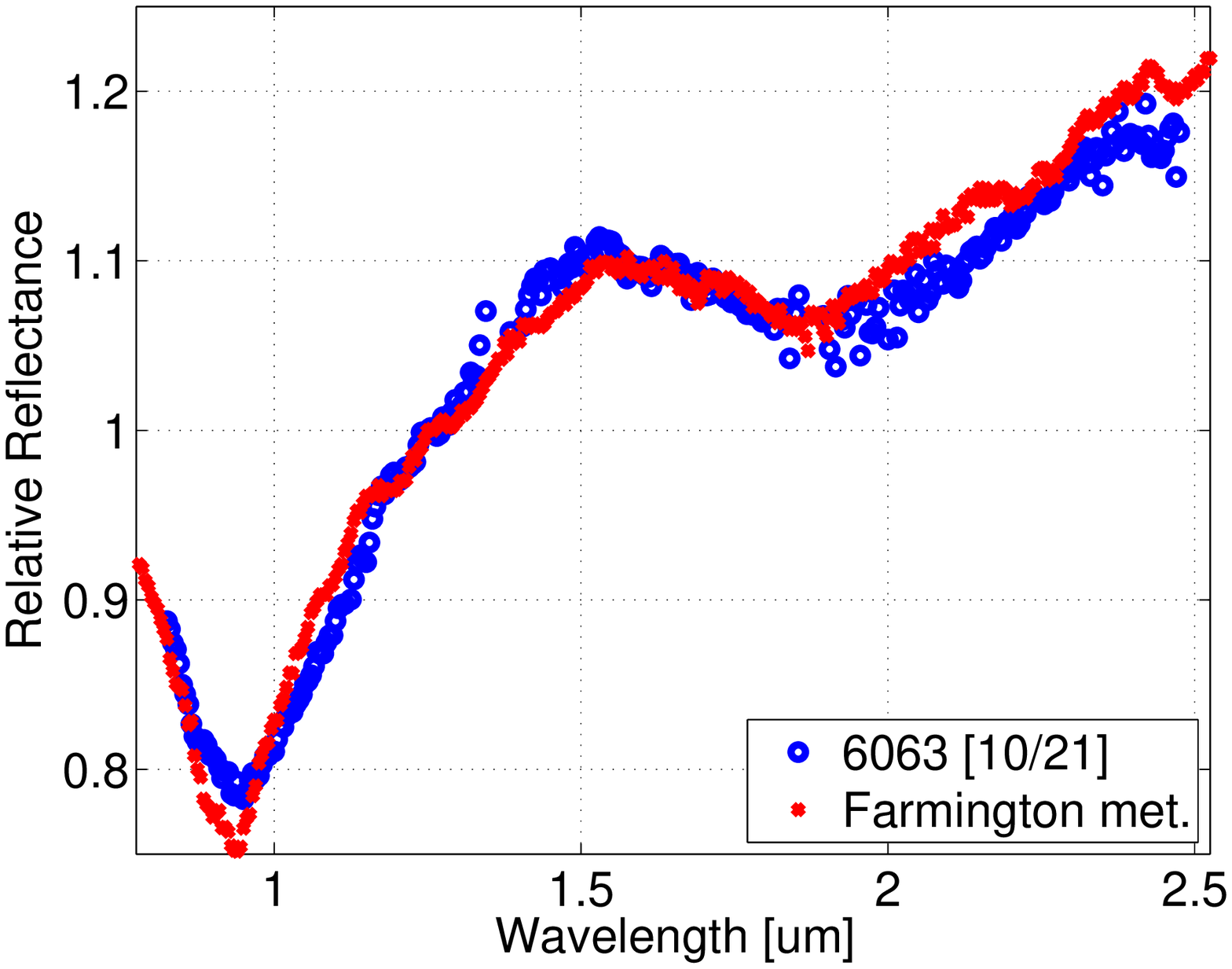}
\end{center}
\caption{Reflectance spectrum of (6063) Jason (spectrum obtained on October 21, 2013) and the closest three matches resulting from meteorite comparison: (a) L6 ordinary chondrite Colby (Wisconsin) (Sample ID: MR-MJG-057); (b)  LL4 ordinary chondrite Y74442 (Sample ID: MB-TXH-086-A); (c) L5 ordinary chondrite Farmington (Sample ID: MH-CMP-003).}
\label{Jason2Met}
\end{figure*}

\begin{figure*}[!ht] 
\begin{center}
\centering
\subfloat[ ]{\label{fig:269690a2met}}\includegraphics[width=6cm]{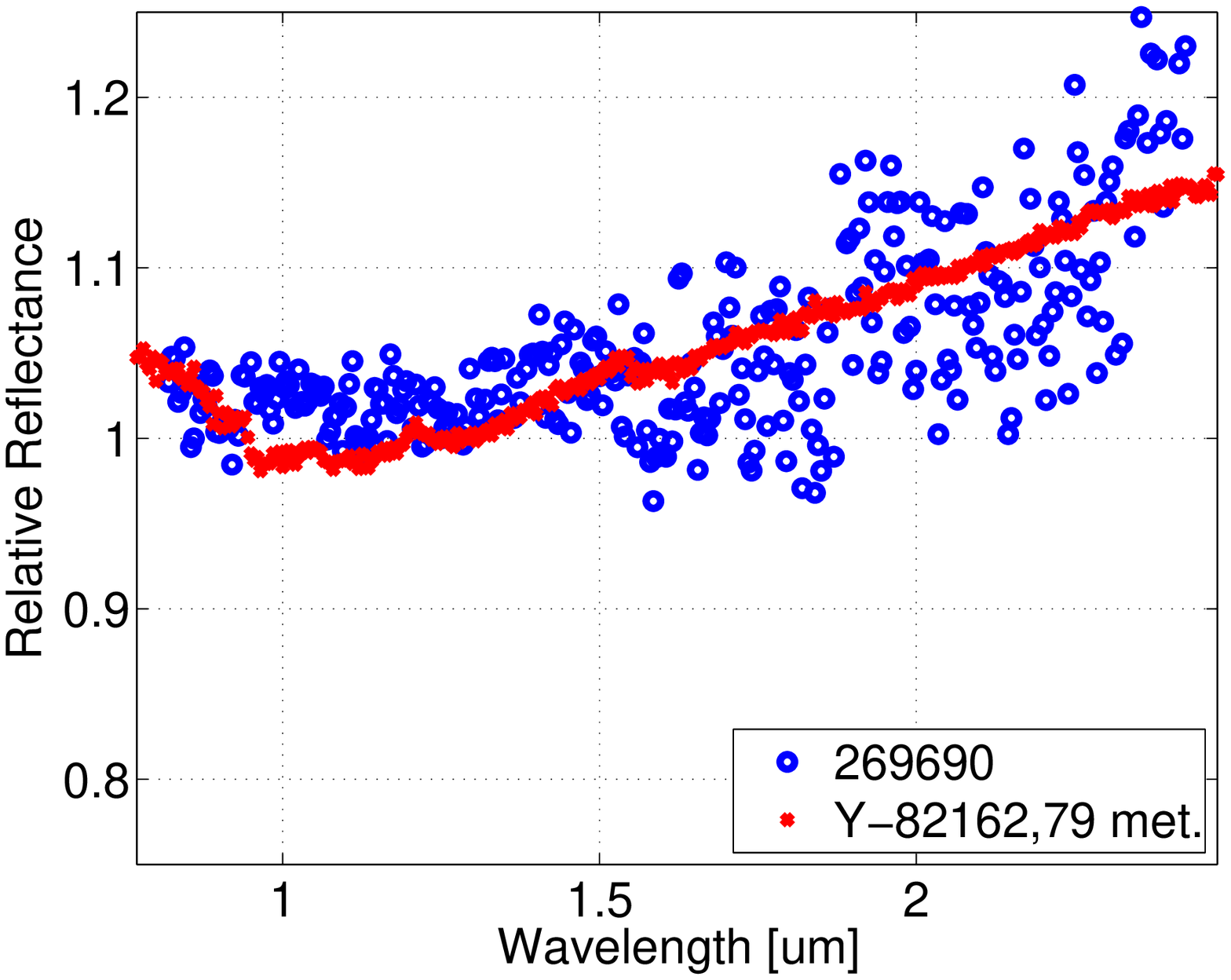}
\subfloat[ ]{\label{fig:269690b2met}}\includegraphics[width=6cm]{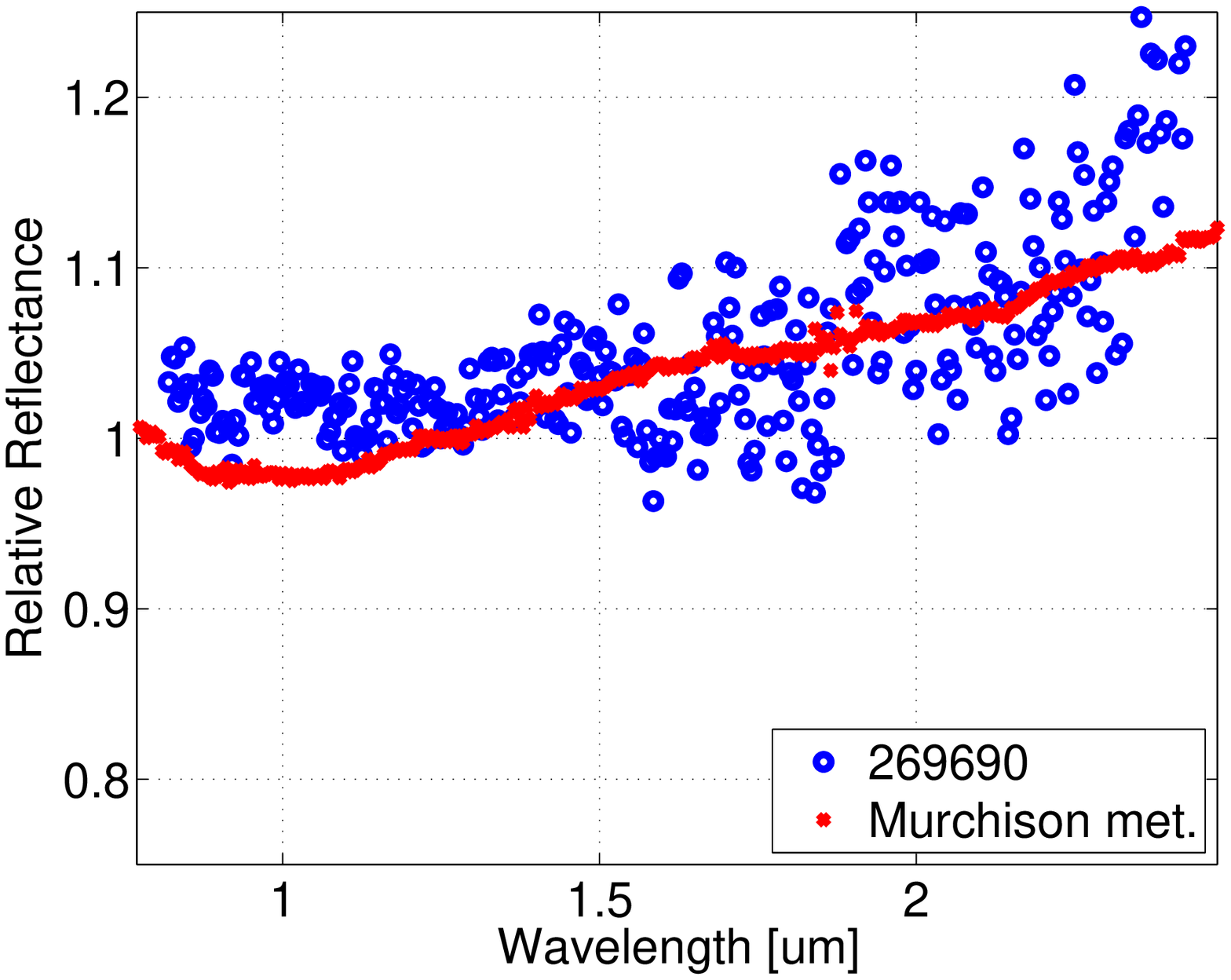}
\subfloat[ ]{\label{fig:269690c2met}}\includegraphics[width=6cm]{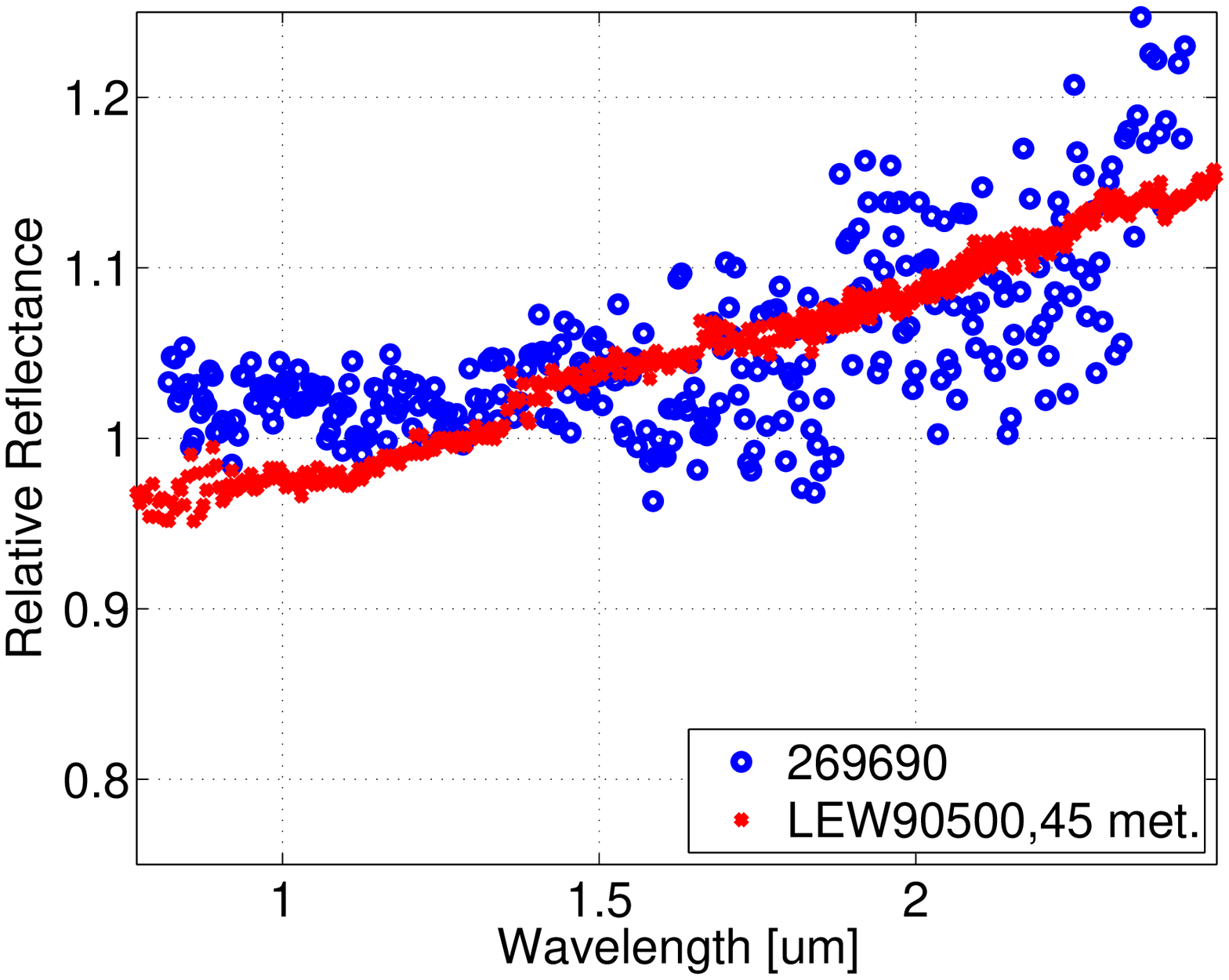}
\end{center}
\caption{Reflectance spectrum of (269690) 1996 RG3 (spectrum obtained on September 21, 2013) and the closest three matches resulting from meteorite comparison: (a) CI Unusual carbonaceous chondrite Y-82162,79 <125 um (Sample ID: MB-CMP-019-1); (b)  CM2 carbonaceous chondrite  Murchison heated at 900C (Sample ID: MB-TXH-064-F); (c) CM2 carbonaceous chondrite LEW90500,45 (Sample ID: MC-RPB-002).}
\label{FG3Met}
\end{figure*}

\clearpage

\section{The parameters of some of the spectra available in the literature}
\label{Anexa2}


\begin{table*}[h]
\caption{Comparison between several visible spectra available in the literature for the NEAs studied in this paper. The asteroid number, the article reference, the observation date, the wavelength interval ($\lambda_{min}$ and $\lambda_{max}$), the maximum position, the slope (computed in the spectral interval 0.55 - 0.9 $\mu m$), the taxonomic classification, and the similarity factor in the common interval with our NIR spectrum are shown.}
\label{LiteratureSpectra}
\centering 
\begin{tabular}{l l c c c c c c c c}
\hline \hline
Object & Reference & Obs. date & $\Phi[\degr]$ & $\lambda_{min}$ & $\lambda_{max}$ & Max. position. & Slope & Tax. & Similarity\\ \hline 
(2201) & \citeads{1996Icar..122..122L} & Oct. 1994 & 7 & 0.369 & 0.972 & 0.659 & -0.364 & - & 0.371 \\ 
       & \citeads{1998Icar..133...69H} & Nov. 1995 & 16 & 0.524 & 0.947 & 0.715 & -0.143 & E & 0.124 \\
       & \citeads{2004Icar..170..259B} & Dec. 1995 & 27 & 0.435 & 0.925 & 0.724 & -0.307 & Sq & 0.013 \\ \hline       
(4183) & \citeads{2004Icar..170..259B} & multiple  & 17,38 & 0.435 & 0.925 & 0.710 & -0.327 & Sq & 0.084 \\ 
       & \citeads{2007Icar..188..175F} & multiple  & 5;22;44 & 0.456 & 0.966 & 0.682 & -0.372 & - & 0.918 \\ \hline       
(5143) & \citeads{2004Icar..170..259B} & Apr. 1997 & 13 & 0.435 & 0.917 & 0.698 & -0.485 & O & 0.135  \\
       & \citeads{1995Icar..115....1X} & Dec. 1991 & 37 & 0.494 & 0.994 & 0.721 & -0.023 & V & -0.301 \\
       & \citeads{2004Icar..169..373L} & - & - & 0.400 & 0.955 & 0.723 & -0.253 & Sk & 0.301\\
\hline
\end{tabular}
\end{table*}

\end{appendix}
\end{document}